# Entwicklung eines Reputationssystems für cyber-physikalische Systeme am Beispiel des inHMotion Forschungsprojektes

Christoph Ponikwar

MASTERARBEIT

eingereicht im
Masterstudiengang

Informatik

an der Hochschule für angewandte Wissenschaften

in München

im Oktober 2015

Diese Arbeit entstand im Rahmen des Gegenstands

Forschungsprojekt inHMotion

im

6. Masterstudiensemester

Betreuer:

Prof. Dr. Hans-Joachim Hof

089 1265 3752
hof@hm.edu
http://muse.bayern/

Zweitprüfer:

Prof. Dr. Christoph Pleier

089 1265 3716
pleier@cs.hm.edu
http://cs.hm.edu/~pleier

# Erklärung

Hiermit erkläre ich, dass ich die Masterarbeit selbständig verfasst, noch nicht anderweitig für Prüfungszwecke vorgelegt, keine anderen als die angegebenen Quellen oder Hilfsmittel benutzt sowie wörtliche und sinngemäße Zitate als solche gekennzeichnet habe.

München, am 30. Oktober 2015

Christoph Ponikwar, 17.04.1989, IG 6, WS2015/16

# Kurzfassung

Durch die stetig fortschreitende Vernetzung, durch Fortschritte in der Technik und durch Initiativen wie Industrie 4.0, SmartGrid, SmartCities, werden Systeme, die bisher weitestgehend autark sind, zu vernetzen Cyber-physical System (CPS). Die Vernetzung birgt nicht nur Vorteile, sondern erleichtert und verstärkt mögliche Angriffe auf Systeme. Eines der grundlegenden Probleme in der Informationstechnologie ist der Aufbau und Erhalt von Vertrauen in diese Systeme. Diese Arbeit befasst sich mit dem Einsatz von Reputationssystemen im Rahmen von CPS, da diese ein Weg sind, um Vertrauenswürdigkeit eines Kommunikationspartners zu bewerten. Dies geschieht am Beispiel des Forschungsprojektes inHMotion der Hochschule München, das im Bereich von Intelligent Transportation System (ITS) forschend tätig ist. Für dieses Projekt wird eine risikobasierte Bedrohungsmodellierung durchgeführt. Basierend auf den erhobenen Bedrohungen werden geeignete Gegenmaßnahmen vorgeschlagen. Einen möglichen Lösungsansatz stellt der Einsatz eines Reputationssystems dar, dieses wird entworfen und prototypisch für eine Simulation implementiert. Mit dem Konzept des Meta-Reputationssystems, das eine Verbindung von bereits existierenden Reputationssystemansätzen und verschiedensten Verifizierungstechniken ermöglicht, ist es möglich, in einer einfachen Simulation erfolgversprechende Resultate zu erzielen. Verbesserungsmöglichkeiten und Kritikpunkte des verfolgten Ansatzes werden abschließend diskutiert.

# Inhaltsverzeichnis

# Akronyme

**3GPP** 3rd Generation Partnership Project. viii, 29, *Glossar:* 3rd Generation Partnership Project

**ACC** Adaptive Cruise Control. viii, 70, *Glossar:* Adaptive Cruise Control

**ASLP** Address Space Layout Permutation. 48

**ASLR** Address Space Layout Randomization. 48

**AuC** Authentication Center. 93

**BDSG** Bundesdatenschutzgesetz. 49, 93

**BNetzA** Bundesnetzagentur. 50

**BTP** Basic Transport Protocol. viii, 7, *Glossar:* Basic Transport Protocol

**CA** Certificate Authority. xi, 46

**CPS** Cyber-physical System. iv, viii, xi, 2, 15, 16, 37, 56, 59, 64, *Glossar:* Cyber-physical System

**CSMA/CA** Carrier Sense Multiple Access / Collision Avoidance. viii, 50, *Glossar:* Carrier Sense Multiple Access / Collision Avoidance

**DoS** Denial of Service. 16, 17, 38, 49, 88

**DSRC** Dedicated Short-Range Communications. viii, *Glossar:* Dedicated Short-Range Communications

**ETSI** European Telecommunications Standards Institute. x, xi, 7

**FCC** Federal Communications Commission. 50

**GeoNet** GeoNetworking. viii, x, 7, *Glossar:* GeoNetworking

**HSM** Hardware Security Module. 93

**IEEE** Institute of Electrical and Electronics Engineers. x, xii, 7, 54

**IP** Internet Protocol. viii, 13, 52, 55, 59, *Glossar:* Internet Protocol

**IPv4** Internet Protocol Version 6. xi, *Glossar:* Internet Protocol

**IPv6** Internet Protocol Version 6. xi, 7, *Glossar:* Internet Protocol

**ITS** Intelligent Transportation System. iv, ix, xii, 2, 15, 29, 37, 67, *Glossar:* Intelligent Transportation System

**LI** Lawful Interception. ix, 29, 106, *Glossar:* Lawful Interception

**MAC** Media Access Control. 13, 59

**MANET** Mobile ad hoc network. 15, 56, 64

**ONP** Overlay Network Protocol. ix, 3, 7, 24–26, 28, 29, 52, 56, 70–86, *Glossar:* Overlay Network Protocol

**OWASP** Open Web Application Security Project. ix, 10, 12, 39, 41, 43, 67, *Glossar:* Open Web Application Security Project

**P2P** Peer-to-Peer. ix, 5, *Glossar:* Peer-to-Peer

**PDU** Packet Data Unit. 54, 58

**PKI** Public Key Infrastructure. ix, 13, 46, *Glossar:* Public Key Infrastructure

**RSU** Roadside Unit. ix, 7, 19, 29, 54, 58, 62, 65, 71, 75, 86, *Glossar:* Roadside Unit

**SAP** Service Access Point. 52

**TCP** Transmission Control Protocol. ix, *Glossar:* Transmission Control Protocol

**TLS** Transport Layer Security. ix, 46, *Glossar:* Transport Layer Security

**UDP** User Datagram Protocol. ix, x, 7, *Glossar:* User Datagram Protocol

**VANET** Vehicular ad hoc network. 13, 15, 16, 18–20, 53, 54, 56, 59, 64, 66

**WAVE** Wireless Access for Vehicular Environments. ix, *Glossar:* Wireless Access for Vehicular Environments

**WoT** Web of Trust. 13

**WSMP** WAVE Short-Message Protocol. ix, 7, *Glossar:* WAVE Short-Message Protocol

# Glossar

**3rd Generation Partnership Project** Organisation zu Standard Entwicklung im Telekommunikationsbereich. viii, 29

**Adaptive Cruise Control** Eine adaptive Geschwindigkeitsregelung, die Abstände zu vorausfahrendem und nachfolgendem Fahrzeug mit in die Geschwindigkeitsregelung mit einbezieht, wird daher auch als Abstandsregeltempomat bezeichnet. viii

**Basic Transport Protocol** Verbindungsloses Transportprotokoll, ähnlich User Datagram Protocol (UDP), verwendet GeoNetworking (GeoNet) um seine Transportdienst zu erbringen. Spezifiziert in European Telecommunications Standards Institute (ETSI) EN 302 636-5-1 [24].. viii, 7

**Carrier Sense Multiple Access / Collision Avoidance** Ein Protokoll das spezifiziert wie auf eine Übertragungsmedium zugegriffen werden kann. Carrier Sense Multiple Access / Collision Avoidance (CSMA/CA) spezifiziert, dass ein Sender auf dem Medium zuerst lauschen muss ob ein belegt oder frei ist und nur wenn es frei ist beginnt er zu senden. Falls es belegt sein sollte wird zur Kollisionsvermeidung ein zufälliges Zeit Intervall lang gewartet, bevor eine erneuter Übertragungsversuch durchgeführt wird.. viii, 50

**Cyber-physical System** Computer Systeme die einen Regelkreislauf mit physikalischen Prozessen bilden. Sensoren liefern Informationen an Computer und diese wiederum benutzen Aktoren um physikalische Prozesse zu steuern.[42]. iv, viii

**Dedicated Short-Range Communications** Generle Bezeichnung für adhoc Kommunikation mit geringer Reichweite, z. B. Bluetooth. In dieser Arbeit ist DSRC gleich zusetzten mit Institute of Electrical and Electronics Engineers (IEEE) 802.11p. Diese Technologie wurde in dem jetzt obsoleten IEEE Std. 802.11p-2010 [34] spezifiziert und als Erweiterung in den IEEE Std. 802.11-2012 [35] integriert.. viii

**GeoNetworking** Protokoll für geografische Adressierung und Routing, mit Unterstützung von 1-zu-1 und 1-zu-n Kommunikation. Spezifiziert in ETSI EN 302 636-3 [25].. viii, x

**Intelligent Transportation System** Systeme die eine Untergruppe der CPS bilden, u. a. im Verkehrsmanagement oder auch im Verkehrsmittel übergreifenden, intermodalen Bereich angesiedelt sind.[43]. iv, ix

**Internet Protocol** Basis Protokoll der Internet Architektur dient zur Adressierung von Teilnehmern. Existiert in erster Version als Internet Protocol Version 6 (IPv4) und als dessen Nachfolger Internet Protocol Version 6 (IPv6).. viii, xi, 13

**Lawful Interception** Gesetzeskonformes Abgreifen von Daten einer Kommunikationsverbindung. ix, 29

**Open Web Application Security Project** Das Open Web Application Security Project (OWASP) ist ein global organisiertes, freies und offenes Projekt, dass sich der Verbesserung der Sicherheit von Anwendungssoftware verschrieben hat. Alle Materialien werden unter freien Open Source Lizenen veröffentlicht. Jeder der möchte kann an diesem Projekt mitarbeiten. Finanziell wird das Projekt durch die OWASP Foundation getragen. Diese ist eine amerikanische 501(c)(3) gemeinnützige Organisation mit dem Hauptsitz in Bel Air, Maryland.. ix, 10

**Overlay Network Protocol** Kurzbezeichnung für das Overlay Netzwerk Protokoll von inHMotion. ix, 3

**Peer-to-Peer** Architektur für verteilte System oder Anwendungen, bei strenger Umsetzung interagieren gleichwertige Teilnehmer untereinander. Auch hierarchischer Aufbau möglich z. B. mit sogenannten Superpeers als Vermittler zwischen den Teilnehmern.. ix, 5

**Public Key Infrastructure** Ein System aus unterschiedlichen Komponenten z. B. Regeln, Soft- oder Hardware, um unter Zuhilfenahme von Kryptographie einer Nutzer Identität, öffentliche Schlüssel zuordnen zu können. Dies geschieht durch Zertifikate und Zertifizierungsstellen, einer sogenannten Certificate Authority (CA).. ix, 13

**Roadside Unit** Ein System das, ähnlich einem drahtlosen Access Point, Kommunikation als Infrastukturdienstleistung anbietet.. ix, 7

**Transmission Control Protocol** Verbindungsorientiertes Transportprotokoll. ix

**Transport Layer Security** Transport Layer Security (TLS) ist ein kryptographisches Protokoll das entworfen wurde um sichere Kommunikation in Computer Netzwerken zu ermöglichen. Spezifiziert im Jahre 1999 [7] erweitert in 2008 [20] und zuletzt aktualisiert 2011 [73].. ix, 46

**User Datagram Protocol** Verbindungsloses Transportprotokoll. ix, x

**WAVE Short-Message Protocol** Kurznachrichten Protokoll für den schnellen Austausch von Informationen über sich ändernde Funkfrequenzen. Spezifiziert in IEEE 1609.3 [37].. ix, 7

**Wireless Access for Vehicular Environments** Ein Ansammlung von Standards IEEE Std. 1609.2-2013[3], IEEE Std. 1609.3-2010[37], IEEE Std. 1609.4-2010 [4], IEEE Std. 1609.11-2010[1], IEEE Std. 1609.12-2012[2] und einem unveröffentlichten IEEE P1609.6®, die in Gesamtheit ein drahtloses Kommunikationssystem für ITS ermöglichen sollen.. ix

# 1. Einleitung

Warten im Stau kostet Nerven, Geld und schadet der Umwelt. Ein Dienst, der seinen Nutzern vorhersagen kann welches Verkehrsmittel individuell unter verschiedenen Gesichtspunkten ideal für die Nutzung ist, könnte dies verhindern. Dabei muss der Benutzer die Möglichkeit haben seinen Ziel-, Abfahrtsort und weitere Parameter, z. B. Ankunftszeit, Kosten oder Grad der Entspannung, in die Empfehlungsfindung einfließen zu lassen.
Ziel dieser Arbeit ist es daher ein System zu entwickeln, das besonders gut auf Anforderungen von cyber-physikalischen Systemen im Bereich Mobilität und Transport abgestimmt ist und durch das mögliche Probleme und Herausforderungen in diesem Bereich erfolgreich gelöst werden.
Da an der Hochschule München bereits ein Projekt zu diesem Themengebiet existiert, wird das Projekt dieser Arbeit zu Grunde gelegt. Es handelt sich dabei um das Projekt inHMotion, das für den Verkehrsteilnehmer u. a. intermodale Kommunikation und eine Bereitstellung von Handlungsempfehlungen zu Verkehrsmitteln ermöglichen soll.
Zunächst werden die Grundlagen im folgenden Kapitel erläutert. Aktuell befindet sich das System des inHMotion Projektes noch in der Entwicklung, deshalb werden Anforderungen in Form von Anwendungsfällen spezifiziert und dokumentiert. Aufgrund der Stofffülle wird hierbei nur ein Anwendungsfall ausführlich erläutert, weitere Fälle können im Anhang der Arbeit nachgelesen werden. Aus diesen Anwendungsfällen wird ein vorläufiger Architekturentwurf als Grundlage für diese Arbeit abgeleitet.
Um das System weiterentwickeln zu können und mögliche Sicherheitslücken bereits während dem Entwurf abdecken zu können, wird im nächsten Schritt eine Bedrohungsanalyse nach Prof. Dr. Sachar Paulus durchgeführt. Aus dieser Analyse ergeben sich Bedrohungen und Sicherheitsanforderungen 4.8, die abgewägt werden und so Aufschluss darüber geben, welche Art System am besten geeignet ist. Ein solches System wird im Hauptteil der Arbeit unter Bestimmung der Voraussetzungen entworfen und simuliert. Abschließend zeigt ein kurzer Ausblick wo noch Raum für weitere wissenschaftliche Forschungen ist.

# 2. Grundlagen

In diesem Kapitel werden das Forschungsprojekt inHMotion, das zu entwickelnde System und das weitere Umfeld des Systems vorgestellt. Ebenso werden Methoden und Herangehensweisen, die in dieser Arbeit verwendet werden, erörtert.

## 2.1 Das inHMotion Forschungsprojekt

Im Jahr 2012 haben die privaten Haushalte in der EU ca. 13.0% [23] ihrer gesamten Ausgaben im Zusammenhang mit Beförderung ausgegeben. Innerhalb der EU lag 2012 die Nutzung von Straßen für den Gütertransport bei 44.9% [23]. Diese Zahlen verdeutlichen, dass das Personen- und Lastentransportwesen ein wichtiger Bestandteil unserer arbeitsteiligen Gesellschaft ist. Güter müssen von den Produzenten zu ihren Abnehmern befördert werden, Arbeitskräfte an ihre Arbeitsstätte. Ineffizienter Verkehr belastet die Umwelt, verschwendet endliche Ressourcen, wie z. B. Treibstoff oder Zeit. Einen besseren Einsatz dieser Ressourcen könnten durch ITS erreicht werden. Die ITS sind eine Untergruppe der CPS. Das inHMotion Projekt ist, zum aktuellen Zeitpunkt, ein Zusammenschluss von vier Professoren an der Hochschule München: Prof. Dr.-Ing. Lars Wischhof, Prof. Dr.-Ing. Hans-Joachim Hof, Prof. Dr. Gerta Köster und Prof. Dr.-Ing. Peter Mandl. Es besteht thematisch aus vier Fragestellungen bzw. Teilprojekte (TP), übernommen aus dem Projektantrag:

- TP1 **Empfehlungssystem**
  Basistechnologien für intermodale Reiseempfehlungen, Beispielapplikation
- TP2 **Overlay-Netzwerk für die Informationsverbreitung**
  Skalierbare Datenverarbeitungsmechanismen über zellulare und Ad-hoc-Netze
- TP3 **IT-Sicherheit**
  Schutz des Systems gegen Missbrauch, Vertrauensmodell
- TP4 **Modellierung und Simulation intermodalen Reisens**

Kombination von Fußgänger- und Fahrzeugbewegungssimulation

Das Overlay Network Protocol (ONP), TP2, bildet die Basis des Systems. Es soll ein Übergangsnetz sein, das von der pragmatischen Welt heute mit den unterschiedlichen Systemen und Technologien zu einer idealen, ad-hoc-kommunikationsfähigen Welt von morgen führen soll. Die Markteinführungshürde für reine Ad-hoc-Ansätze soll durch dieses Netz genommen werden, indem es auch ohne Ad-hoc-Kommunikation funktionieren wird. Ebenso muss der andere Extremfall von reiner Ad-hoc-Kommunikation problemlos funktionieren. Intermodalität soll durch ONP gefördert werden, dazu soll es Plattform unabhängig sein, zudem soll es u. a. auf einem Smartphone, einem Auto, einer Bahn oder einem Verkehrsleitsystem einsetzbar sein. Für Entwickler soll ONP die Komplexität der zugrundeliegenden Technologien reduziert werden, u. a. soll ein Entwickler nicht entscheiden müssen welches Netzwerk bzw. Medium für eine bestimmte Kommunikation genutzt werden soll. Die Informationen sollen in ONP mit Ortsbezug „leben", das heißt Daten sollen gespeichert und wenn nützlich weitergegeben werden (Store-and-Forward Prinzip). Durch situationsbezogene Aggregation und Überlastkontrolle (Congestion Control) soll Skalierbarkeit vorhanden sein. Eine erfolgreiche Markteinführung würde verhindert werden, wenn z. B. jeder Autohersteller das ONP nur als Basis nehmen würde und spezielle Erweiterungen vornehmen würde, so dass andere Teilnehmer das Protokoll nicht mehr verstehen können. Es ist sehr wichtig, dass das Protokoll herstellerübergreifend ist und herstellerspezifische Dienste auf der Anwendungsebene abgewickelt werden, nicht im Protokoll.

Das Empfehlungssystem für intermodale Reiseempfehlungen, TP1, ist stark abhängig von den Fähigkeiten des ONP. Der Fokus dieser Arbeit liegt auf der Sicherheit des ONP, TP3, das Empfehlungssystem wird an geeigneten Stellen gesondert betrachtet.

## 2.2 Anwendungsfälle

Im Rahmen des inHMotion Projekts wurde initial eine Geschichte, hauptsächlich durch die Kollegen Dan Lüdtke und Prof. Dr. Lars Wischhof, entwickelt. Aus dieser Geschichte werden einzelne Anwendungsfälle extrahiert und genauer beschrieben. Die Anwendungsfälle sind frei nach den „Use Case Fundamentals" von Alistair Cockburn [18], [19] modelliert. Ein beispielhafter Anwendungsfall ist in der Tabelle 2.1 zu sehen. Die Anwendungsfälle dienen zum einen dazu eine Systemarchitektur 2.1 entwerfen zu können, die diese erfüllen kann und zum anderen dazu die Interaktionen der Nutzer mit dem System zu verdeutlichen. Der Fokus der Anwendungsfälle liegt auf den Fähigkeiten des Systems (Empfehlungssystem und Overlay Netzwerk), die wiederum gleichzeitig die Möglichkeiten der Nutzer und die Nützlichkeit des Systems darstellen.

| Anwendungsfall | Pünktlich ankommen |
|---|---|
| Ziel: | Pünktlich am Reiseziel ankommen |
| Level: | Systemüberblick |
| Vorbedingung: | Nutzer hat ein Smartphone mit ONP-Software und ein Profil beim Empfehlungsdienst |
| Akteur: | Empfehlungssystemnutzer |
| Ereignis: | Ein Stau hat sich auf der gewöhnlichen Route des Nutzers gebildet |
| Ergebnis: | Nutzer steht früher auf, kommt rechtzeitig an sein Reiseziel |
| Beschreibung: | Der Nutzer wird durch das System früher geweckt als durch den regulären Wecker, da das System einen Stau auf der gewöhnlichen Reiseroute des Nutzers (z.B.: in die Arbeit) erkannt hat. Er kann entscheiden, ob er der Empfehlung folgt oder nicht. Möglich wird die Benachrichtigung, da das System lokale Verkehrsinformationen erhält und die tägliche Routine eines Nutzers erlernt hat oder diese vorgegeben wurde. |
| Fehlerfall: | Falsche Verkehrsinformationen, es existiert kein Stau mehr. |

**Tabelle 2.1:** Beispiel Anwendungsfall

Das Ziel dieses Anwendungsfalles ist es, dass der Nutzer pünktlich an seinem Reiseziel ankommt. Die Betrachtungsweise des Anwendungsfalles ist eine globale Systemsicht, die sich nicht in technischen Details verliert, sondern das System im Überblick betrachtet. Damit dieser Anwendungsfall anwendbar ist, muss der Nutzer ein Smartphone mit der nötigen ONP Software und ein Profil bei dem Empfehlungssystem besitzen, dies macht den Nutzer zu einem Empfehlungssystemnutzer. Dieser Anwendungsfall wird ausgelöst durch ein besonderes Ereignis, denn auf der gewöhnlichen Route des Nutzer hat sich ein Stau gebildet. Der Erfolg des Anwendungsfalles besteht darin, das der Nutzer rechtzeitig geweckt wird, so dass er pünktlich an seinem Ziel ankommt. Die Beschreibung enthält einzelne Teilschritte oder Abfolgen, die ein Realisierung des Anwendungsfalles ermöglichen. Das System erlernt die Route des Nutzer, erhält lokale Verkehrsinformationen und kann den Nutzer bei Bedarf wecken. Bei detaillierteren technischen Anwendungsfällen für die Softwareentwicklung würde jeder Teilschritt in einen eigenen Anwendungsfall ausgegliedert werden und exakter beschrieben werden müssen. Wenn nicht mehr aktuelle oder falsche Verkehrsinformationen verwendet werden, z. B. es existiert kein Stau mehr, ist dies ein möglicher Fehlerfall in dem der Nutzer nicht geweckt werden dürfte.
Alle weiteren Anwendungsfälle, die genutzt wurden um eine Architektur des Systems zu entwerfen, sind im Anhang unter A.1 zu finden.

## 2.3 Architektur des zu betrachtenden Systems

Aus den zuvor angesprochenen Anwendungsfällen entsteht eine grober Architekturentwurf, bei dem das inHMotion System ein heterogenes Netzwerk, bestehend aus ad-hoc und zentralen Bereichen darstellt. Es ist durch die Verbindung von Peer-to-Peer (P2P) und traditionellen Client-Server Ansätzen eine leichte, allerdings flache, Hierarchie vorhanden. Die Architektur besteht aus folgenden Komponenten, betroffene Anwendungsfälle werden referenziert:

- Endnutzergeräte
  - Smartphone (A.1.1, A.1.5, A.1.11, A.1.12, A.1.13, A.1.14, A.1.15)
  - Fahrzeuge (Auto, Motorrad, Bus, Bahn, Tram, usw.) (A.1.2, A.1.3, A.1.14, A.1.5, A.1.6, A.1.7, A.1.8, A.1.9, A.1.10, A.1.16, A.1.17, A.1.18)
- Infrastruktur
  - Verkehrsleitsysteme u. a. Ampeln (A.1.3, A.1.7, A.1.18)
  - Parkleitsysteme u. a. Parkhäuser (A.1.4)
- Dienstanbieter
  - Fahrzeughersteller (siehe Endnutzergeräte, Punkt Fahrzeuge)
  - Navigationsdienste (A.1.9, A.1.15)
  - Empfehlungsdienste (A.1.1, A.1.11, A.1.12, A.1.13, A.1.14, A.1.15, A.1.16, A.1.17, A.1.18)
- Plattform AppStores für Endnutzergeräte u. a. NaviApp (A.1.9, A.1.15)
- Drahtlose Kommunikationstechnologien (Alle Anwendungsfälle)
  - zelluar: LTE, LTE D2D
  - ad-hoc: 802.11p

Diese Architektur ist bewusst flach gehalten und besteht im Groben aus nur drei interagierenden Komponenten dem Endgerät, dem Dienstanbieter und dem AppStore als Quelle der Anwendung. Dienste und AppStore werden über zellulare Verbindungen genutzt und der direkte lokale Informationsaustausch über ad-hoc Kommunikation. Das Architekturdiagramm 2.1 stellt nur die informationstechnischen Komponenten dar, die in dem Gesamtsystem interagieren. Vorhandene Zwischensysteme, wie das Netzwerk eines Mobilfunkanbieters, werden nicht betrachtet.

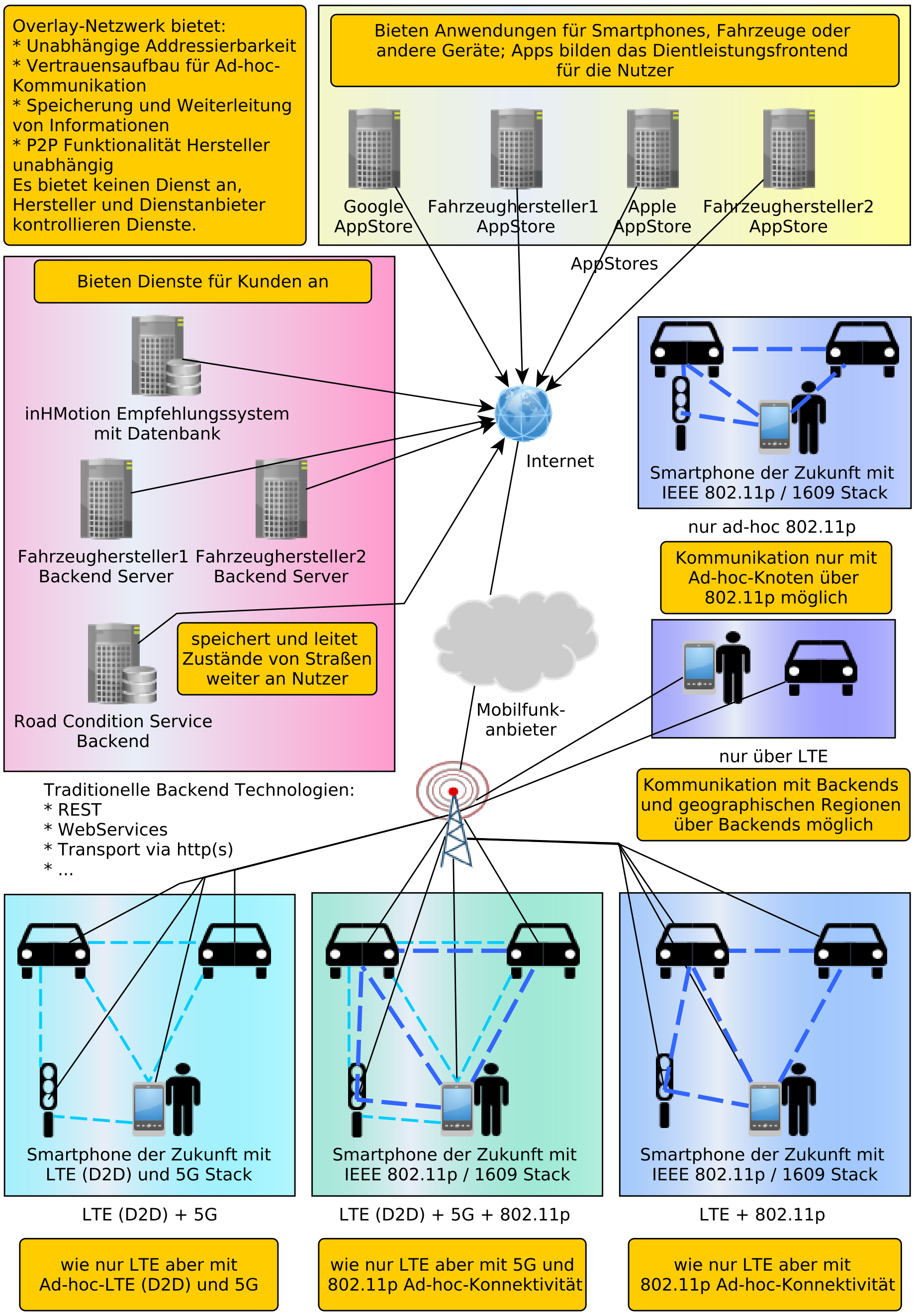

**Abbildung 2.1:** Grobe Systemarchitektur

Ein möglicher Netzwerk Stack, wie er auf einem Endnutzer Gerät vorhanden sein könnte, ist in Diagramm 2.2 dargestellt. Auf der linken Seite der Grafik befinden sich zum Vergleich zwei Netzwerkschichtenmodelle, das OSI und das TCP/IP Modell [6]. Der rechte Teil der Grafik stellt einen möglichen Netzwerkschichtenaufbau eines Smartphones, Fahrzeugs oder einer Roadside Unit (RSU), der Zukunft dar. Die tieferen Schichten sind zweigeteilt, in ad-hoc und in zellulare Kommunikationstechnologien. Auf Seiten der Ad-hoc-Kommunikation findet sich der IEEE 802.11p Standard [33], der als Ergänzung in den IEEE 802.11-2013 Standard aufgenommen wurde. Der zellulare Teil ist nur in verkürzter Form dargestellt, da in diesem Bereich häufig mit Tunnelprotokollen gearbeitet wird und in dieser Betrachtung die Nutzdaten von Interesse sind.

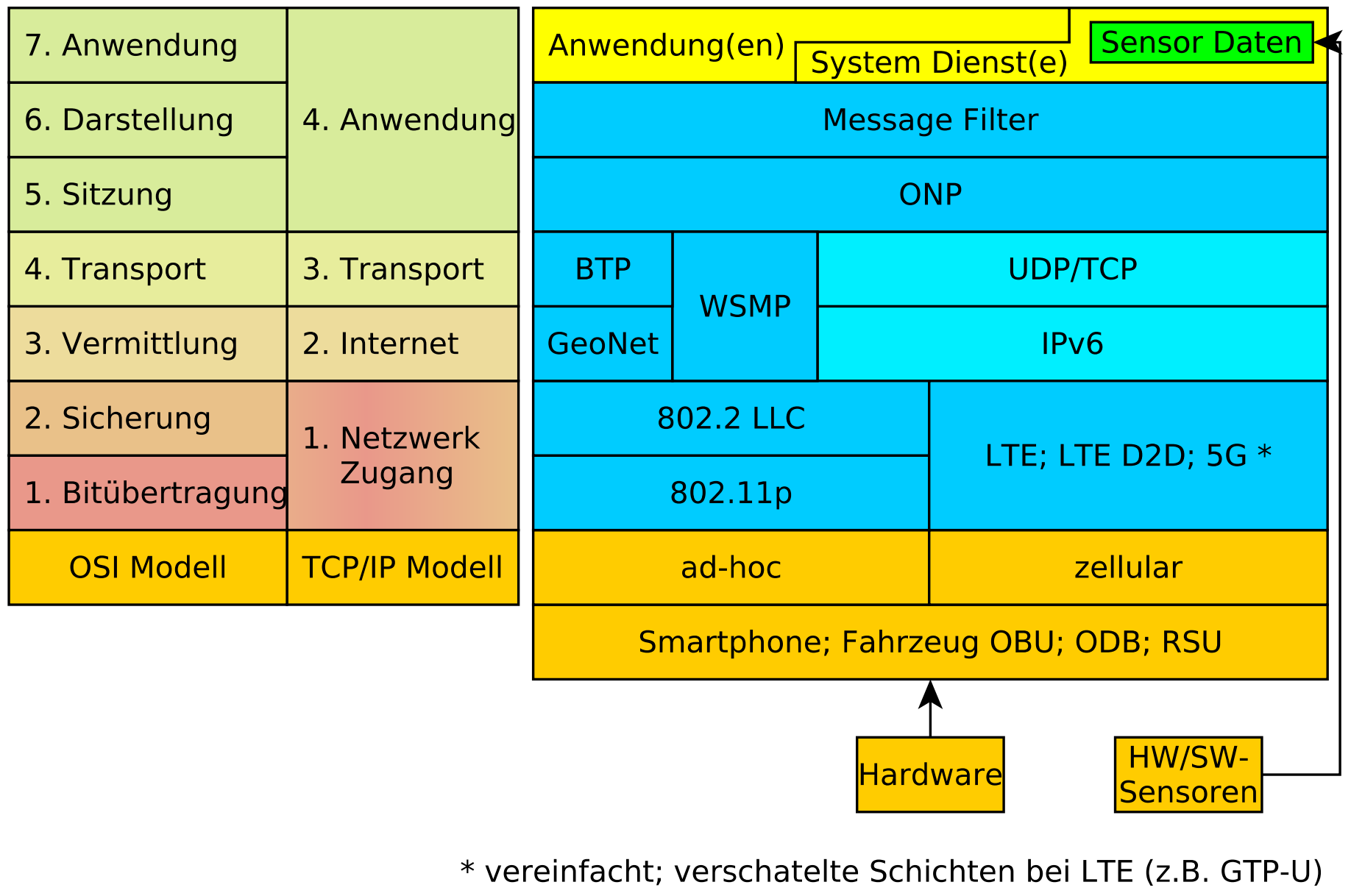

**Abbildung 2.2:** Netzwerk-Schicht-Architektur

Ab der Vermittlungsschicht, gemäß OSI, bzw. der Internetschicht, gemäß TCP/IP, verschwimmt die klare Trennung der Technologien. Zwar bietet der Ad-hoc-Bereich noch zwei konkurrierende Standards, den IEEE 1609 WAVE Short-Message Protocol (WSMP) [36] und den ETSI GeoNet/Basic Transport Protocol (BTP) [25] an, allerdings unterstützen beide auch IPv6/UDP genauso wie die zellularen Technologien. Oberhalb der Transportschicht setzt das ONP an, das alle darunterliegenden Schichten zusammenführt.

Der Message-Filter ist ein Platzhalter und die noch zu entwerfende Komponente, die für eine Vorverarbeitung der Informationen zuständig ist und weitere Funktionalität bereitstellen kann, wie z. B. den Zugriff auf Sensordaten oder Netzwerkinformationen. Für die oberste Schicht, die Anwendungen und Systemdienste beherbergt, werden zuvor beschriebene Informationen durch den Message-Filter bereitgestellt.

Die Anwendungen aus den AppStores sind das Frontend, das der Nutzer verwendet, um Zugang zu den Diensten zu bekommen. Diese werden durch die Backends der Dienstanbieter bereitgestellt. Im Folgenden werden nun Methoden vorgestellt, die bei der Bedrohungsmodellierung eingesetzt wurden, um Bedrohungen zu identifizieren und um diese zu bewerten.

## 2.4 Bedrohungsidentifizierung mit Angriffsbäumen

Angriffsbäume [67] sind ein formale Methode, mit der sich Bedrohungen und Realisierung dieser durch Angriffe strukturiert darstellen lassen. Prinzipiell stellt die Wurzel eines Angriffsbaums eine Bedrohung bzw. das Ziel des Angreifers dar, die durch Angriffe, ihre Kindknoten und Blätter realisiert werden kann.

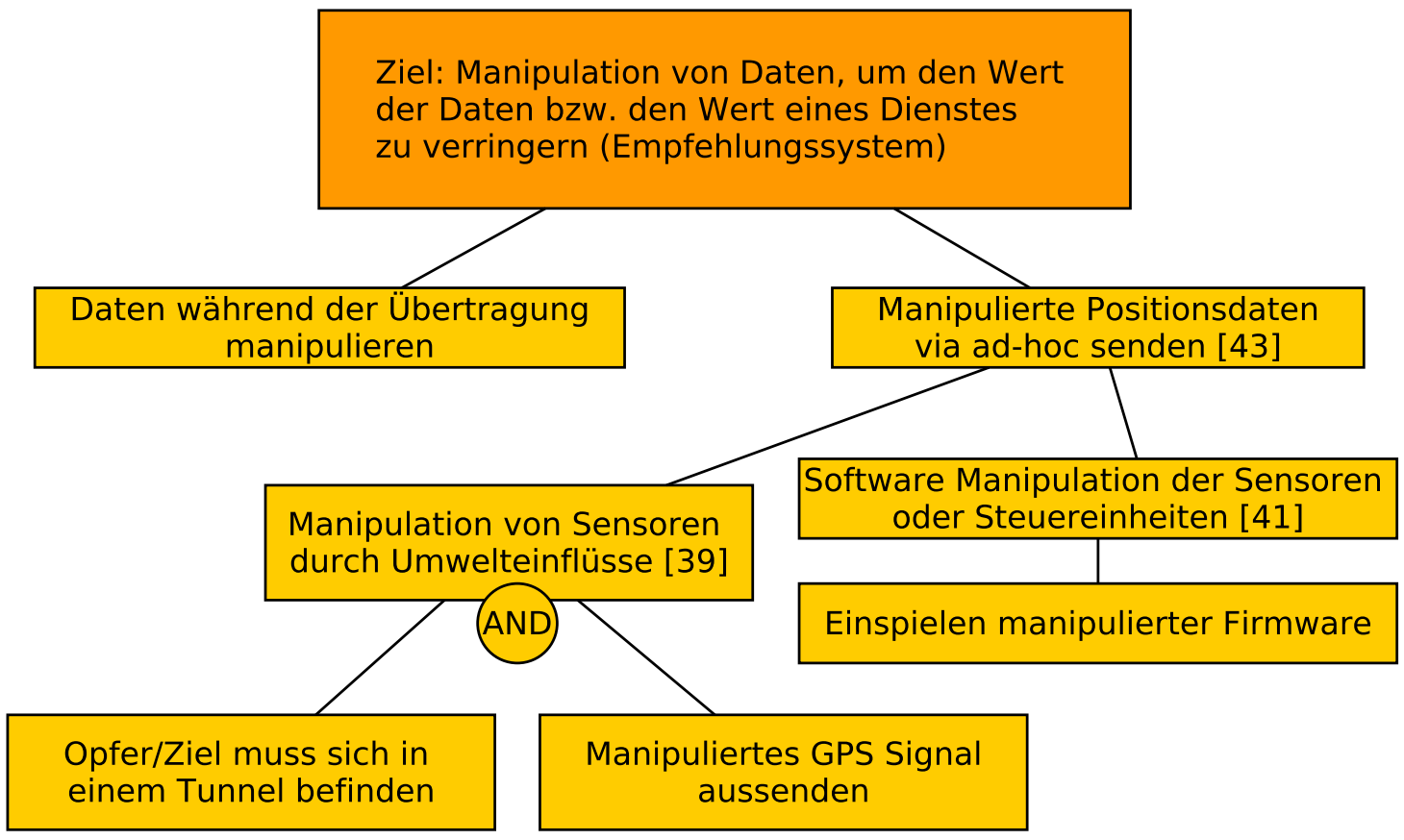

**Abbildung 2.3:** Beispiel Angriffsbaum: Verringerung des Informationswertes

Die Abbildung 2.3 stellt einen stark vereinfachten Angriffsbaum dar, dessen Bedrohung oder Ziel es ist, den Wert von Informationen zu verringern um einem System (hier Empfehlungssystem) zu schaden. Im Folgenden wird derselbe Baum in einer textuellen Repräsentation als durchnummerierte Liste dargestellt. Jeder Kindknoten oder jedes Blatt stellt einen Angriff dar und ist grundsätzlich als logisch „oder“ verknüpft zu betrachten. Dies bedeutet, dass entweder das Blatt „Daten während der Übertragung manipulieren“ oder ein Pfad aus dem Teilbaum „Manipulierte Positionsdaten via ad-hoc senden [43]“ umgesetzt werden muss, um den Wert von Daten zu beeinflussen. Dieser Teilbaum enthält eine Besonderheit, denn um „Manipulation von Sensoren durch Umwelteinflüsse [39]“ zu erfüllen, müssen beide Blätter „Opfer/Ziel muss sich in einem Tunnel befinden“ und „Manipuliertes GPS Signal aussenden“ erfüllt sein. Diese logische „und“ Verknüpfung wird in der Grafik durch ein Kreis mit der Beschriftung „AND“ gekennzeichnet, in der Textform wird an die Elemente ein „(AND)“ angehängt.

Ziel: Manipulation von Daten, um den Wert der Daten bzw. den Wert eines Dienstes zu verringern (Empfehlungssystem):

1. Daten während der Übertragung manipulieren
2. Manipulierte Positionsdaten via ad-hoc senden [43]
   (a) Manipulation von Sensoren durch Umwelteinflüsse [39]
      i. Opfer/Ziel muss sich in einem Tunnel befinden (AND)
      ii. Manipuliertes GPS Signal aussenden (AND)
   (b) Software Manipulation der Sensoren oder Steuereinheiten [41]
      i. Einspielen manipulierter Firmware

Nach der Konstruktion eines Angriffsbaumes können weitere Schritte angefügt werden, die zu einer Bewertung oder auch zu einer Gewichtung von den einzelnen Angriffen führt. Je nach gewählter Metrik können qualitative Aussagen getroffen werden, z. B. wie wahrscheinlich ein Angriffspfad ist, ob ein Angriffspfad überhaupt realisierbar ist oder auch wie hoch ein möglicher Schaden ausfallen würde. Weitere Details zu Angriffsbäumen im Allgemeinen und zu möglichen Metriken kann bei Bruce Schneier [67] nachgelesen werden. Da in dieser Arbeit eine andere Methode der Bewertung gewählt wird, wird hier nicht weiter auf das Vorgehen mit Angriffsbäumen eingegangen.

## 2.5 Bedrohungsbewertung mit dem OWASP Risk Rating

Das OWASP Risk Rating [51] ist eine Risikobewertungsmethode bei der jede einzelne Bedrohung in 16 Faktoren bewertet wird. Diese sind wiederum in 4 Kategorien unterteilt. Ein Faktor kann die Werte[1] von 1 bis 9 annehmen [55], wobei 9 der schlechteste Fall aus Sicht des technisch und finanziell Verantwortlichen, z. B. eines Herstellers oder Anbieters, ist. Da die Open Web Application Security Project (OWASP) aus einem internationalen Umfeld kommt, sind die Begrifflichkeiten in der Regel in englischer Sprache und werden hier beibehalten, um Inkonsistenzen und Missverständnisse zu vermeiden.

[1]Die OWASP [51] verwendet aktuell 0 bis 9, in Absprache mit dem Betreuer wurde auf die von Prof. Paulus [55] beschrieben Bereich 1 bis 9 zurückgegriffen.

Die 4 Kategorien lauten: „Threat Agent, Vulnerability, Technical Impact [und] Business Impact [51]“. Die einzelnen Faktoren werden durch ihre spezifische Fragestellung definiert und anhand von Bewertungsbeispielen erläutert. Diese Beispiele wurden von [51] übernommen und sind lediglich ins Deutsche übersetzt worden.

- **Threat Agent Factors** Als Erstes definiert man die Faktoren bezüglich des Angreifers, um darüber die Eintrittswahrscheinlichkeit (likelihood) zu erfassen. Für den Angreifer wird immer der „schlechteste Fall [51]“ angenommen.
  - **Skill Level** Wie groß ist das technische Know-How der Angreifer?
    *z. B.* IT-Security Experte (1), keine technischen Fähigkeiten (9)
  - **Motive** Wie stark ist die Motivation der Angreifer?
    *z. B.* geringe (1), große (9) Belohnung
  - **Opportunity** Welche Mittel sind nötig um diese Schwachstelle auszunützen?
    *z. B.* direkter Zugang und teure Ressourcen (0), kein Zugang und keine Ressourcen (9)
  - **Size** Wie groß ist die Gruppe der Angreifer?
    *z. B.* Entwickler (2), anonyme Internetnutzer (9)
- **Vulnerability Factors** Anhand von dem zuvor angenommen Angreifer wird nun die Eintrittswahrscheinlichkeit in Bezug zur Schwachstelle betrachtet.
  - **Ease of discovery** Wie einfach kann die Schwachstelle entdeckt werden?
    *z. B.* praktisch unmöglich (1), durch automatisierte Werkzeuge (9)
  - **Ease of exploit** Wie einfach kann die Schwachstelle ausgenutzt werden?
    *z. B.* theoretisch (1), durch automatisierte Werkzeuge (9)
  - **Awareness** Wie bekannt ist diese Schwachstelle in der Gruppe der Angreifer?
    *z. B.* unbekannt (1), öffentlich bekannt (9)
  - **Intusion Detection** Wie wahrscheinlich wird ein Angriff erkannt?
    *z. B.* aktive Erkennung (1), keine Aufzeichnungen (9)
- **Technical Impact Factors** Die technischen Auswirkungen werden über den Grad des Verlustes von Vertraulichkeit, Integrität, Verfügbarkeit und Nachvollziehbarkeit definiert.

- **Loss of confidentiality** Wie viele und wie sensibel sind die Daten, die veröffentlicht würden?
  *z. B.* wenige nicht sensible Daten (2), alle Daten (9)
- **Loss of integrity** Wie viele und wie stark werden Daten beschädigt?
  *z. B.* wenige leicht beschädigt (1), alle Daten vollständig beschädigt (9)
- **Loss of availability** Wie viele Dienste gehen verloren und wie wichtig sind diese?
  *z. B.* wenige sekundär Dienste (1), alle Dienste vollständig verloren (9)
- **Loss of accountability** Sind die Aktionen der Angreifer nachverfolgbar?
  *z. B.* vollständig nachvollziehbar (1), vollständig anonym (9)

- **Business Impact Factors** Die Auswirkungen auf das Geschäft eines betroffenen Unternehmens sind sehr speziell für das jeweilige Unternehmen. Daher wird dieser Wert über vier Bereiche definiert, die auf viele Geschäfte zutreffen: einen finanziellen Schaden, Beschädigung des Ansehens, Nichteinhalten von gesetzlichen Bestimmungen und Verletzung der Privatsphäre.
  - **Financial damage** Wie groß ist der finanzielle Schaden, wenn die Schwachstelle ausgenutzt wird?
    *z. B.* weniger, als es kostet sie zu beheben (1), Bankrott (9)
  - **Reputation damage** Wie groß ist der Schaden für das Ansehen des Unternehmens, wenn die Schwachstelle ausgenutzt wird?
    *z. B.* geringer Schaden (1), Beschädigung des Markennamens (9)
  - **Non-compliance** Wie groß ist die Gefährdung durch das Nichteinhalten von Bestimmungen?
    *z. B.* geringfügige Verletzung (2), Finden öffentlicher Beachtung durch einen Verstoß (7)
  - **Privacy violation** Wie viel personenbezogene Daten werden veröffentlicht?
    *z. B.* Daten eines einzigen Individuums (3), Millionen von betroffenen Personen (9)

Um das Gesamtrisiko abzuschätzen, können, laut [51], eine informelle oder eine reproduzierbare Methode eingesetzt werden. Neben der Nachvollziehbarkeit, ist die Möglichkeit für den jeweiligen Wert, eine fundierte Argumentationsgrundlage zu bieten bzgl. der getroffenen Annahmen, ein Vorteile der reproduzierbaren Methode, Da diese Eigenschaften sehr wünschenswert sind, wird in dieser Arbeit ausschließlich auf die reproduzierbare Methode fokussiert.

Dazu wird nun der Durchschnitt über die 8 Werte von „Threat agent factors [und] Vulnerability factors“ berechnet. Ebenso wird mit den jeweils 4 Werten von „Technical Impact“ und „Business Impact“ verfahren. Nun hat man 3 Werte die zwischen 0 und 9 liegen, diese werden nun gemäß der folgenden Tabelle 2.2 zunächst in Bereiche und daraufhin in Begriffe übersetzt.

| **Likelihood and Impact Levels** | |
|---|---|
| $0 <= x < 3$ | LOW |
| $3 <= x < 6$ | MEDIUM |
| $6 <= x <= 9$ | HIGH |

**Tabelle 2.2:** Wertebereich von Likelihood und Impact zu Bezeichnungen

Wurden die Bezeichnungen ermittelt, lässt die folgende Risikomatrix 2.3 eine Bestimmung des Gesamtrisikos über die Eintrittswahrscheinlichkeit (Likelihood) und Auswirkungen (Impact) zu. Sowohl Prof. Paulus [55, S. 108] als auch die OWASP [51] empfiehlt den Business Impact als primären Index in die Matrix zu wählen. Wenn keine verlässlichen Schätzungen über die Auswirkungen auf das Geschäft getroffen werden können, so sollte der Technical Impact [51] benutzt werden.

| | | **Overall Risk Severity** | | |
|---|---|---|---|---|
| **Impact** | HIGH | Medium | High | Critical |
| | MEDIUM | Low | Medium | High |
| | LOW | Note | Low | Medium |
| | | LOW | MEDIUM | HIGH |
| | | **Likelihood** | | |

**Tabelle 2.3:** Gesamtrisiko Matrix mit Likelihood und Impact als Indices

Auf diese Weise kann jeder Bedrohung ein Gesamtrisiko oder „Kritikalität[55, S. 109]“ zugeordnet werden. Gegen welche Bedrohungen vorzugehen ist, liegt ganz in der Entscheidungsmacht der Risikoträger, es wird lediglich empfohlen die größten Risiken zuerst zu beheben [51].

## 2.6 Grundbegriff: Identitätsmanagement

Der in dieser Arbeit verwendete Begriff der Identität bezieht sich auf Informationen, die in Rechnernetzwerken und -systemen, Personen, Gruppen von Personen (Organisationen) und Maschinen repräsentieren. Diese können Bedeutungen im rechtlichen, sozialen oder technischen Umfeld eines Rechners haben. Im Zusammenhang mit Vehicular ad hoc network (VANET)s kann eine einzige Identität unterschiedliche Bedeutungen in allen drei Einflussbereichen haben. Ein digitales Nummernschild stellt eine rechtliche Identität dar, die bei Verkehrsvergehen für die Straf- oder zivilrechtliche Verfolgung genutzt werden kann. Dieses könnte auch bei einem hypothetischen Dienst als soziales Netzwerk von Fahrzeugen, bei dem die Fahrer den Fahrstil des vorausfahrenden Fahrzeuges bewerten, verwendet werden. Das digitale Nummernschild könnte in einem Computernetzwerk als Adresse genutzt werden, damit gegenseitige Kommunikation von Maschinen, in diesem Fall Fahrzeugen, möglich wird. Erschwerend kommt hinzu, dass Identitäten unterschiedliche Eigenschaften haben können, z. B. ob, wie und durch wen eine Verifizierung dieser möglich ist. Im folgenden wird eine exemplarische Liste aufgeführt von Identitäten, wie sie in einem VANET eingesetzt werden könnten:

- Identitäten zentral vergeben, lokal/entfernt verifizierbar, einzigartig
  - Digitales Nummernschild (behördlich vergeben, kryptographisches Material: Zertifikat, privater und öffentlicher Schlüssel)
  - Public Key Infrastructure (PKI) eines beliebigen Anbieters (kryptographisches Material: Zertifikat, privater und öffentlicher Schlüssel, gültig in globalem Vertrauenskontext oder nur im Kontext des Anbieters)
  - Overlay Netzwerk PKI
- Identitäten zentral vergeben, lokal/entfernt verifizierbar, einzigartig in dem Systemkontext, Verwendung in systemfremdem Kontext
  - Soziale Identitäten von Drittanbietern (Facebook, Twitter, Google)
- Identitäten dezentral vergeben, lokal/entfernt verifizierbar, nicht oder nur lokal einzigartig
  - selbst erzeugtes kryptographisches Material, gegenseitige Vertrauensbekundung durch Web of Trust (WoT), o. Ä.
  - Overlay Netzwerk mit Cluster-Identitäten, siehe [9]
- Identitäten dezentral vergeben, nicht verifizierbar, nicht einzigartig
  - Media Access Control (MAC) Adressen
  - Internet Protocol (IP) Adressen
  - Overlay Netzwerk Pseudonyme
  - Overlay Netzwerk Anonyme

## 2.7 Grundbegriff: Vertrauen

Für Vertrauen und Reputation werden die Definitionen nach Jøsang *et al.*[40] verwendet. Gemäß diesen Definitionen existieren zwei verschiedene Arten von

Vertrauen:

- **Vertrauen in die Zuverlässigkeit**: Vertrauen ist eine subjektive Wahrscheinlichkeit eines Individuums A, mit der es erwartetet, dass ein weiteres Individuum B, eine Handlung durchführt, auf der das eigene Wohlergehen von A abhängt.
- **Vertrauen in Entscheidungen**: Vertrauen ist das Ausmaß, in dem ein Teilnehmer bereit ist, in einer gegebenen Situation, sich auf etwas oder jemanden anderes zu verlassen und dies mit einem relativen Gefühl von Sicherheit, obwohl negative Konsequenzen möglich sind.

Die Reputation oder auch das Ansehen wird wie folgt definiert:

- **Reputation**: Reputation ist das, was generell gesagt oder angenommen wird über den Charakter oder das Ansehen einer Person oder eines Gegenstandes.

Gemäß diesen Definitionen werden die zuvor aufgeführten Begriffe hier verwendet.

# 3. Stand der Forschung

Das Anwendungsfeld der CPS ist groß, da viele verschiedene Arten von physikalischen Prozessen existieren, die bereits durch Computer über einen Regelkreislauf gesteuert werden oder dies in Zukunft tun könnten. Als Beispiele für Anwendungsbereiche sind Energiegewinnung/-verbrauch, Gesundheitssektor, Produktion von Waren, militärische Anwendungen, Infrastruktur oder der Transport- und Beförderungssektor zu nennen. Letztere werden auch als ITS (intelligente Transport Systeme) bezeichnet. Ein solches System könnte aus Verkehrsteilnehmern, selbst oder pilotiert fahrenden Bussen, Transportern oder PKWs bestehen, die Ampelphasen beeinflussen oder die Routenplanung der Verkehrsauslastung anpassen. Bei den CPS könnte ein System entweder selbständig entscheiden oder eine Person trifft letztendlich eine bestimmte Entscheidung. Die Vernetzung der einzelnen Komponenten, hier Teilnehmer, ist der entscheidende Vorteil gegenüber bereits eingesetzten Verkehrsleitsystemen. Eine Ampelsteuerung, die durch Bilderkennung oder Induktionsschleifen das Verkehrsaufkommen registriert und dementsprechend die Phasen schaltet, hat nur beschränkten ein Fluss auf die Großverkehrslage. Trotz zentraler Verkehrsflusssteuerung erreichen Verkehrsflussinformationen den Verkehrsteilnehmer meistens erst, wenn dieser bereits im Stau oder zäh fließenden Verkehr steckt. In einem idealen ITS würden Staus der Vergangenheit angehören, da jeder Teilnehmer beispielsweise die Verkehrsflussdaten zentral oder dezentral abrufen und die Route und Reisegeschwindigkeit anpassen kann.

Findet die Kommunikation spontan mit anderen Verkehrsteilnehmern über sogenannte Ad-hoc-Kommunikationstechnologien statt, spricht man von einem VANET. Beschränkt man sich nicht nur auf kommunizierende Fahrzeuge oder Infrastruktur aus dem Transportbereich, sondern bezieht auch mobile Endgeräte, wie z. B. Smartphones mit ein, so ist das resultierende Netzwerk ein Mobile ad hoc network (MANET).

Die Anforderungen an die Betriebssicherheit, im Sinne von „Safety“, bei CPS sind sehr hoch, da durch Fehlfunktionen physikalische Prozesse beeinflusst werden. Ebenso sind die Anforderungen an die Sicherheit, im Sinne von „Security“, an CPS sehr hoch, da fehlende Sicherheit auch die Betriebssicherheit beeinflussen kann. Dies wird deutlich indem Systeme nicht nur fehlerfrei arbeiten, sondern auch resistent gegen Manipulation oder Angriffe sein müssen. Wenn ein Angreifer die Sicherheit eines Systems beeinträchtigen kann, z. B. schafft dieser es beliebigen Programmcode auszuführen oder sorgt durch Überlastung für einen Ausfall, so ist auch die Betriebssicherheit betroffen.

## 3.1 Sicherheit in VANETs

Durch die Fokussierung dieser Arbeit auf das Umfeld von VANETs, können spezifischere Beispiele, Anforderungen, Bedrohungen oder Lösungen definiert werden, als wenn das breite Feld der CPS den Ausgangspunkt darstellt. Die klassischen drei Schutzwerte: Vertraulichkeit, Integrität und Verfügbarkeit, lassen sich im VANET Umfeld auf sieben erweitern [82]:

- Authentifizierung
- Integrität und Konsistenz
- Vertraulichkeit
- Verfügbarkeit
- Zugriffs- / Zugangskontrolle
- Nichtabstreitbarkeit
- Datenschutz / Schutz der Privatsphäre

Diesen Schutzwerten steht eine große Anzahl von Angriffen entgegen, die diese Werte bedrohen. Eine kurze Auflistung einiger Kategorien von Angriffen, die in VANETs auftreten können, [32], [45], [82] oder [28], wird im Folgenden gegeben:

- Vortäuschen von Identitäten
- Manipulation von Daten (oder falsche Informationen)
- Angriffe auf das Routing (oder Grey/Black Hole)
- Sybil Angriffe
- Eclipse Angriffe
- Wormhole Angriffe
- Denial of Service (DoS) Angriffe
- Timing Angriffe
- Schadsoftware und Spam
- Privatsphärenverletzung
- Hardware Manipulation

Für eine Auswahl von Angriffen können verschiedene Lösungsansätze in drei Kategorien eingeteilt werden, diese sind *zentral*, *dezentral* oder *hybrid* [58]. Nach einer Definition der Kategorien werden verschiedene Lösungsansätze untersucht und kategorisiert. Eine kurzer Überblick über die untersuchten Ansätze ist in folgender Tabelle zu begutachten:

| Sicherheitsrisiken | zentral | dezentral | hybrid |
|---|---|---|---|
| Vortäuschen von Identitäten | [69],[32],[60] | [26],[8] | [74] |
| Manipulation von Daten | [44],[69],[59] | [26] | [83] |
| Angriffe auf das Routing | [22],[48],[84] | [30],[14] | [60] |
| Sybil Angriffe | [57] | [80],[54] | [53] |
| Eclipse Angriffe | [75] | [80] | - |
| Wormhole Angriffe | - | [64],[22] | - |
| DoS Angriffe | [60] | [27] | - |
| Privatsphärenverletzung | [69],[17],[74],[75],[76] | [44] | [9],[85] |

**Tabelle 3.1:** Übersicht über Lösung zu einigen Angriffen

## 3.2 Reputationssysteme

In der Informationstechnik sammelt, verteilt und aggregiert ein Reputationssystem Feedback über zurückliegende Transaktionen zwischen Teilnehmern eines Systems [62]. Allgemeiner verfolgen Reputationssysteme drei Hauptziele [61]:

1. Bereitstellung von Informationen, um zwischen vertrauenswürdigen und nicht vertrauenswürdigen Teilnehmern unterscheiden zu können.
2. Förderung von vertrauenswürdigem Verhalten von Teilnehmern.
3. Abschreckung von nicht vertrauenswürdigen Teilnehmern bzgl. der Teilnahme an einem Prozess.

Das Bereitstellen von Informationen geschieht bei Reputationssystemen durch Berechnung von Vertrauens- oder Reputationswerten [40]. Informationen, die in eine Berechnung einfließen, können aus erster oder zweiter Hand stammen. Informationen aus erster Hand werden auch als private Informationen bezeichnet, ihnen wird oft ein höherer Wert zugemessen. Ist die Information lediglich aus zweiter Hand so wird dies als öffentliche Information bezeichnet.
Die Berechnung kann auf unterschiedliche Art und Weisen stattfinden [40]:

- Einfache Summierung oder Durchschnittsbildung von Bewertungen, z. B. [61]
- Basierend auf dem bayesscher Wahrscheinlichkeitsbegriff, z. B. [39]
- Diskrete Vertrauensmodelle als Berechnungsgrundlage, z. B. [15]
- Evidenztheorie als Basis, z. B. [38]
- Fuzzylogik oder auch unscharfe Logik, z. B. [63]
- Flussmodelle oder Vertrauensketten, z. B. [52]

## 3.3 Reputationssysteme für VANETs

In der Literatur sind bereits einige Vorschläge von Reputationssystemen speziell für den Einsatz in einem VANET Umfeld vorhanden. Das „Vehicle Behavior Analysis and Evaluation Scheme (VEBAS)“ [65] System fällt Entscheidungen darüber, ob ein Teilnehmer neutral, vertrauenswürdig oder nicht vertrauenswürdig ist, basierend auf Beobachtung und Analyse von Verhalten der Teilnehmer. Der Fokus liegt hierbei auf der sensorbasierten Verifikation von Positionsinformationen.

Ebenso beobachten sich auch die Teilnehmer des „ERS“ [46] Ansatzes gegenseitig. Allerdings ist das zentrale Element des Ansatzes die Verarbeitung von Verkehrsereignissen, die sich in einem VANET ereignen. Dazu wird eine Art Gedächtnis über zurückliegende Ereignisse aufrechterhalten und wenn neue Ereignisse auftreten mit diesen verglichen. Basierend auf diesem Ver-gleich und dem Abgleich mit Sensordaten werden Reputationswerte für jedes Ereignis berechnet.

Durch den Austausch von Meinungen zu anderen Teilnehmern und der Verarbeitung dieser versucht das „Vehicular Ad-Hoc Reputation System“ (VARS) [21] eine Reputationsbewertung zu erreichen. Dem beschriebenen Ansatz fehlt es an Authentizitäts- und Integritätsschutz, wodurch Manipulationen oder das Weglassen von Bewertungen ein Problem darstellt.

Eine Alternative ist ein zentralisierter Ansatz mit einem offline Reputationsdienst [16], der sporadisch durch einen Proxy Server mit den Teilnehmern kommuniziert und basierend auf Feedback nicht vertrauenswürdige Teilnehmer identifiziert. Jedem Teilnehmer wird von dem zentralen Dienst einen Reputationsausweis ausgestellt. Dieser dient als gültige Reputation für den Teilnehmer. Wird ein Teilnehmer als nicht vertrauenswürdig erkannt, so erhält dieser keinen neuen Reputationsausweis.

Mit dem „Centralized Reputation System (CRS)“ [11] wird ebenfalls ein zentralisierter Ansatz favorisiert. Die Teilnehmer erfragen Reputationswerte bei den RSUs, diese wiederum werden von dem zentralen Datenbanksystem versorgt, das die Reputation von allen Teilnehmern bereithält.

## 3.4 Informationsverifikation in VANETs

Oft ist in der Kommunikation der Kommunikationspartner ein wichtiges Detail, aber der eigentliche Wert liegt in den Informationen, die kommuniziert werden. Dies trifft auch auf die VANET Umgebung zu, denn ob ein Fahrzeug ein bestimmtes Nummernschild hat ist für die Information, dass es in ca. 500 Metern auf ein Stauende auffährt, zweitrangig. Ob diese Information vertrauenswürdig und wahrheitsgemäß ist, ist wichtig, um die richtigen Konsequenzen und Entscheidungen zu treffen. Durch die Verifikation von Informationen oder die Plausibilitätsüberprüfung kann Vertrauen in eine Information hergestellt werden. Ein Verifikation kann auf zwei Arten erfolgen:

- Lokale Verifikation von Informationen
- Entfernte Verifikation durch Überprüfung von Dritten

Bei der lokalen Verifikation werden Informationen innerhalb desselben Systems oder Komponenten geprüft ohne, dass eine entfernte Ressource, wie bei entfernter Verifikation, abgefragt werden muss. Ein Beispiel hierfür ist die Positionsverifikation durch Variationen in den gemessenen Empfangsstärken [13]. Als entfernte Verifikation ist zu verstehen, dass andere Teilnehmer unter weiterem Kommunikationsaufwand zu Hilfe genommen werden. Dies kann passiv, durch periodisches Aussenden und Empfangen von Empfangsleistungsinformationen von Nachbarknoten [80] oder auch aktiv, ebenfalls zur Positionsbestimmung, durch verifizierte Multilateration [32], erfolgen. Durch die Verwendung von Partikelfiltern, bekannt aus der bayesscheren Wahrscheinlichkeitstheorie, und verschiedensten Sensoren können Positionsveri-fikationen durchgeführt werden [12]. Das „Plausibility Validation Network (PVN)“ [47] kombiniert ebenfalls Sensordaten mit empfangenen Nachrichten und versucht diese anhand eines Plausibilitätsnetzwerks zu bewerten. Dieses besteht aus Regeln und einzelnen Datenelementen der Nachrichten. Ein grundsätzlich ähnlicher Ansatz ist, dass jeder Teilnehmer in einem VANET ein eigenes Modell von seiner Umgebung hat, gegen das dieser Informationen validiert [26].

# 4. Bedrohungsmodellierung

Dieses Kapitel beschäftigt sich mit der Bedrohungsmodellierung des Gesamtsystems. Nach der Beschreibung der Vorgehensweise wird diese umgesetzt und die einzelnen Teilschritte ausgeführt. Die Ergebnisse der einzelnen Schritte werden im Folgenden präsentiert. Teilergebnisse oder detailliertere Ergebnisse sind in den Anhängen A bzw. B zu finden.

## 4.1 Vorgehensweise zur Bedrohungsmodellierung

Eine Bedrohungsmodellierung dient nach Paulus [55] dazu Werte zu identifizieren, um eine risikobasierte Einschätzung durchführen zu können. Diese wiederum soll dazu genutzt werden, um Priorisierung von Investitionen in Sicherheit festzulegen [55]. Während der Durchführung versucht man ein Verständnis der Angriffsmechanismen zu erlangen und zusätzliche Testfälle zu erstellen. Die oft subjektiven Anforderungen sollen dadurch zu objektiveren und messbareren Anforderungen werden [55]. Nach Prof. Dr. Sachar Paulus „ist die Bedrohungsmodellierung die einzige sicherheitsrelevante Aktivität, die [...] versucht[,] Designfehler zu identifizieren“ [55]. Eine Bedrohungsanalyse nach Prof. Paulus beinhaltet folgende Teilschritte (übernommen aus [55]):

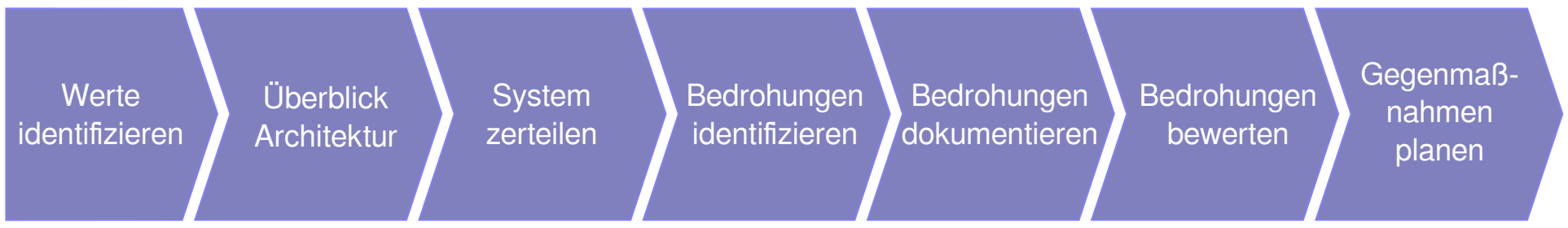

**Abbildung 4.1:** Prozessschritte der Bedrohungsmodellierung (übernommen aus [55])

Im ersten Schritt, dem Identifizieren von Werten, Akteuren und Operationen empfiehlt Prof. Paulus in seinem Buch eine Bedrohungsmodellierung, die sich an Werten orientiert [55]. Dazu ist es erforderlich Werte, Akteure und Operationen, die mit dem zu entwickelnden System in Verbindung stehen, zu identifizieren. Zur Unterstützung dienen einige Leitfragen mit deren Hilfe man Werte und Akteure identifizieren kann. Unter anderem werden folgende

Fragen vorgestellt: „Was ist für den Kunden am wichtigsten [...]? [55]“ oder „Wer hat Interesse an den identifizierten Werten? [55]“.

Im zweiten Schritt spricht Prof. Paulus von einem „Henne-Ei-Problem [55]“, da sich Sicherheitsvorgaben auf eine Software- oder Systemarchitektur beziehen und die Qualität der jeweiligen Architektur gleichzeitig davon abhängt, wie gut die Sicherheitsvorgaben in der Architektur berücksichtigt werden [55]. Wobei einige Sicherheitsanforderungen möglicherweise erst bei der Durchführung der Bedrohungsmodellierung zu Tage treten. Deswegen ist diese besonders wichtig. Um eine Bedrohungsmodellierung durchführen zu können, muss zumindest eine grobe Architektur vorhanden sein. Wenn sie auch nur dazu dient, über Angriffe und mögliche vorhandene Schwachstellen diskutieren zu können [55].

Der dritte Schritt beseht daraus Vertrauensgrenzen zu erkennen und zu definieren. Interessant in diesem Zusammenhang sind insbesondere die Datenhaltung und die Datenflüsse eines Systems, da sich aus diesen sogenannte Vertrauensgrenzen ableiten lassen [55]. Diese Vertrauensgrenzen sind Schnittstellen zwischen einzelnen Komponenten eines Systems, einer Software oder sogar einzelne Programmteile. Über diese Grenzen hinweg sollten Daten nur eingeschränkt fließen, weil sich die Kommunikationspartner nicht blind vertrauen [55] sollten. Die richtige Granularität hat man erreicht, wenn Blöcke der Datenhaltung nicht weiter sinnvoll zerteilt werden können [55].

Im vierten Schritt müssen nun Bedrohungen identifiziert werden, dazu bieten sich verschiedene Methoden an. Prof. Paulus stellt folgende Werkzeuge und Techniken vor [55]:

- Brainstorming
- Checklisten
- Datenflussdiagramme
- Angriffsbäume
- Automatische Werkzeuge

Während Bedrohungen identifiziert werden, sollten diese direkt dokumentiert werden, dies stellt den fünften Schritt dar. Eine Tabellenkalkulation ist zu empfehlen, da die Informationen für die Bedrohungen gesammelt, übersichtlich dargestellt und bearbeitet werden können. Als die wichtigsten festzuhaltenden Informationen werden genannt:

- die Bedrohung selbst,
- die möglichen Angreifer,
- die Schnittstelle, an der der Angriff erfolgt,
- der Entdecker der Bedrohung und
- die Methode der Identifizierung.

Aus Ressourcengründen ist es selten möglich für jede Bedrohung geeignete Maßnahmen umzusetzen [55], daher empfiehlt sich eine Bewertung der Bedrohungen in Verbindung mit einem risikobasierten Vorgehen. Die Bewertung der Bedrohungen bildet den sechsten Schritt des Vorgehens. Der weitverbreitete Ansatz des Schätzens von Eintrittswahrscheinlichkeit und Schadenshöhe, die mangels statistischem Material, selten auf genauen Daten fußt und daher unzuverlässig ist, sollte nicht eingesetzt werden [55]. Stattdessen sollte DREAD [29] wegen seiner geringen Komplexität oder das OWASP Risk Rating [51] eingesetzt werden. Das OWASP Risk Rating ist eine Methode, die zum Zeitpunkt des Erscheinens des Buches, im Jahre 2011 nach Prof. Paulus [55], „den besten Kompromiss aus Genauigkeit und Einfachheit darstellt". Dieser Argumentation folgend wird in dieser Arbeit das OWASP Risk Rating eingesetzt.

Der letzte Schritt der Bedrohungsmodellierung besteht darin, aus den bewerteten und nach der Kritikalität geordneten Bedrohungen, Sicherheitsanforderungen abzuleiten. Von diesen Sicherheitsanforderungen ausgehend plant man wiederum Gegenmaßnahmen, um den Bedrohungen zu begegnen. Zielführend hierbei ist die Untersuchung der Ursache von Bedrohungen und wie diese Ursache kontrollierbar wird. Möglicherweise müssen neue Vertrauensgrenzen oder größere Architekturänderungen vorgenommen werden, um Bedrohungen beherrschbar zu machen. Ein wichtiger Punkt ist die genaue Untersuchung der Gegenmaßnahmen und ihrer Auswirkungen auf das System, denn sie sollte keine neuen Schwachstellen einführen [55].

In dieser Arbeit dient die Bedrohungsmodellierung dazu die möglichen Bedrohungen zu identifizieren, um anhand dieser die Anforderungen an ein geeignetes Reputationssystem zu entwickeln. Die Bedrohungsmodellierung befasst sich mit dem Gesamtsystem und fokussiert im Speziellen auf dem zugrunde liegenden Overlay-Netzwerk. Einzelne Aspekte, die nicht im Fokus liegen, werden gesondert betrachtet, wenn es dem Gesamtverständnis zuträglich ist.

## 4.2 Werte, Akteure und Operationen

Der erste Schritt in der Bedrohungsmodellierung ist einer der wichtigsten, da basierend auf den Werten, Akteuren und Operationen, Bedrohungen identifiziert und bewertet werden. Das Übersehen von Werten kann dazu führen, dass diese von keinen resultierenden Sicherheitsanforderungen berücksichtigt werden und in Folge ist es nicht verwunderlich, wenn es an entsprechenden Schutzmaßnahmen mangelt.

### 4.2.1 Werte des Systems

Obwohl Prof. Paulus in seinem Buch zunächst die Akteure ausführt, wird hier mit den Werten begonnen, da dieses Vorgehen die Definition der Akteu-re erleichtert. Die Werte, die durch ein System verarbeitet, erarbeitet oder bedient werden stehen im Fokus [55].

### „Was ist für Sie als Hersteller am wichtigsten? [55]“

Das inHMotion Projekt tritt nicht als Hersteller des ONP und des Empfehlungssystems auf, sondern es wird als Betreiber der intermodalen intelligenten Transporttechnologie ein Automobilhersteller oder ein Konsortium aus diesen angenommen. Ein Zusammenschluss von Autobauern ist angebracht, da einige der Anwendungsfälle, z. B. A.1.2, A.1.6 oder A.1.8, ihr volles Potenzial bei einer großen Verbreitung von ONP entfalten. Innerhalb des Konsortiums wird nicht mehr unterschieden, sondern es stellt eine Einheit und Interessensgemeinschaft dar, in der Interessenskonflikte innerhalb aufgelöst werden. Daher wird nicht mehr unterschieden zwischen einem Fahrzeughersteller und einem Konsortium als Betreiber.

Der Betreiber hat ein hohes Interesse daran, dass *Vertrauen und Ansehen* in das Netzwerk und die angebotenen Dienste hoch sind. Um bei der Entwicklung eines tragfähigen Geschäftsplans größere Freiheiten zu haben, ist ein wichtiger Aspekt die *Minimierung der Betriebskosten* für Dienste und für das Netzwerk selbst. Es ist wichtig ein *funktionierendes, intelligentes Empfehlungssystem* als Dienst zu betreiben und ein zuverlässiges ONP zu entwickeln, das *wenig Infrastruktur* benötigt, eine *hohe Marktdurchdringung* erreichen kann, ein *vielseitiges Angebot von Diensten* ermöglicht und *Technologie Transparenz* im Bereich zellular bzw. Ad-hoc-Kommunikation bietet. Gleichzeitig muss es auch ein *hohes Maß an Betriebssicherheit* bieten, so dass auch sicherheitskritische Funktionalität über das System abgebildet werden kann. Das bedeutet, dass die *Fehlerfreiheit der Software* bzw. bei auftretenden Fehlern eine *sichere Fehlerbehandlung* gegeben sein muss. Im Besonderen darf das *Leib und Leben der Nutzer* nicht gefährdet werden.

### Was ist für die Nutzer eines Dienstes am wichtigsten?

Die *Verfügbarkeit eines Dienstes* ist wichtig, besonders wenn in Stoßzeiten ein Einsatz zu erwarten ist, z. B. durch Pendler. Fallen Dienste aus, wenn ein Nutzer diese in Anspruch nehmen möchte oder die *Verfügbarkeit der Anwendung* auf der, vom Nutzer eingesetzten, Plattform ist nicht gegeben, werden sich diese Endnutzer umorientieren. Zwar ist man schon fast daran gewöhnt, dass persönliche Daten durch Dritte gestohlen, verkauft oder veröffentlicht werden, trotzdem liegt die Vermutung nahe, dass das *Vertrauen in den Dienst* erheblich leidet, sollten vertrauliche Daten, wie z. B. Anschrift der Kunden, Kontoinformationen oder andere personenbezogene Daten, entwendet und missbraucht werden. Gerade deswegen ist die *Vertraulichkeit der personenbezogenen Daten* für die Nutzer wichtig.

Für Endkunden sind Dienste wie das Empfehlungssystem der entscheidende Faktor. Das Kommunikationsnetzwerk ist der Mechanismus, über den die Kunden oder Partner den Endkunden ihre Dienste anbieten. Diese Dienste werden u. a. durch die Qualität der Daten, die das System liefert, verbessert. Die Befriedigung der Bedürfnisse dieser Endkunden liegt im Zuständigkeitsbereich der Dienstanbieter selbst, da ONP nur eine Kommunikationsplattform für diese Dienste bereitstellt. Wenn ONP nicht funktioniert und eine Anwendung oder Dienstleistung wiederum nur eine eingeschränkte oder gar keine Funktionalität bietet, trifft der Unmut in erster Linie den Dienstanbieter und erst in zweiter Instanz die Hersteller. Daraus resultieren trotzdem weitere Anforderungen, die ein Nutzer eines Dienstes an das System hat. Die Nutzung eines möglicherweise bezahlten Dienstes darf *keiner negativen Beeinflussung* durch das ONP-Kommunikationsnetzwerk unterliegen. Der Endkunde wird erwarten, dass die *Datenverarbeitung gemäß dem Bundesdatenschutzgesetz* erfolgt, dies bedeutet, dass insbesondere die *Vertraulichkeit der Kommunikation* durch das ONP gewahrt wird.

Jeder Dienst kann über die allgemeinen Werte hinaus noch eigene mit sich bringen, dies wird exemplarisch an dem Empfehlungsdienst dargestellt. Ein Empfehlungssystem muss *zuverlässige Handlungsempfehlungen* geben. Dabei ist sehr entscheidend, dass die Qualität einer Handlungsempfehlung sehr hoch ist. Eine Empfehlung muss z. B. präzise sein, einen Mehrwert bieten und auch umsetzbar sein. Wenn Empfehlungen unbrauchbar sind, wird sich das Interesse der Endnutzer schnell verflüchtigen. Wenn Bewegungsprofile angelegt werden, z. B. um Nutzer proaktiv vor Verzögerungen auf ihrem Arbeitsweg hinzuweisen, so sind diese als sensible personenbezogene Daten zu sehen. Für den Nutzer sind somit die *Vertraulichkeit der Ortsinformationen, der Historie dieser und die Vorhersagen über zukünftige Ortsinformationen* entscheidend. Die *Verfügbarkeit der Ortsinformationen* für die Verwendung im Empfehlungssystem muss gegeben sein. Bei Gefährdung von *Leib und Leben der Nutzer* durch das Empfehlungssystem kann davon ausgegangen werden, dass das Interesse an einer Nutzung erlischt.

#### Was ist für die Kunden, die Nutzer des Overlay-Netzwerkes, am wichtigsten?

Für Geräte- und Komponentenhersteller ist eine *hohe Effizienz*, vor allem im Bereich Energie oder Datenverbrauch speziell bei mobilen Geräten, von Interesse. Ebenso sollten notwendige Softwarekomponenten für die jeweilige Plattform *einfache und möglichst fehlerfreie Implementierbarkeit* bieten. Eine *Priorisierung der Nachrichten*, z. B. von Gefahrenmeldungen, muss möglich sein. Verschiedene Dienste benötigen *unterschiedliche Niveaus von Vertraulichkeit, Integrität, Authentizität, Verfügbarkeit und Datenschutz*. Der Einsatz von ONP darf nicht zu *Rufschädigung des Dienstanbieters* führen, z. B. bei Fehlfunktionen im Overlay. Ebenso darf die Nutzung von ONP nicht das *Leib und Leben der Nutzer* gefährden. Dies nicht nur aus ethisch moralischen Gründen, sondern auch als Schutz vor Schadensersatzforderungen und Gewährleistungsansprüchen. *Keine ungewollte gegenseitige Beeinflussung von verschiedenen Diensten* sollte möglich sein. Sicherstellen von *Nichtabstreitbarkeit bei sicherheitskritischen Aktionen* oder Aktionen mit monetären Konsequenzen, z. B. streckenabhängige Maut. Die *Qualität der Informationen*, die durch den Einsatz von ONP gewonnen werden können, wie z. B. Geschwindigkeit, Position oder Verkehrsdichte, ist wichtig für die weitere Verarbeitung.

### Zusammenfassung der Werte

In der folgenden Tabelle 4.1 werden die zuvor erarbeiteten Werte übersichtlich zusammengefasst.

| | Werte |
|---|---|
| spez. Dienstnutzer | • zuverlässige Handlungsempfehlungen<br>• Vertraulichkeit der Ortsinformationen, der Historie dieser und die Vorhersagen über zukünftige Ortsinformationen<br>• Verfügbarkeit der Ortsinformationen<br>• Schutz von Leib und Leben bei Nutzung |
| allg. Dienstnutzer | • Verfügbarkeit des Dienstes<br>• Verfügbarkeit der Anwendung<br>• Vertrauen in den Dienst<br>• Vertraulichkeit der personenbezogenen Daten<br>• Keine negative Beeinflussung durch Overlay<br>• Datenverarbeitung gemäß dem Bundesdatenschutzgesetz<br>• Vertraulichkeit der Kommunikation |
| Overlaynutzer | • hohe Effizienz<br>• einfache und möglichst fehlerfreie Implementierbarkeit<br>• Priorisierung der Nachrichten<br>• unterschiedliche Niveaus von Vertraulichkeit, Integrität, Authentizität, Verfügbarkeit und Datenschutz<br>• Keine Rufschädigung des Dienstanbieters<br>• Brückentechnologie<br>• Leib und Leben der Nutzer, auch aus Gewährleistungsgründen und Schutz vor Schadensersatzansprüchen<br>• Keine ungewollte gegenseitige Beeinflussung von verschiedenen Diensten<br>• Nichtabstreitbarkeit bei sicherheitskritischen Aktionen<br>• Qualität der Informationen |
| Hersteller | • Vertrauen und Ansehen<br>• Minimierung der Betriebskosten<br>• hohe Marktdurchdringung<br>• vielseitiges Angebot von Diensten<br>• Technologie Transparenz<br>• Hohes Maß an Betriebssicherheit<br>• Fehlerfreiheit der Software<br>• sichere Fehlerbehandlung<br>• Leib und Leben der Nutzer<br>• wenig Infrastruktur<br>• funktionierendes, intelligentes Empfehlungssystem |

**Tabelle 4.1:** Zusammenfassung der Werte

### 4.2.2 Akteure des Systems

Um die Suche nach den Akteuren zu erleichtern, gibt Prof. Paulus auch hier einige Fragen und Hilfestellungen. Im Folgenden werden diese um eigene Fragen erweitert, beantwortet und abschließend zusammengefasst.

#### Wer wird den Dienst (Empfehlungsdienst) benutzen?

Das Empfehlungssystem soll basierend auf den ONP-Informationen in Verbindung mit u. a. Fahrplänen, Vorschläge machen für intermodale Reisepläne. Ein Dienst Backend aggregiert die zentral abrufbaren Informationen, z. B. Bus- oder Bahnverbindungen und übermittelt sie der Anwendung auf dem Endnutzergerät. Dort kombiniert das Empfehlungssystem die zentralen mit den lokalen Informationen und spricht eine Empfehlung an den Nutzer aus. Die Anwendung führt möglicherweise weitere Kommunikation, die allerdings für diese Fragestellung unerheblich ist. Die Anwendung und das Backend stellen somit ein verteiltes Empfehlungssystem dar. Nutzer für dieses System sind im Allgemeinen *Reisende*, z. B. regelmäßige *Pendler*, die u. a. öffentliche oder private Verkehrsmittel optimiert nutzen wollen. Damit der Dienst Vorschläge machen kann, die den persönlichen Vorstellungen und Wünschen entsprechen, kann der Nutzer Profile anlegen. Dies ermöglicht auch die Optimierung von regelmäßigen Fahrten, z. B. in die Arbeit, so kann der Dienst noch vor Reiseantritt reagieren, z. B. auf spontan auftretende Verkehrslageänderung und den Nutzer auf einen geänderten Zeitplan hinweisen. Ein *Fahrzeugführer* könnte die Anwendung für die Routenplanung nutzen, um Kosten, Zeit oder Wegstrecke einzusparen. Das System wird die gefahrenen Routen und das Streckenprofil auf jeden Fall anonymisiert abspeichern, um Prognosen auf historischen Daten zu ermöglichen. Sollten Nutzerprofile angelegt werden, könnte der *Fahrzeughalter* sein Fahrtenbuch von diesem System führen lassen. Der Besitzer des Fahrzeuges könnte auch für jeden Fahrer ein Profil anlegen, um dessen Vorlieben, z. B. Fahrstil mit in die personalisierte Empfehlung einfließen zu lassen. *Fahrzeugvermietungen* könnten diese Profile auf ihren Fahrzeugen nutzen, um Fahrverhalten ihrer Kunden bewerten und z. B. Aufpreise bei schlechtem und Rabatte bei vorbildlich defensivem Fahrstil geben zu können.

#### Wer wird das Overlay benutzen?

Das ONP ist eine Basisfunktionalität, die direkt nur von den *Fahrzeugherstellern*, den *Zulieferern*, *Dienstleistern*, *Netzwerkausrüstern* oder *Herstellern von Endgeräten* benutzt wird, die es implementieren bzw. einsetzten. Der Endnutzer verwendet die Anwendungen der *Dienstanbieter*, wie z. B. das Empfehlungssystem. Andere Benutzer des ONP wären z. B. auch *Werbeagenturen für ortsbezogene Werbung*. Durch die technischen Gegebenheiten der Ad-hoc-Kommunikation, die Knappheit von Sendezeiten und Bandbrei-ten wäre ein Dienst zur ortsbezogenen Werbung ein eher niedrig priorisierter Dienst. Das Sammeln von Daten zum Fahrverhalten könnte als weiterer Dienst auch für *Autoversicherungen* interessant sein. Der *Staat*, z. B. in Form des Straßenverkehrsamtes, könnte an dem ONP interessiert sein. Zum einen um Verkehrswarnungen direkt an Betroffene weiterzugeben. Zum anderen könnte der *Staat*, vertreten durch die Verkehrsleitzentrale, ONP auch nutzen, um den Verkehrsfluss besser steuern zu können, z. B. durch RSUs, die Stauwarnungen schon auf Zubringerstraßen an die Verkehrsteilnehmer weitergeben. Diese Informationen könnten von dem Empfehlungssystem für eine Empfehlung zum Ausweichen auf eine Umfahrung o. Ä. genutzt werden. Ebenso könnte der *Staat* über ONP auch ein streckenabhängiges Mautsystem realisieren.

#### „Wer kann die Werte beschädigen? [55]“

Um diese Frage beantworten zu können müssen zunächst Werte identifiziert werden, wie im Abschnitt 4.2.1 geschehen. Daraus resultieren unterschiedliche Gruppen von Angreifern, die Interesse an den verschiedenen Werten haben. Wenn Dienstanbieter, wie z. B. der Empfehlungsdienst, personenbezogene Daten speichern, könnte die *organisierte Kriminalität*, z. B. Identitätsdiebe, die Daten entwenden und diese auf dem Schwarzmarkt in monetäre Werte umsetzten. Denkbar wäre auch, dass in die Systeme der Dienstleister oder der Endnutzer eingedrungen wird, um diese Ressourcen für andere Aktivitäten zu missbrauchen. Ein mögliches Szenario wäre der Verkauf der Kontrolle über diese Systeme oder wertvolle Informationen, z. B. geistiges Eigentum, das auf diesen Systemen gespeichert sein könnte. Neben den Verbrechern die abseits der gesetzlichen Grenzen agieren, existiert eine diskussionsfähige Grauzone, in der sowohl *eigene wie auch fremde Geheimdienste* operieren. Diese setzen ähnliche Methoden wie Kriminelle ein, um ihre eigenen Interessen zu verfolgen. Ein denkbares Interesse wäre die generelle Überwachung von Einzelnen durch die Bewegungsprofile oder die Korrelation von Profilen, um Gruppenbewegungen zu identifizieren. Die Interessensgruppe der Geheimdienste ist kritisch zu betrachten, da diese durch ihre Eigenschaften, sehr ausdauernd, zielstrebig, motiviert und Ressourcen stark zu sein, einen übermächtigen Angreifer darstellen. Als Unternehmen, das Zugang zum Wirtschaftsraum eines Landes haben möchte, muss man sich den dortigen Gesetzten gemäß verhalten und fällt somit in den Einflussbereich des jeweiligen *Staates*. Dieser könnte durch Gesetzte einen Dienstleister dazu verpflichten Informationen zur Strafverfolgung zu sammeln oder vorzuhalten. Im zellularen Bereich, z. B. bei LTE, ist dies durch einen Standard 3rd Generation Partnership Project (3GPP) 33.107[5], der die Architektur und Funktionen von Lawful Interception (LI) beschreibt, geregelt. Eine weitere Gruppe von Angreifern könnten *Konkurrenten* auf dem Markt der ITS oder *Telekommunikationsanbieter* sein, die eigene Dienste priorisieren gegenüber fremden. *Unruhestifter*, die politisch, persönlich motiviert sind oder auch aus nicht nachvollziehbaren Gründen handeln und Schaden anrichten. Eine weitere Gruppierung sind die *Nutzer der Dienste*, die einen kostenpflichtigen Dienst kostenlos nutzen wollen und dafür Schwachstellen auf technischer Ebene, z. B. Nutzer einer gestohlenen Identität (Seriennummer), oder auf Geschäftsebene, z. B. durch Angabe von falschen Kontodaten oder durch beliebig oft durchgeführte Neuregistrierung jedes Mal ein Begrüßungsgeschenk erhalten, ausnutzen.

Wie diese Angreifer ihre unterschiedlichen Ziele erreichen könnten wird in Abschnitt 4.5 ausgeführt. Eine Zusammenfassung der angreifenden Akteure und deren eigenen Interessen wird in Tabelle 4.2 gegeben.

| | Akteure | Interessen |
|---|---|---|
| Angreifer | • Geheimdienste<br>• organisierte Kriminalität<br>• Staat<br>• Konkurrenten<br>• Telekommunikationsanbieter<br>• Kleinkriminelle<br>• Unruhestifter<br>• Nutzer der Dienste | • Monetäre Vorteile (Betrug, Verkauf von Daten oder Ressourcen, Benachteiligung des Systems, um seine eigene Lösung besser zu verkaufen, usw.)<br>• Totale Überwachung (Verbrechensprävention, Kontrolle durch Omnipräsenz)<br>• Gezielte Überwachung (Strafverfolgung, systematische Schikane)<br>• politische oder persönliche Motivation<br>• Stören oder Zerstören, um des Aktes willen ohne intrinsische Motivation |

**Tabelle 4.2:** Zusammenfassung von Akteuren und Interessen der Angreifer

### Zusammenfassung von Werten und Akteuren

Die in den vorherigen Abschnitten beschriebenen Werte und Akteure werden in der folgenden Tabelle 4.3 zusammengefasst. Die Zusammenfassung der Interessen und möglichen Akteure als Angreifer sind in der separaten Tabelle 4.2 aufgeführt.

| | Akteure | Werte |
|---|---|---|
| spez. Dienstnutzer | • Reisende/Pendler<br>• Fahrzeugführer<br>• Fahrzeughalter<br>• Fahrzeugvermietung | • zuverlässige Handlungsempfehlungen<br>• Vertraulichkeit der Ortsinformationen, der Historie dieser und die Vorhersagen über zukünftige Ortsinformationen<br>• Verfügbarkeit der Ortsinformationen<br>• Schutz von Leib und Leben bei Nutzung |
| allg. Dienstnutzer | Dienstspezifisch | • Verfügbarkeit des Dienstes<br>• Verfügbarkeit der Anwendung<br>• Vertrauen in den Dienst<br>• Vertraulichkeit der personenbezogenen Daten<br>• Keine negative Beeinflussung durch Overlay<br>• Datenverarbeitung gemäß dem Bundesdatenschutzgesetz<br>• Vertraulichkeit der Kommunikation |
| Overlaynutzer | • Fahrzeughersteller<br>• Zulieferer<br>• Netzwerkausrüster<br>• Endgerätehersteller<br>• Dienstleister<br>• Dienstanbieter<br>• Werbeagenturen<br>• Autoversicherungen<br>• Staat | • hohe Effizienz<br>• einfache und möglichst fehlerfreie Implementierbarkeit<br>• Priorisierung der Nachrichten<br>• unterschiedliche Niveaus von Vertraulichkeit, Integrität, Authentizität, Verfügbarkeit und Datenschutz<br>• Keine Rufschädigung des Dienstanbieters<br>• Brückentechnologie<br>• Leib und Leben der Nutzer, auch aus Gewährleistungsgründen und Schutz vor Schadensersatzansprüchen<br>• Keine ungewollte gegenseitige Beeinflussung von verschiedenen Diensten<br>• Nichtabstreitbarkeit bei sicherheitskritischen Aktionen<br>• Qualität der Informationen |
| Hersteller | Fahrzeughersteller (Konsortium) | • Vertrauen und Ansehen<br>• Minimierung der Betriebskosten<br>• hohe Marktdurchdringung<br>• vielseitiges Angebot von Diensten<br>• Technologie Transparenz<br>• Hohes Maß an Betriebssicherheit<br>• Fehlerfreiheit der Software<br>• sichere Fehlerbehandlung<br>• Leib und Leben der Nutzer<br>• wenig Infrastruktur<br>• funktionierendes, intelligentes Empfehlungssystem |

**Tabelle 4.3:** Zusammenfassung Akteure und Werte

### 4.2.3 Operationen und Aktionen des Systems

Operationen und Aktionen beschreiben Zusammenhänge zwischen Akteuren und Werten. Diese Beziehungen sind wichtig, um zu erkennen wie ein System mit bzw. auf den Werten des Systems oder seiner Benutzer arbeitet. Diese Phase dient ebenfalls dazu Werte zu identifizieren, die außerhalb des Einflussbereiches des Systems liegen.

#### Welche Operationen sind auf den Werten möglich?

Operationen oder Aktionen sind Aktivitäten, die einen Wert oder den Zustand eines Wertes verändern. In der folgenden Tabelle 4.4 sind gemeinsame Werte zusammengefasst und mit möglichen Aktionen versehen. Die ersten vier Spalten geben an, welche Akteure betroffen sind.

| H | O | A | S | Werte | Aktionen |
|---|---|---|---|---|---|
| ✓ | ✓ | × | ✓ | zuverlässige *Handlungsempfehlung* | werden angefragt, berechnet (erbracht), genutzt (befolgt), verworfen, bezahlt |
| ✓ | ✓ | ✓ | ✓ | Vertraulichkeit der *Profildaten* | diese werden erstellt, gelöscht, exportiert, zur Authentifizierung und Autorisierung verwendet, zur Bildung einer Identität genutzt, zur Vorlieben- und Verhaltensanalyse mit historischen Nutzerverhaltensdaten befüllt, durch Nutzer aktualisiert, zur Speicherung der personalisierten Diensteinstellung benötigt |
| × | ✓ | ✓ | × | Vertraulichkeit der *Nachrichten* | diese werden definiert, abgesendet, empfangen, verworfen, aggregiert, priorisiert, wiederholt/weitergeleitet, analysiert/bewertet, validiert, mit Information gefüllt |
| ✓ | ✓ | × | ✓ | Qualität der *Informationen* | diese werden z. B. durch externe Datenquellen, Fahrplanauskunft, gesammelt, generiert z. B. durch Sensoren, erstellt z. B. durch Nutzer, Dienstleister, Straßenbauamt bei Baustellen |
| ✓ | ✓ | ✓ | ✓ | *Verfügbarkeit* der Informationen, Dienste | kann nicht gegeben, eingeschränkt, ausreichend vorhanden sein |
| ✓ | ✓ | × | ✓ | Betriebssicherheit, *Leib und Leben der Nutzer* | muss geschützt werden, kann verloren gehen, ist das höchste Gut des Menschen |
| ✓ | ✓ | ✓ | × | *Vertrauen und Ansehen* | wird gewonnen, verloren, missbraucht |
| × | ✓ | ✓ | × | negative *Beeinflussung* | wird erzeugt, kann ausbleiben, zunehmen durch starke Nutzung |
| ✓ | ✓ | × | ✓ | *Ressourcen*, wie z. B. Datenvolumen, Bandbreite, Energie | muss erzeugt, verbraucht (genutzt), bezahlt werden |
| H = Hersteller, O = Overlaynutzer, A = allg. Dienstnutzer, S = spez. Dienstnutzer | | | | | |

**Tabelle 4.4:** Zusammenfassung Akteure, Werte, Operationen

## 4.3 Ein Architekturentwurf

Grundsätzlich steht man in diesem Schritt vor einem Henne-Ei-Problem [55], denn eine gute System- oder Softwarearchitektur berücksichtigt Sicherheitsanforderungen. Wobei einige Sicherheitsvorgaben möglicherweise erst bei der Durchführung der Bedrohungsmodellierung zu Tage treten. Aber um eine Bedrohungsmodellierung durchführen zu können, benötigt man zumindestens eine grobe Architektur, um Angriffe und mögliche vorhandene Schwachstellen diskutieren zu können.

Ein erster grober Architekturentwurf wurde im Abschnitt 2.3 vorgestellt, dieser leitet sich aus den zuvor beschriebenen Anforderungen unter A.1 ab. Der Prozess der Bedrohungsmodellierung ist idealerweise ein iterativer Prozess und wird bei Neuerungen oder Veränderungen im Projekt aktualisiert.

Wichtige Informationen bezüglich der Architektur sind die Datenhaltung und Datenflüsse, da sich aus diesen sogenannte „Vertrauensgrenzen [55]“ ableiten lassen. Diese Vertrauensgrenzen sind Schnittstellen zwischen einzelnen Komponenten eines Systems, einer Software oder sogar einzelner Programmteile. Über diese Grenzen hinweg fließen Daten nur eingeschränkt, weil sich die Kommunikationspartner nicht bedingungslos vertrauen.

Sicherheit und Vertrauen erhielten erst später in der Pionierzeit der Computer- und Netzwerktechnologien die nötige Aufmerksamkeit, siehe [71]. Dies wird im Folgenden am Beispiel eines Firmennetzwerkes verdeutlicht; denn zunächst sicherte man ein Firmennetz durch eine einzige Vertrauensgrenze, einen Perimeterschutz, eine Firewall. Dies bedeutet ein Angreifer musste nur ein Hindernis überwinden und er befand sich im vertrauenswürdigen Teil des Netzes und konnte meist ungehindert sein Unwesen treiben. Heutzutage sollte man eher von einzelnen Vertrauensinseln ausgehen mit sehr vielen Vertrauensgrenzen, die anderen Systemteilen nur bedingt vertrauen. Dies erhöht zwar die Komplexität und den Verwaltungsaufwand, allerdings müssen Angreifer mehrere Hürden überwinden und dies am besten immer von Neuem, wenn sie an Informationen innerhalb einer Vertrauensinsel gelangen wollen. Vertrauensgrenzen sind daher sehr wichtig für die Sicherheitsarchitektur eines Systems, Netzes oder einer Software.

Die vorhandene Grobarchitektur wird in folgender Abbildung 4.2 um Vertrauensgrenzen erweitert, jede rechteckige Komponente stellt eine Vertrauensgrenze dar.

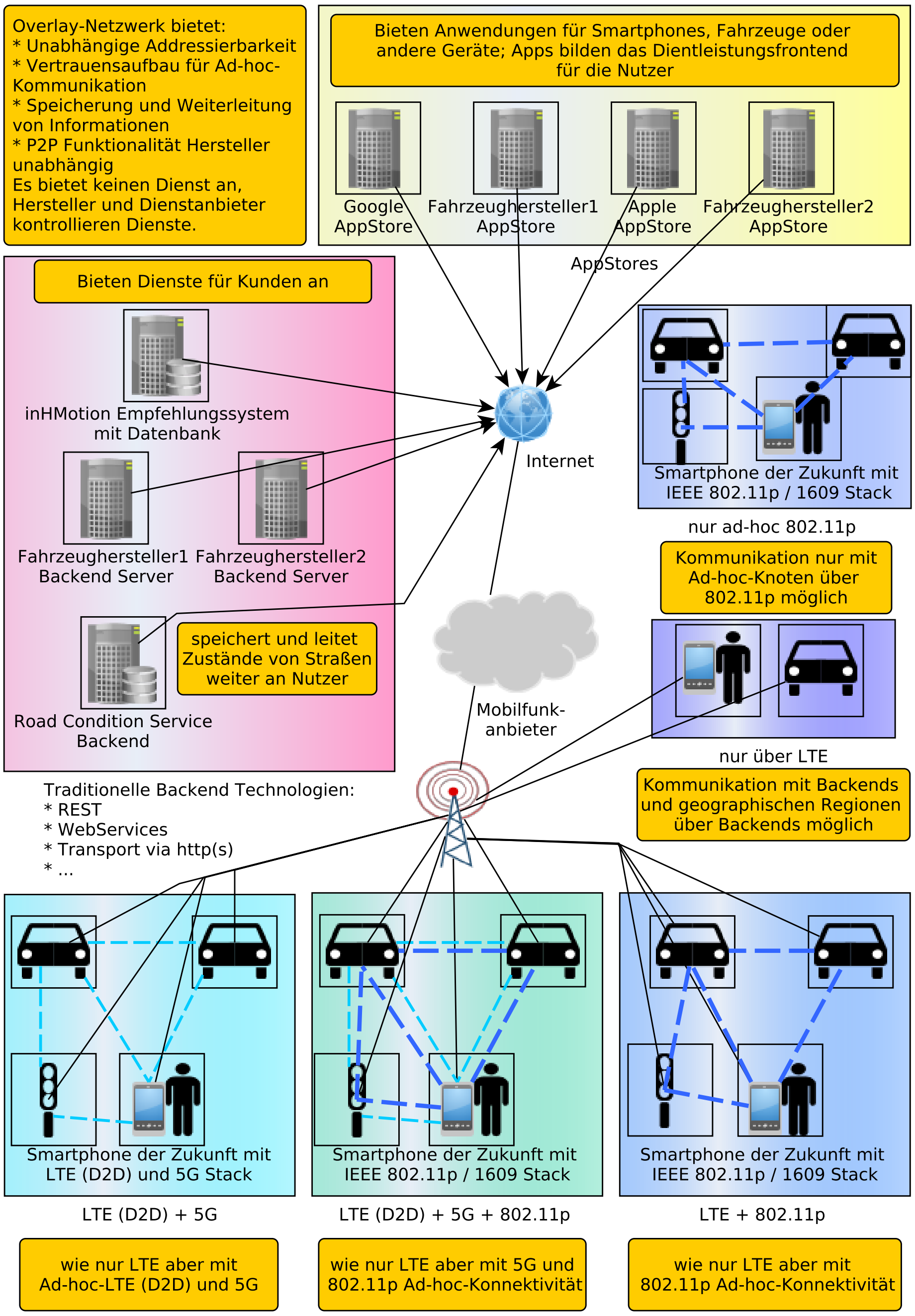

**Abbildung 4.2:** Grobe Systemarchitektur mit Vertrauensgrenzen

## 4.4 System zerteilen

Nach dem Überblick über die Systemarchitektur folgt die Überprüfung, ob die aktuelle Architektur noch weiter zerteilt werden kann oder muss. Die richtige Granularität wurde erreicht, wenn eine weitere sinnvolle Unterteilung der Datenhaltungsblöcke nicht mehr möglich ist [55]. Die in Abbildung 4.2 dargestellte Architektur und ihre Vertrauensgrenzen können noch weiter zerlegt werden, da noch keine Datenhaltung und Datenflüsse ersichtlich sind.

Dazu wird eine detailliertere Architektur entworfen, diese fasst einige Komponenten, wie z. B. AppStores, Dienstanbieter oder die unterschiedlichen Konnektivitätsszenarien, zusammen und stellt dafür in der Grafik 4.3 Datenhaltung und Datenflüsse dar. Die Pfeilspitze deutet an die Komponente, an die Daten fließen, die sogenannte Senke. Die Komponente, aus der Daten fließen, ist die Quelle. Diese Nomenklatur wird in dieser Arbeit beibehalten. Jede rechteckige oder zylinderförmige Komponente stellt eine Vertrauensinsel dar. Elemente in einer solchen Insel sind innerhalb einer Vertrauensgrenze und genießen ein gleich hohes Vertrauensverhältnis. Vertrauensgrenzen können verschachtelt sein, wie z. B. der ONP-Message-Filter, der innerhalb eines ONP fähigen Gerätes liegt. Jede Beziehung über eine Insel hinaus, meist durch eine Interaktion, wie z. B. Datenflüsse, stellt eine Überschreitung von einer Vertrauensgrenze dar. An dieser muss entschieden werden, ob eine Interaktion, z. B. autorisiert ist oder zu verarbeitende Daten integer und valide sind. Basierend auf dem aktuellen Grobkonzept und der bereits präsentierten Zerteilung im Datenfluss und dem Vertrauensgrenzen-Diagramm 4.3, wird keine weitere Zerlegung vorgenommen. Dieser Schritt kann wiederholt werden, sollten weitere Detailkonzepte bezüglich eingesetzter Komponenten oder Datenhaltung erarbeitet werden.

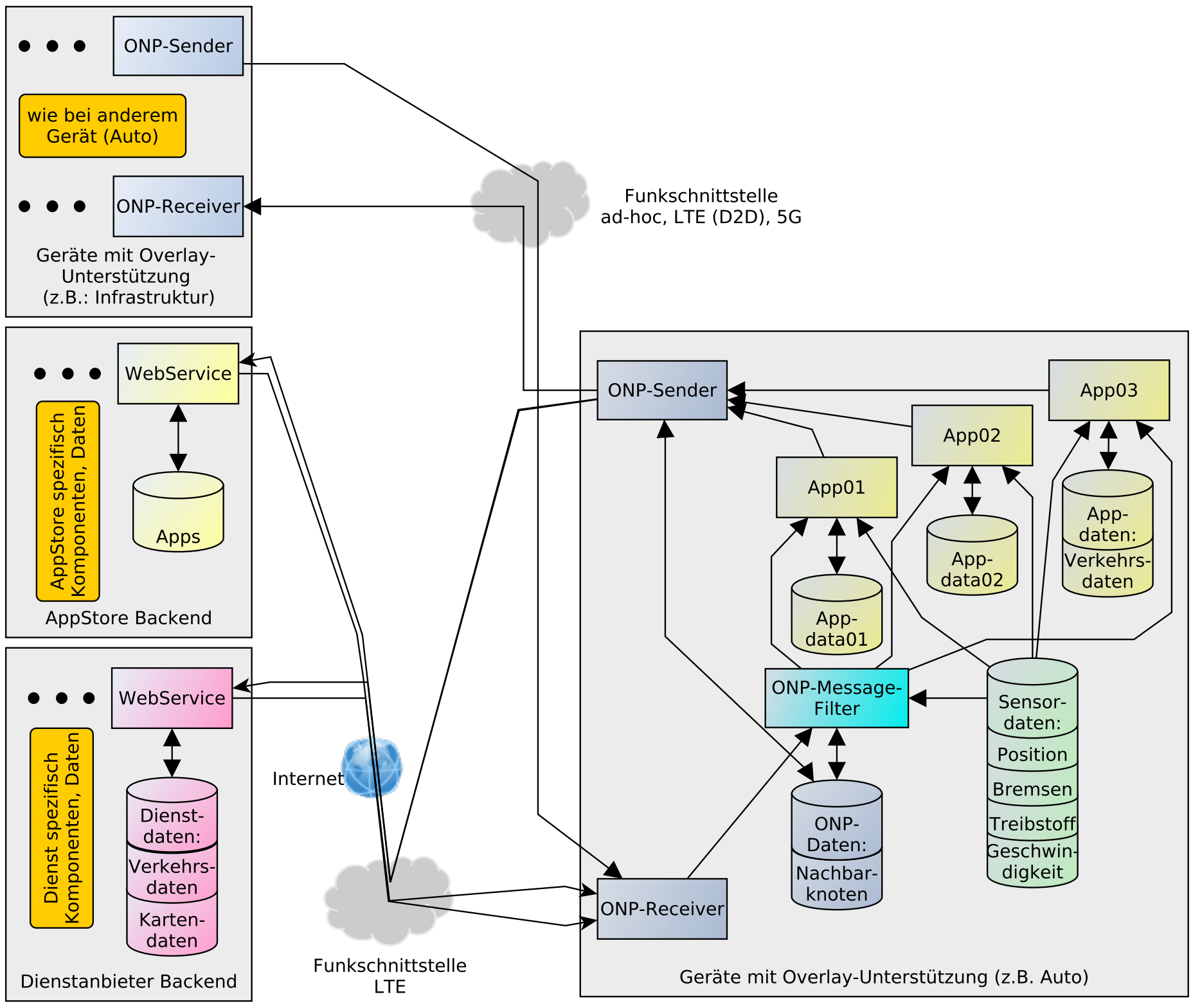

**Abbildung 4.3:** Datenflüsse in abstrahierter Systemarchitektur

## 4.5 Bedrohungen identifizieren

Diese Arbeit verwendet u. a. die Brainstorming Methode und Angriffsbäume, in Verbindung mit zusätzlicher Recherche über bekannte Bedrohungen im Bereich CPS, ITS, Ad-hoc-Kommunikation, Overlay-Netzwerke und Erfahrung im Bereich des Penetration Testing. Um Bedrohungen, Angriffspunkte und Schwachstellen zu finden und zu identifizieren, wird die Architektur mit der Einstellung des Angreifers untersucht. Dabei hilft die Suche nach Aktionen, die Werte der unterschiedlichen Akteure, z. B. Endkunde, Dienstleister, gefährden. Die Ergebnisse dieser Phase werden in der Nächsten dokumentiert, daher wird an dieser Stelle nur ein kleiner Auszug aus denkbaren Angriffen gegeben.

### Welche Aktionen gefährden Werte?

Um die Suche nach Bedrohungen zu erleichtern hilft die Beantwortung folgender Frage: Welche Aktionen gefährden Werte? Angreifer gefährden Werte, die ...

- die Qualität der Daten und damit den Wert manipulieren wollen, könnten dies tun durch ...
  - Nicht-Weiterleiten von Paketen bzw. Daten.
  - Manipulation der Aggregation von Daten.
  - Aussenden von manipulierten Informationen.
- den Wert der Dienstleistungen mindern wollen, könnten dies tun durch ...
  - Generierung von Anfragen, die unnötig Ressourcen des Dienstes verbrauchen.
  - aggressive Sendeleistung, die die beschränkte Bandbreite des Netzwerks verbrauchen.
  - Manipulation von Anfragen, um sich Vorteile bei Diensten zu erschleichen.
- das Ansehen und Vertrauen der Akteure in das System schwächen wollen, könnten dies tun durch ...
  - Entwendung der Nutzerprofile, um diese zu veröffentlichen oder zur Verfolgung oder Überwachung der Endnutzer zu verwenden.
  - eigene Dienste, die alle Anderen beeinträchtigen, durch Ressourcenverbrauch, oder die sich selbst zu Lasten Anderer bereichern.
  - Manipulation von Vertrauen innerhalb des Netzwerkes zu ihren Gunsten.
- das Leib und Leben der Nutzer gefährden möchten, könnten dies tun durch ...
  - Manipulation von sicherheitskritischen Sensoren.
  - Manipulation von sicherheitskritischer Software.
  - Aussenden von manipulierten Informationen.

**Wie können Werte gefährdet werden?**

Um dieses kreative Vorgehen durch einen strukturierten Ansatz zu unterstützen, wird das Konstruieren von Angriffsbäumen gewählt. Wie das Konzept der Angriffsbäume anzuwenden ist, wird im Kapitel 2 im Abschnitt 2.4 beschrieben. Im Folgenden wird ein Angriffsbaum dargestellt, der das Ziel verfolgt:
*Einen ordnungsgemäßen Betrieb zu behindern DoS.*

1. Überfluten mit Datenpaketen im zellularen Netzwerk B.7
2. Überfluten mit Datenpaketen im Ad-hoc-Netzwerk B.6
3. Störsender im zellularen Frequenzbereich B.5
4. Störsender im Ad-hoc-Frequenzbereich B.4
5. Dienste mit hohem Ressourcenverbrauch anbieten B.46

6. Gezieltes Unterdrücken von einzelnen Informationen
   (a) Gezieltes Unterdrücken von einzelnen Informationen via Ad-hoc-Kommunikation B.8
   (b) Gezieltes Unterdrücken von einzelnen Informationen via zellularer Kommunikation B.9
7. Beeinflussung von Softwarekomponenten durch manipulierte andere Komponenten B.48

Die folgenden Werte sind durch dieses Ziel und die einzelnen Bedrohungen betroffen:

- allg. Dienstnutzer:
  - Verfügbarkeit des Dienstes
  - Vertrauen in den Dienst
- Overlaynutzer:
  - hohe Effizienz
  - Priorisierung der Nachrichten
  - Rufschädigung des Dienstanbieters
  - Keine ungewollte gegenseitige Beeinflussung von verschiedenen Diensten
- Hersteller:
  - wenig Infrastruktur
  - Minimierung der Betriebskosten
  - Vertrauen und Ansehen des Herstellers

Die im Angriffsbaum angegebenen Referenzen verweisen im Anhang B auf die jeweilige Bedrohung. An der obigen Auflistung ist erkennbar, dass Bedrohungen auf unterschiedlichen logischen Ebenen identifiziert werden können. Die Bedrohung 6., das gezielte Unterdrücken von Informationen, besitzt selbst keine eigene Bedrohungsbewertung durch das OWASP Risk Rating. Diese Bewertung lässt sich durch Übernehmen der maximalen Werte der Kindknoten ableiten. Durch diese Verfahren kann sichergestellt werden, dass es zu jeder Bedrohung der ersten logischen Ebene eine Bewertung gibt. Dadurch kann ein gewisses Maß an Vollständigkeit der Betrachtung gewährleistet werden. Die vollständige Auflistung aller konstruierten Angriffsbäume ist im Anhang unter A.2 zu finden.

## 4.6 Bedrohungen dokumentieren

Ein wichtiger Schritt ist es nun die identifizierten Bedrohungen zu dokumentieren. Dazu wird wie empfohlen eine Tabellenkalkulation verwendet. Um Schnittstellen angeben zu können, werden in Abbildung 4.4 zusätzlich die Schnittstellen zwischen den einzelnen Komponenten benannt. Eine voll-ständige Auflistung aller Bedrohungen ist im Anhang B zu finden. Für jede Bedrohung wird eine eigene Tabelle mit allen relevanten Informationen aufgeführt. Diese Darstellung wurde gewählt, da sich eine Tabellenkalkulation nur bedingt für die Darstellung der Informationen in dieser Arbeit eignet, siehe 4.5. Auf der beiliegenden CD ist das originale Dokument beigefügt.

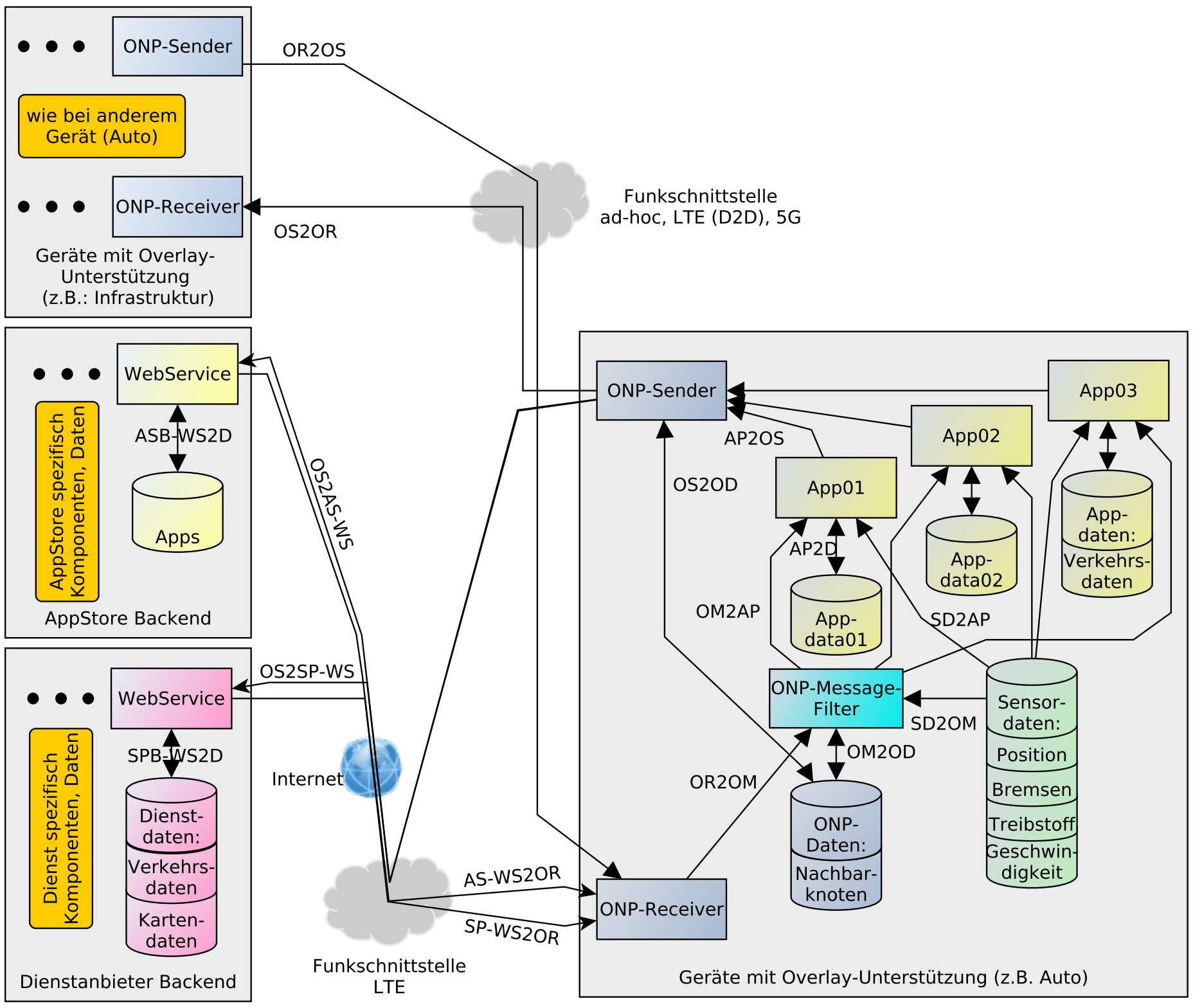

**Abbildung 4.4:** Datenflüsse in Systemarchitektur inkl. Schnittstellennamen

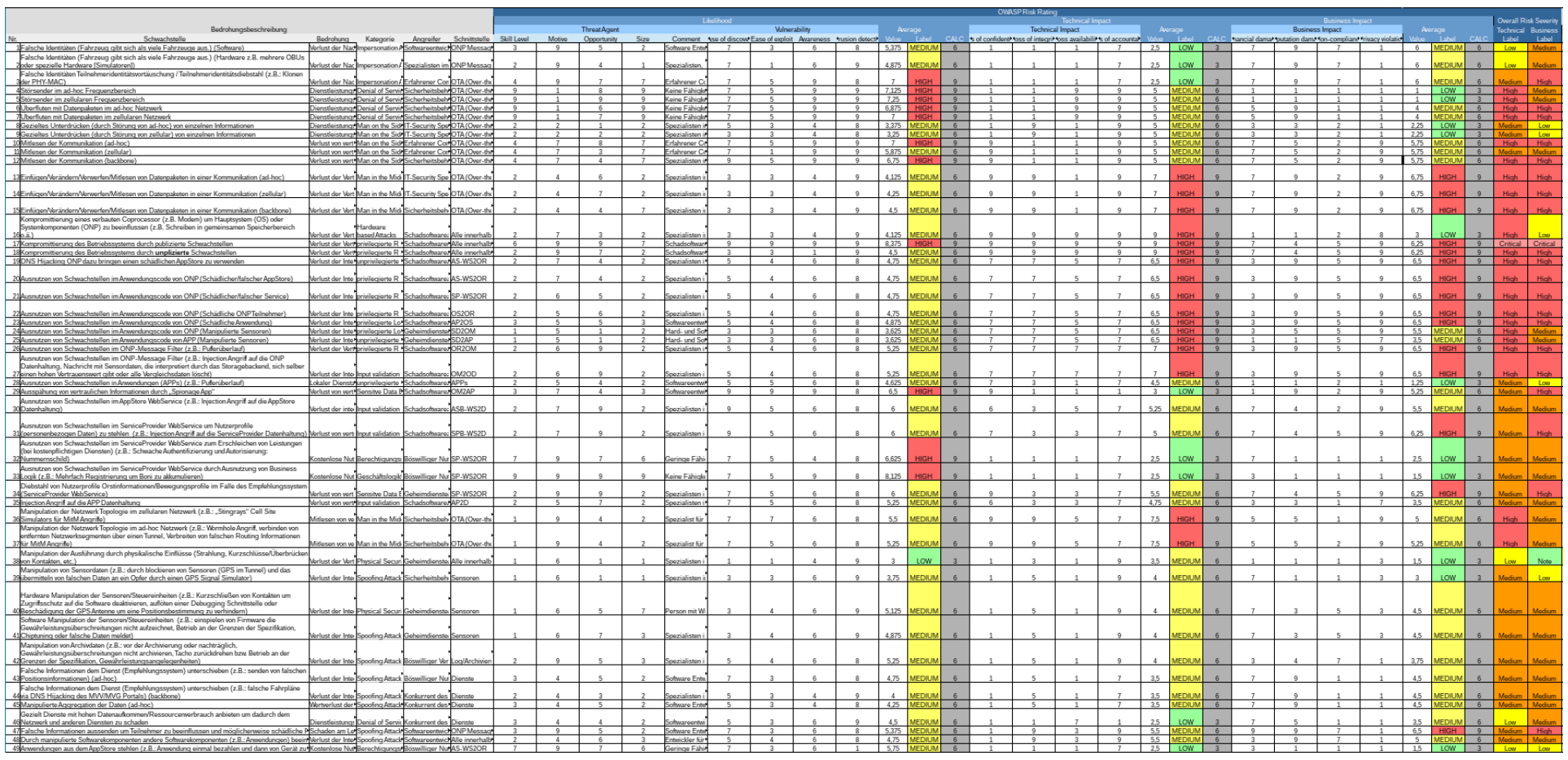

**Abbildung 4.5:** Bildschirmfoto mit Auszügen aus der Tabellenkalkulation zur Bedrohungsmodellierung

## 4.7 Bedrohungen bewerten

Wie bereits zuvor beschrieben ist es aus Ressourcengründen selten möglich für jede Bedrohung geeignete Maßnahmen umzusetzen, daher muss eine Risikobewertung der Bedrohungen stattfinden. Anhand dieser Bewertung können gezielt Maßnahmen definiert und umgesetzt werden, um das Gesamtrisiko zu senken. Als Bewertungsmethode wird das OWASP Risk Rating eingesetzt. Das Vorgehen nach OWASP Risk Rating wird in Kapitel 2 im Abschnitt 2.5 beschrieben.
Im Folgenden werden die Bedrohungen aufgeführt, die als Gesamtrisiko den Wert „hoch“ oder „kritisch“ tragen.

Die Bedrohungen, bei denen das Gesamtrisiko von der technischen und der wirtschaftlichen „hoch“ oder „kritisch“ ist:

- B.6 Überfluten mit Datenpaketen im Ad-hoc-Netzwerk
- B.7 Überfluten mit Datenpaketen im zellularen Netzwerk
- B.10 Mitlesen der Kommunikation (ad-hoc)
- B.12 Mitlesen der Kommunikation (backbone)
- B.13 Einfügen/Verändern/Verwerfen/Mitlesen von Datenpaketen in einer Kommunikation (ad-hoc)
- B.14 Einfügen/Verändern/Verwerfen/Mitlesen von Datenpaketen in einer Kommunikation (zellular)
- B.15 Einfügen/Verändern/Verwerfen/Mitlesen von Datenpaketen in einer Kommunikation (backbone)

- B.17 Kompromittierung des Betriebssystems durch publizierte Schwachstellen
- B.18 Kompromittierung des Betriebssystems durch unpublizierte Schwachstellen
- B.19 DNS-Hijacking, ONP dazu bringen einen schädlichen AppStore zu verwenden
- B.20 Ausnutzen von Schwachstellen im Anwendungscode von ONP (Schädlicher/falscher AppStore)
- B.21 Ausnutzen von Schwachstellen im Anwendungscode von ONP (Schädlicher/falscher Service)
- B.22 Ausnutzen von Schwachstellen im Anwendungscode von ONP (Schädliche ONP-Teilnehmer)
- B.23 Ausnutzen von Schwachstellen im Anwendungscode von ONP (Schädliche Anwendung)
- B.26 Ausnutzen von Schwachstellen im ONP-Message-Filter (z.B. Pufferüberlauf)
- B.27 Ausnutzen von Schwachstellen im ONP-Message-Filter (z.B. Injection Angriff auf die ONP-Datenhaltung, Nachricht mit Sensordaten, die interpretiert durch das Storagebackend, sich selber einen hohen Vertrauenswert gibt oder alle Vergleichsdaten löscht)
- B.31 Ausnutzen von Schwachstellen im ServiceProvider-WebService, um Nutzerprofile (personenbezogene Daten) zu stehlen (z.B. Injection Angriff auf die ServiceProvider-Datenhaltung)

Die Bedrohungen, bei denen das Gesamtrisiko aus wirtschaftlicher Sicht „hoch“ ist und die nicht schon in der vorherigen Auflistung enthalten sind:

- B.3 Falsche Identitäten Teilnehmeridentitätsvortäuschung / Teilnehmeridentitätsdiebstahl (z.B.: Klonen der PHY-MAC)
- B.29 Ausspähung von vertraulichen Informationen durch „Spionage App“
- B.34 Diebstahl von Nutzerprofilen Orstinformationen/Bewegungsprofile im Falle des Empfehlungssystems (ServiceProvider WebService)
- B.47 Falsche Informationen aussenden, um Teilnehmer zu beeinflussen und möglicherweise schädliche Reaktionen der Nutzer oder von sicherheitskritischen Anwendungen zu provozieren

Die Bedrohungen, bei denen das Gesamtrisiko aus technischer Sicht „hoch“ ist und nicht bereits zuvor aufgelistet wurden:

- B.4 Störsender im Ad-hoc-Frequenzbereich
- B.5 Störsender im zellularen Frequenzbereich
- B.16 Kompromittierung eines verbauten Coprocessors (z.B. Modem), um Hauptsystem (OS) oder Systemkomponenten (ONP) zu beeinflussen (z.B. Schreiben in gemeinsamen Speicherbereich o. Ä.)

- B.24 Ausnutzen von Schwachstellen im Anwendungscode von ONP (Manipulierte Sensoren)
- B.25 Ausnutzen von Schwachstellen im Anwendungscode von APP (Manipulierte Sensoren)
- B.36 Manipulation der Netzwerktopologie im zellularen Netzwerk (z.B.: „Stingrays“ Cell Site Simulators für MitM Angriffe)
- B.37 Manipulation der Netzwerktopologie im Ad-hoc-Netzwerk (z.B.: Wormhole Angriff, verbinden von entfernten Netzwerksegmenten über einen Tunnel, Verbreiten von falschen Routing-Informationen für MitM Angriffe)

Bei Einsatz des OWASP Risk Rating in der betrieblichen Praxis und bei ausreichend fundierten statistischen Informationen zu den geschäftlichen Auswirkungen (Business Impact), sollten nur die Bedrohungen mit dem wirtschaftlichen Gesamtrisiko „hoch“ oder „kritisch“ [51] betrachtet werden. Da diese Informationen nicht vorliegen und die Abschätzung der geschäftlichen Auswirkungen auf einer Vorstellung einer Interessengemeinschaft beruht, werden auch die Bedrohungen aus technischer Sicht betrachtet.

## 4.8 Sicherheitsanforderungen

Aus den zuvor aufgeführten Bedrohungen werden folgende Sicherheitsanforderungen, die den Bedrohungen entgegenwirken sollen, abgeleitet:

1. Transparenz der Qualität von Identitäten
2. Bereitstellung von Identitäten unterschiedlicher Qualität je nach Anwendung
3. Schadensbegrenzung bei schädlicher Kommunikation inkl. der Verfügbarkeit von Informationen, z.B. schädliche Veränderungen der Netzwerktopologie (ad-hoc + zellular)
4. Erkennung von schädlichen Kommunikationsteilnehmern (ad-hoc, zellular)
5. Gewährleisten von vertraulicher Kommunikation (Ende-zu-Ende, Link)
6. Gewährleisten von Integrität und Authentizität der Kommunikation (ad-hoc, zellular und backbone)
7. Schädliche Anwendungen müssen verhindert oder in ihren Möglichkeiten eingeschränkt werden, z.B. dürfen Anwendungen andere Anwendungen und Betriebssystemkomponenten nicht manipulieren können.
8. Softwarekomponenten müssen aktualisierbar sein.
9. Softwarekomponenten müssen Schutzvorkehrungen gegen unbekannte Schwachstellen enthalten.
10. Authentizität und Integrität von nachgeladenem ausführbarem Code muss gewährgeleistet sein.

11. Zugriff auf vertrauliche Informationen darf nur für berechtigte Personen/Komponenten möglich sein.
12. Verwendung von weniger störsenderanfälligen Kommunikationsschnittstellen.
13. Auf sichere Programmierung sollte geachtet werden.
14. Gegenseitige Beeinflussung von Hardwarekomponenten sollte minimiert werden.

In folgender Tabelle 4.5 ist die Zuordnung von Bedrohung zu abgeleiteter Sicherheitsanforderung dargestellt.

| *Bedrohungen* | *Anforderungen* |
|---|---|
| B.3 | 1, 2 |
| B.4, B.5 | 12 |
| B.6, B.7 | 4 |
| B.10, B.12 | 5 |
| B.13, B.14, B.15, B.47 | 6 |
| B.16 | 14 |
| B.17 | 8 |
| B.18 | 9 |
| B.19 | 10 |
| B.20, B.21, B.22, B.23, B.24, B.25, B.26, B.27 | 13 |
| B.29 | 7 |
| B.31, B.34 | 11 |
| B.36, B.37 | 3 |

**Tabelle 4.5:** Zuordnung Bedrohung zu Sicherheitsanforderung

Da die Bedrohungen erfolgreich identifiziert und Sicherheitsanforderungen erhoben wurden, kann nun die Umsetzung der Anforderungen innerhalb des Systementwurfs erfolgen. Damit befasst sich eingehend das folgende Kapitel.

# 5. Gegenmaßnahmen

Das zuvor beschriebene Vorgehen nach Paulus [55] sieht das Definieren von Gegenmaßnahmen als den letzten Schritt einer Bedrohungsmodellierung vor. Die Betrachtung der Gegenmaßnahmen wird in dieser Arbeit in einem eigenen Kapitel behandelt und dient lediglich der Trennung von Problemen und deren Lösung, dies stellt keinen Bruch zum Vorgehen nach Paulus [55] dar. Zunächst werde für jede Sicherheitsanforderung Lösungsansätze beschrieben. Die Lösungsansätze sind als Vorschläge zu verstehen, nicht als eine finale Festlegung auf bestimmte Verfahren oder Vorgehen. Eine definitive Auswahl von Methoden und Techniken hängt von Eigenschaften und Zielen ab, die aus Komplexitätsgründen hier nicht behandelt werden können. Eine bestimmte Lösungsmöglichkeit, die Einführung eines Reputationssystems, wird detaillierter ausgeführt, da diese eine ganze Gruppe der Sicherheitsanforderungen betrifft. Diese Lösung wird zunächst beschrieben, simuliert und abschließend diskutiert.

## 5.1 Lösungsansätze für Sicherheitsanfoderungen

Die Reihenfolge der Sicherheitsanforderungen, wie sie in diesem Kapitel abgearbeitet werden, hat keine Bedeutung und wurde lediglich aus stilistischen Gründen so gewählt. Jede Sicherheitsanforderung wird in ihrem eigenen Abschnitt kurz erläutert. Es wird zu jeder Anforderung mindestens ein Lösungsvorschlag gegeben und gegebenenfalls kurz diskutiert. Diese Lösungsvorschläge sind Maßnahmen, die durch das inHMotion Projekt umgesetzt werden könnten, um den Anforderungen gerecht zu werden. Ebenso obliegt es dem Projekt Maßnahmen oder Verantwortung für diese auf andere Personen zu übertragen, z. B. die Erfüllung der Anforderungen an die Hardwaresicherheit könnte an den Produzenten delegiert werden. Im Folgenden werden nun die Sicherheitsanforderungen aus der Auflistung im Abschnitt 4.8 behandelt.

### 5.1.1 Gewährleisten von vertraulicher Kommunikation

Diese Sicherheitsanforderung gibt vor, dass vertrauliche Kommunikation möglich ist und auch gewährleistet werden kann. Je nach Art der Kommunikation existieren bereits verschiedene Lösungsansätze. Um vertrauliche Kommunikation gewährleisten zu können, muss darauf geachtet werden, eine sichere Implementierung und Konfiguration zu wählen, bzw. das Aktualisieren oder Austauschen schnell und einfach zu gestalten. Als Lösung für zentralisierte Kommunikation, klassische Client-zu-Server-Kommunikation, ist z. B. Transport Layer Security (TLS) [7]. Bei dezentral organisierter Kommunikation könnte „SECSPP“ [44], das auf zentrale Vertrauensanker setzt oder ein Verfahren, das auf Clusterbildung [9] und unterschiedlichen Schutzlevels basiert, eingesetzt werden. Diese Schutzlevel erstrecken sich von keiner Verschlüsselung bis hin zu dezentraler PKI, inklusive Signaturen von dezentralen CAs, den Clusterköpfen. Abhängig von den Vertraulichkeitsanforderungen, z. B. Ende-zu-Ende, muss das passende Verfahren gewählt werden. Dies bedeutet, dass – gerade wenn die Kommunikation über einen zentralen Dienst geführt wird – nicht nur der Kanal zu diesem Dienst abgesichert sein muss, sondern es muss zusätzlich, um die Ende-zu-Ende Anforderung zu erfüllen, sichergestellt werden, dass nur derjenige die Information nutzen kann, für den sie bestimmt war.

### 5.1.2 Gewährleisten von Integrität und Authentizität der Kommunikation

Die meisten Ansätze, die im vorherigen Abschnitt 5.1.1 angesprochen wurden, enthalten Maßnahmen, um Integrität und Authentizität bereitzustellen. Abhängig davon, ob symmetrische oder asymmetrische Verfahren eingesetzt werden, bieten sich hierfür unterschiedliche Verfahren an, siehe [66]. Zum Beispiel existieren einige Betriebsmodi für Block Chiffren, um die Effizienz zu steigern, die Authentizität, Integrität und Vertraulichkeit zugleich gewährleisten zu können, z. B. „EAX“ [10], „CCM“ [79] oder „GCM“ [50]. Daher ist es empfehlenswert einen ganzheitlichen Ansatz, der sowohl 5.1.1 wie auch 5.1.2 in einem Zug befriedigen kann, zu wählen.

### 5.1.3 Schädliche Anwendungen verhindern und Möglichkeiten einschränken

Um schädliche Anwendungen auf den Endgeräten der Nutzer zu verhindern, existiert das Model der geschützten Bereiche zum Bezug von Software, auch „AppStores“ genannt. Alle großen Hersteller von mobilen Plattformen setzen dieses Model mit variierendem Erfolg ein. Denn der Erfolg dieses Ansatzes liegt in der Qualität der Untersuchung von Anwendungen, die über den „AppStore“ vertrieben werden sollen. Natürlich müssen die tatsächlichen Übertragungswege, über die ausführbarer Code nachgeladen wird, gemäß Anforderung 5.1.2 geschützt sein, bzw. eine Überprüfung der Integrität und Authentizität muss lokal durchführbar sein. Andere Modelle, wie Paketquellen o. Ä., setzen ebenso auf Singnaturen, allerdings wird dort dem sogenannten Paket Maintainer vertraut, dass er keinen Schadcode einbaut bzw. dass der Entwickler der Software dies nicht getan hat. Mehr zu diesem Problem in Abschnitt 5.1.6.
Um die Möglichkeiten von schädlicher Software einzuschränken, bietet sich ein Verfahren an, das als „Sandboxing“ bezeichnet wird. Unter dem Begriff lassen sich verschiedenste Technologien fassen, dennoch ist das Ziel das Gleiche. Das Ziel ist es, alle Anwendungen, inklusive fehlerhafte oder böswillige, als nicht vertrauenswürdigen Code zu betrachten und diese losgelöst vom Betriebssystem zur Ausführung zu bringen, ohne dabei die Sicherheit und Stabilität des Systems zu beeinträchtigen. Dem so ausgeführten Code werden nur nötige Ressourcen zur Verfügung gestellt, wie es zuvor durch Nutzer oder den Hersteller vorgegeben wurde.

### 5.1.4 Softwarekomponenten müssen aktualisierbar sein

Selbst wenn durch mathematische Verifizierung sicher gestellt wurde, dass der geschriebene Code fehlerfrei ist, ist dies noch kein Beweis dafür, dass die Hardware ebenso fehlerfrei ist. Durch Fehlerkorrektur oder auch durch neue Anforderungen, also Erweiterungen, kann es der Austausch von Softwarekomponenten nötig werden. Hierzu müssen Komponenten möglicherweise spezielle Mechanismen bereitstellen, wie z. B. das Externalisieren von Zuständen, so dass die Ersatzkomponente den Zustand wieder internalisieren kann und keine Informationen verloren gehen. Ebenso wäre es denkbar, einen Ansatz zu verfolgen wie das „Live-Patching“, das im laufenden Betrieb Softwarekomponenten im Speicher austauscht. Die Verteilung von Aktualisierungen wäre durch regelmäßiges Nachfragen oder durch Benachrichtigungen mittels eines „AppStores“ realisierbar, siehe 5.1.3.

### 5.1.5 Schutzvorkehrungen gegen unbekannte Schwachstellen in Softwarekomponenten

Unbekannte Schwachstellen, im Sinne dieser Anforderung, sind bisher unentdeckte bzw. nicht publizierte Schwachstellen verursacht durch Fehler bei der Softwareentwicklung. Grundsätzlich sollte die Ursache der Fehler behoben werden. Da sich dies allerdings als nicht trivial herausstellt, wie in Abschnitt 5.1.9 beschrieben, sollten geeignete Maßnahmen ergriffen werden, um proaktiv eine Kompromittierung zu verhindern. Ein System sollte nicht nur gegen die Fehler an sich geschützt werden, sondern auch gegen die erfolgreiche Ausnutzung von Schwachstellen. Um die Ausnutzbarkeit von Schwachstellen zu erschweren, bieten sich unterschiedliche Techniken, abhängig von der eingesetzten Plattform, an. Eine Möglichkeit ist der Einsatz von Hard- und Software, die die Hardwarehersteller abhängigen Technologien „NX Bit (No-eXecute), XD Bit (eXecute Disable), XN (eXecute Never)“ unterstützt, bei der durch Markierung von Speicherbereichen die Ausführung des Speiche-rinhalts unterbunden wird. Oder Techniken wie Address Space Layout Randomization (ASLR) [70] oder Address Space Layout Permutation (ASLP) [41], die durch Verwürfeln der Speicheradressbereiche das gezielte Anspringen oder Überschreiben von Routinen erschweren. Ausdrücklich nicht betrachtet werden Schwachstellen, die durch den Softwareentwickler oder Hardwareentwickler als Hintertüren in einem System versteckt werden. Für diese Art von gezielt eingebrachten Schwachstellen müssen andere Maßnahmen oder Vorkehrungen getroffen werden.

### 5.1.6 Authentizität und Integrität von nachgeladenem ausführbarem Code

Da Software aktualisiert (siehe 5.1.4) oder ersetzt werden muss, muss dies auf eine Art und Weise geschehen, die sicherstellt, dass nur Software installiert werden kann, deren Integrität überprüft und Authentizität verifiziert wurde. Falls dies nicht der Fall ist, stellt dieser Prozess eine Schwachstelle dar, über die eine Plattform angreifbar wird. Wendet man das Konzept der AppStores (5.1.3) an, so wäre eine Lösung der Einsatz von signierten Softwarepaketen, wodurch bei Überprüfung dieser Authentizität und Integrität sichergestellt werden kann. Diese Singnaturüberprüfung könnte auch ohne einen zentralen AppStore realisiert werden, z. B. mit mehreren Paketquellen, die jeweils selbstssignierte Pakete bereitstellt. Beide Ansätze schützen den Pfad vom AppStore oder der Paketquelle zum Endgerät, aber nicht den Pfad von einem Entwickler zu einem AppStore oder der Paketquelle. Dieses Problem stellt ein fundamentales Vertrauensproblem in der Informatik dar, das bis auf die grundlegenden Werkzeuge eines Entwicklers zurückgeführt werden kann, z. B. den Compiler. Als der erste, der diese Problematik beschrieben hat, ist Ken Thompson [71] zu nennen. Durch das Modifizieren eines Compilers baut dieser während dem Übersetzen von einem legitimen Quellcode in ausführbaren Code – intransparent für den Entwickler – Hintertüren in das Programm ein. Gegen diese Art der Hintertüren, ob sie nun wissentlich durch den Entwickler eingebaut werden oder der Entwickler selber Opfer eines Angriffs geworden ist, kann der Einsatz von Signaturen nicht schützen. Hierfür werden u. a. heuristische Verfahren oder Verfahren basierend auf der Verhaltensanalyse von Anwendungen vorgeschlagen [86].

### 5.1.7 Zugriff auf vertrauliche Informationen nur für berechtigte Personen/Komponenten

Vertrauliche Informationen sollten in der Regel auch vertraulich bleiben, um dies umzusetzen bedarf es gewisser Schutzmaßnahmen. Denn wenn vertrauliche Daten von nicht berechtigten Personen oder anderen Komponenten eines Systems abgefragt werden können, dann ist die Vertraulichkeit einer Information nicht mehr gegeben. In dieser Aussage verbergen sich be-reits weitere Annahmen. Erstens müssen Daten klassifiziert werden, ob sie überhaupt vertraulich sind oder nicht, und zweitens muss ein Mechanismus existieren, um eine Zugriffsberechtigung abzubilden. Dies trifft sowohl auf Daten in Transit wie auch auf abgespeicherte Daten zu. Um die Zugriffsproblematik in den Griff zu bekommen, ist eine weitverbreitete Methode das Einsetzen und Durchsetzten eines Berechtigungssystems, das anhand von definierten Berechtigungen Identitäten, z. B. Personen oder Anwendungen, Zugriff gewährt oder unterbindet. Dazu muss eine Authentifizierung der entsprechenden Identitäten gewährleistet sein. Als Klassifizierung von Informationen könnte die im Bundesdatenschutzgesetz (BDSG) gebräuchliche Definition von personenbezogenen Daten genutzt werden, oder auch der Grundsatz, dass jede Kommunikation vertraulich ist, denn nur der Kontext entscheidet, ab wann eine Information nicht vertraulich ist. Hierzu ein Beispiel, als personenbezogene Daten können das Adressbuch oder die Positionsdaten auf einem Endgerät gelten. Wenn ein Anwender einen Eintrag aus dem Adressbuch mit einer Person teilen möchte, so muss dieser Eintrag für den Empfänger lesbar sein. Für alle anderen Unbeteiligten hingegen muss die Information verborgen bleiben. Wenn eine Datensicherung des Adressbuches als Dienst oder wie im Anwendungsfall A.1.16 beschrieben, die Positionsdaten zur Führung eines automatischen Fahrtenbuches benötigt werden, so kann die Speicherung der Daten auf Seiten des Anbieters auf unterschiedliche Arten erfolgen. Einmal wäre es möglich, die Daten ohne jeglichen Schutz auf der Infrastruktur des Anbieters abzulegen oder aber die Daten werden bereits vom Endgerät aus so verarbeitet, dass der Anbieter nicht an die Informationen gelangen kann. Hierfür existiert keine allgemeine Lösung, je nachdem ob und wie ein Dienst Daten verarbeitet, müssen hier unterschiedliche Maßnahmen getroffen werden. Grundsätzlich sollte ein Dienst nur die nötigen Daten sammeln, verarbeiten und aus Selbstschutz, um z. B. nicht ein lukratives Angriffsziel darzustellen, das Speichern von Informationen minimieren und diese so ablegen, dass der Zugriff nur mit einem Geheimnis des Nutzers möglich ist. Ein Geheimnis eines Nutzer kann als Schlüssel bezeichnet werden. Ein Schlüsselmanagementsystem übernimmt die Verwaltung von Schlüsseln und muss zwangsläufig durch ein Berechtigungssystem kontrolliert werden, damit nur berechtigte Entitäten die Schlüssel einsetzen können.

### 5.1.8 Verwendung von weniger störsenderanfälligen Kommunikationsschnittstellen

Alle Nutzer von elektromagnetischen drahtlosen Kommunikationstechnologien teilen sich dasselbe physikalische Medium. Auf diese Weise entfällt der Zugangsschutz zum Medium ganz im Gegensatz zu kabelgebundenen Technologien. Daher sind drahtlose Kommunikationstechnologien anfälliger für DoS Angriffe, z. B. mit Hilfe von Störsendern. Störsender lassen sich in fünf verschiedene Störsenderkategorien unterteilen [81], den *ständig sendenden*, *täuschenden*, *zufällig sendenden*, *reaktiven*, oder den *intelligenten Störsender*. Das Besondere am intelligenten Störsender ist, dass er durch Störungen auf der physikalischen Ebene gezielt höhere Ebenen, wie Transport- oder Anwendungsprotokolle, angreift. Die anderen eher simplen Methoden der Kommunikationsstörung beruhen auf unterschiedlichen Zeitpunkten oder der Dauer der Störung. Der für Ad-hoc-Kommunikation vorgeschlagene Standard 802.11p [33] setzt Carrier Sense Multiple Access / Collision Avoidance (CSMA/CA) ein, um den Zugriff auf das Medium zu regeln. Durch das Lauschen und Warten auf einen freien Kanal auf dem Übertragungsmedium ist dieser besonders anfällig für Störsender. Als Lösungen werden das Erhöhen der Sendeleistung, der Einsatz von Frequenzspreizungstechniken, Vorwärtsfehlerkorrektur, gerichtete Antennen, oder versteckte Kommunikationskanäle genannt [56]. Wobei lediglich die zuletzt genannten Lösungen Erleichterung für CSMA/CA- Technologien bringen können. Es wäre wünschenswert, drahtlose Kommunikationstechnologien zur Verfügung zu haben, die immun gegen Störsender sind. Allerdings stellt das geteilte Medium ein Problem da, das in der Realität oft durch organisatorische oder regulatorische Maßnahmen wie Frequenzzuteilung oder Zulassung von Sende- und Empfangsgeräte, durch Organisationen wie Bundesnetzagentur (BNetzA) oder Federal Communications Commission (FCC), umgangen bzw. begrenzt wird.

### 5.1.9 Auf sichere Programmierung sollte geachtet werden

Diese Anforderung ist sehr weit gefasst und betrifft jede Art von Software. Software, die zur Erhöhung der Sicherheit eingesetzt werden soll, sollte gerade aus diesem Grund ein Vorbild bei der Vermeidung von Softwareschwachstellen sein. Die Lösungsansätze reichen hier von der Verwendung von Standards zur sicheren Programmierung [68], über den Einsatz von Werkzeugen zur statischen oder dynamischen Programmcodeanalyse, bis hin zur Anwendung von Methoden zur formalen Verifizierbarkeit von Algorithmen und Programmabläufen. Aus diesem breiten Repertoire sollten, abgesehen von den Grundlagen, wie dem Befolgen von Programmierrichtlinien zur sicheren Programmierung, so viele qualitätssichernde Maßnahmen ergriffen werden wie möglich. Auf grundlegendes Wissen des Personals im Bereich sichere Programmierung ist Wert zu legen und gezielte Weiterbildung in diesem Bereich ist empfehlenswert. Dies ist eine indirekte Anforderung an das System und eher eine direkte Anforderung an den Entwicklungsprozess, mit dem ein solches System implementiert wird.

### 5.1.10 Minimierung gegenseitiger Beeinflussung von Hardwarekomponenten

Die Basis eines Systems bildet die Hardware, auf der die Systemsoftware ausgeführt wird. Existieren Schwachstellen auf dieser Ebene, so ist es schwer ein darüber liegendes Softwaresystem abzusichern. Eine der Bedrohungen ist, dass sich Komponenten gegenseitig beeinflussen oder manipulieren können. Wenn in Systemen verschiedene Prozessoren verbaut sind, z. B. ein Applikations- und ein Kommunikationsprozessor, so geschieht dies, um Benutzungsregeln des verwendeten Mediums durchzusetzen. Dies ist ein legitimer Weg, um Schadsoftware oder böswillige Nutzer davon abzuhalten durch Manipulation der Systemsoftware, die Spezifikation der drahtlosen Kommunikation zu überschreiten und damit andere Teilnehmer zu schädigen. Für die Sicherheit ist allerdings ganz entscheidend wie die Kommunikation zwischen diesen Prozessoren geregelt ist. Denn Schwachstellen [77] in dem Kommunikationsprozessor können unterschiedliche Auswirkungen auf die Sicherheit des Gesamtsystems haben. Kommunizieren diese über dedizierte Busse oder über spezielle Speicherbereiche, so ist diese Architektur weniger anfällig für gegenseitige Manipulation als würde der gesamte Speicher vollständig gemeinsam genutzt werden [77]. Die beschriebenen Architekturen finden sich u. a. in verschiedenen Smartphones wieder.

## 5.2 Lösungsansätze für Vertrauensbildung – Entwurf eines Reputationssystems

Um die verbliebenen vier Anforderungen 1, 2, 3 und 4 erfüllen zu können, sind spezielle Maßnahmen erforderlich. Gerade bei der Erkennung von schädlicher Kommunikation und der Schadensbegrenzung sind konventionelle Maßnahmen nicht ausreichend. Angreifer versuchen oft bestehende Vertrauensbeziehungen auszunutzen, um ihren Einflussbereich von einem System zum Nächsten auszuweiten. Die grundsätzliche Frage über den Aufbau und die Erhaltung von Vertrauen versuchen Reputationssysteme zu beantworten. Die drei Hauptziele von Reputationssystemen, siehe Abschnitt 3.2, entsprechen sehr genau den Anforderungen 3 und 4. Reputationssysteme stellen Informationen über die Vertrauenswürdigkeit von Teilnehmern und auch Angreifern zur Verfügung, sie versuchen ehrliches Verhalten zu fördern und Angreifer oder schädliche Teilnehmer zu entmutigen. Daher wird in den folgenden Abschnitten ein Reputationssystem konzipiert und dessen Leistungsfähigkeit mit einer Simulation untersucht. Lösungen für 1 und 2 werden unmittelbar im nächsten Abschnitt aufgezeigt.

### 5.2.1 Voraussetzungen und Annahmen

Gemäß der Sicherheitsanforderung 5 (Gewährleisten von vertraulicher Kommunikation (Ende-zu-Ende, Link)) muss entweder der Message Filter, die ONP-Schicht oder eine Schicht darunter diese Ende-zu-Ende Verbindung bereitstellen. Denn wenn dies nicht der Fall ist, z. B. Ende-zu-Ende auf Anwendungsebene, so kann das Reputationssystem keine oder nur eine einge-schränkte Vertrauensbewertung durchführen, da die nötigen Informationen nicht zur Verfügung stehen. Der Einsatz von Informationsbewertung oder Datenverifizierung ist dann nicht mehr möglich.

Die kurze Aufzählung aus dem Grundlagenkapitel zum Thema Identitätsmanagement 2.6 zeigt, dass der Themenbereich um den Einsatz von Identitäten ein sehr weitläufiges Betätigungsfeld bietet. Um diesem und den Anforderungen 1 und 2 gerecht zu werden, muss ein eigenes Identitätsmanagementsystem konzipiert und eingesetzt werden, das sowohl eigene wie fremde Identitäten verwaltet und zur Verfügung stellt. Eine exakte Spezifizierung der Anforderungen, Entwurf und Implementierung eines Systems zur Verwaltung von Identitäten wird Grundlage einer weiteren Publikation sein. Das Reputationssystem verwendet das Identitätsmanagementsystem, um Identitäten überprüfen zu können.

### 5.2.2 Konzeptionierung des Reputationssystems

Das Reputationssystem ist eine Komponente des *Message Filters*, der oberhalb der *ONP*-Schicht agiert, siehe Abbildung 5.1. Der Name *Message Filter* wird von der allgemeinen Bezeichnung IP-Filter, auch Firewall genannt, abgeleitet. Dies geschieht vor dem Hintergrund, dass die reine Funktionalität über die eines Reputationssystems hinausreicht, z. B. bei zu geringer Reputation könnten Nachrichten, ähnlich einem IP-Filter, verworfen oder einer gesonderten Behandlung unterzogen werden. In der Grafik 5.1 wird jede

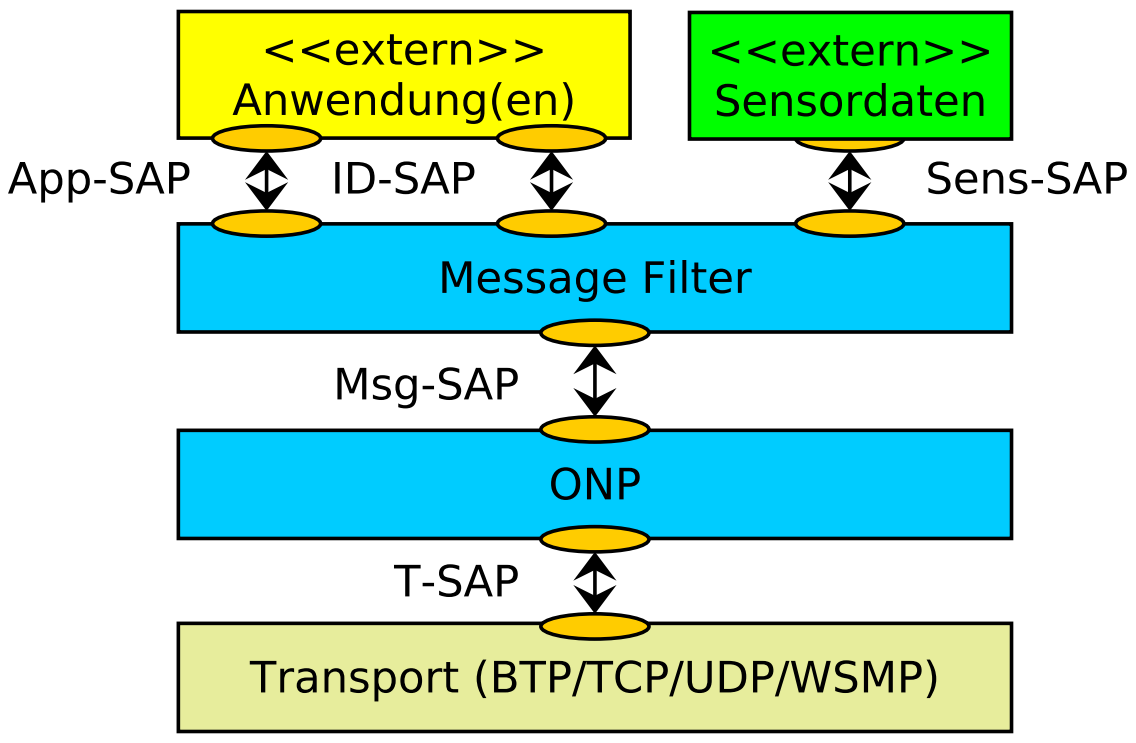

**Abbildung 5.1:** Grobe Schichtarchitektur mit SAPs (Service Access Points)

Schnittstelle zwischen den einzelnen Schichten als ein Service Access Point (SAP), der eine bestimmte Funktionalität abbildet, dargestellt. Der Nachrichtenaustausch zwischen Anwendungen und *Message Filter* findet über den *App-SAP* statt. Darüber hinaus können Anwendungen über den *ID-SAP* beim *Message Filter* auch Identitäten anfragen oder ablegen. Eine weitere

Funktionalität ist die Verarbeitung von Sensordaten, die über den *Sens-SAP* ausgetauscht werden können. Die Nachrichtenübermittlung zwischen *Message Filter* und *ONP* findet über den *Msg-SAP* statt. Die Transportschicht, die aus verschiedenen Technologien besteht, wird von der Overlay-Netzwerkschicht über den *T-SAP* angebunden. Das konzipierte Reputations-

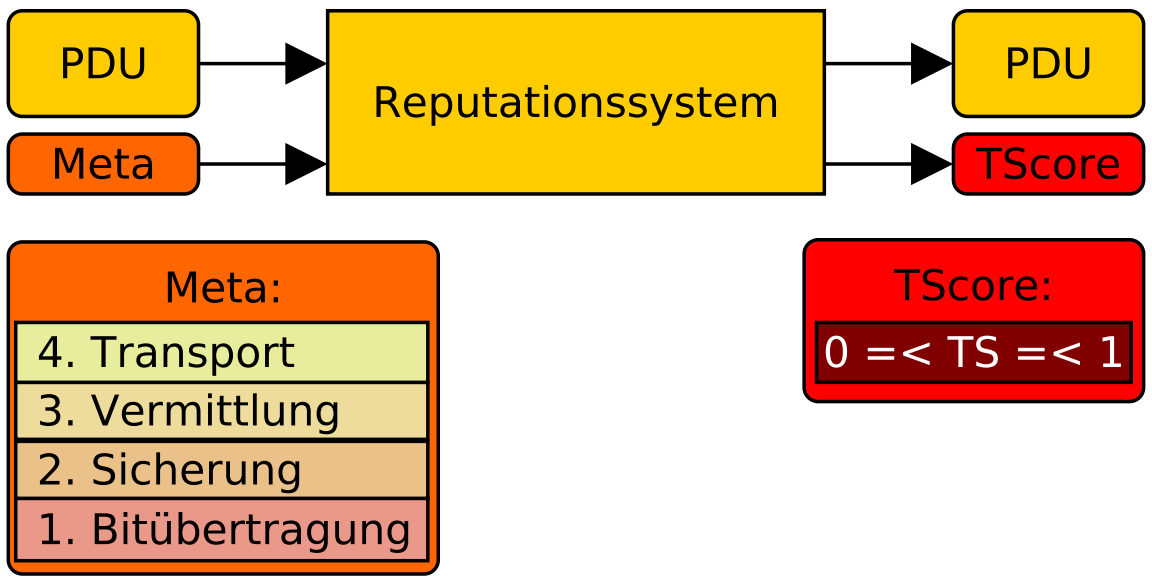

**Abbildung 5.2:** Überblick über die Funktionsweise des Reputationssystems

system hat dieselben grundlegenden Ziele wie in 3.2 aufgeführt. Das heißt, es berechnet aus Informationen einen Wert über Vertrauen oder Ansehen eines Teilnehmers oder von diesem ausgehende Informationen. Hierfür verwendet das Reputationssystem zusätzliche Informationen, die in der Abbildung 5.2 als *Meta* bezeichnet werden. Die Analyse der Informationen und der Daten bezieht sich immer auf eine eingehende Nachricht, ebenso der berechnete Reputationswert. Dieser Wert, in der Grafik 5.2 als *TScore* bezeichnet, liegt zwischen 0 und 1, wobei 1 gleichbedeutend ist mit vollstem Vertrauen und 0 gänzlichem Misstrauen entspricht. Der berechnete Vertrauenswert liegt zwischen diesen beiden Extrema. Bei genauer Betrachtung des konzipierten

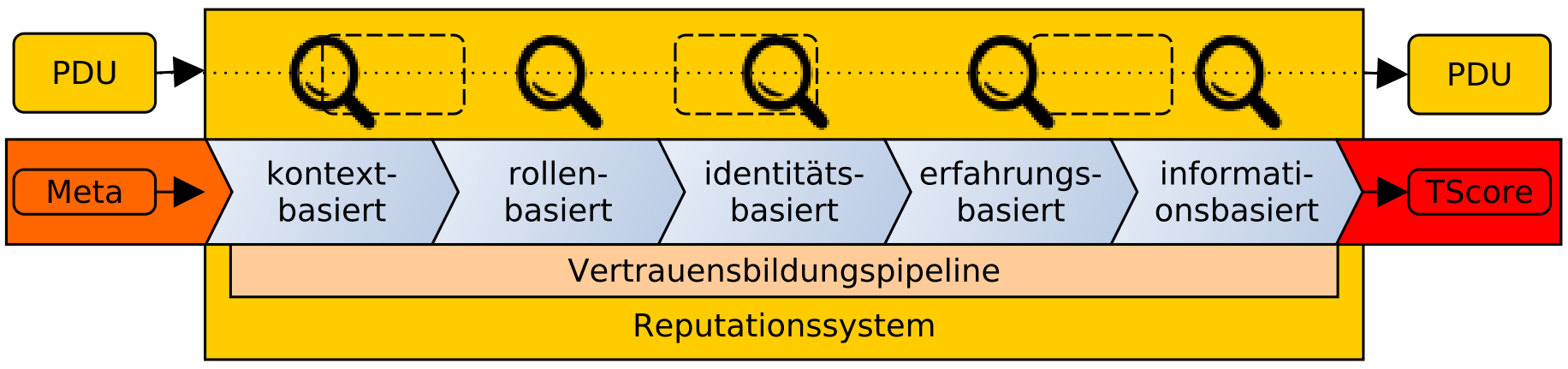

**Abbildung 5.3:** Vertrauensbildungspipeline des Reputationssystems

Reputationssystems fällt auf, dass dieses eine Art Meta-Reputationssystem darstellt, da es den Einsatz von bereits vorgeschlagenen Reputationssystemen, siehe 3.3, unterstützt. Dies wird durch die Organisation der Vertrauens- oder Reputationsbildung in eine sogenannte *Vertrauensbildungspipeline*, dar-gestellt in Diagramm 5.3 ermöglicht. Diese setzt sich aus einzelnen Teilschritten zusammen, die entweder kontext-, rollen-, identitäts-, erfahrungs- oder informationsbasiert arbeiten. Einige der im Kapitel 3.3 beschriebenen existierenden Ansätze für Reputationssysteme im VANET Bereich, könnten hier in der Vertrauensbildungspipeline eingesetzt werden, z. B. [65], [21]. Im Folgenden werden die einzelnen Teilschritte, deren Möglichkeiten und welchen Beitrag diese zur Reputationsbestimmung leisten können, beschrieben.

**Kontext:**

Der Kontext einer Packet Data Unit (PDU) kann durch unterschiedlichste Parameter bestimmt werden. In einem VANET kann ein Kontext aus den untersten Netzwerkschichten extrahiert werden, denn abhängig ob die Kommunikation sicherheitskritisch oder eine Dienstleitung ist, werden diese u. a. in verschiedenen Funkkanälen übertragen bei IEEE 802.11p [33]. Generell kann ein Kontext u. a. durch die Wahl des Kommunikationsweges bestimmt werden, z. B. via Infrarot, Bluetooth, allgemein zellulare oder ad-hoc Technologien, die sich letztendlich in der Vertrauenswürdigkeit unterscheiden. Ein Kontext kann auch durch die Zugehörigkeit einer PDU zu einem Datenstrom sein, der durch Mechanismen auf höherer Ebene (z. B. Transport, Sitzung oder Anwendungsschicht) etabliert wird. Ein Kontext könnte auch eine Beschreibung von Parametern sein, die ähnlich wie eine Security Association im IPsec Protokoll, Informationen über Identitäten, kryptographische Primitiven oder die gesamten Parameter des Durchstichs durch alle Netzwerkschichten, enthält.

**Rolle:**

Rollen existieren, sobald nicht alle Teilnehmer einer Kommunikation exakt gleichgestellt sind, sondern es Knoten gibt, die spezielle Funktionen oder Eigenschaften haben. Wenn ein Nutzer mit einem Endgerät, ein gewöhnlicher Computer, eine Webseite abrufen möchte, so nehmen an dieser Kommunikation mehrere Systeme, die verschiedene Rollen innehaben, teil. Ein Nutzer könnte die Rolle Fahrzeug oder mobiles Endgerät besitzen, wohingegen ein Rettungs- oder Einsatzfahrzeug die Rolle eines Fahrzeuges im Notfalleinsatz einnehmen könnte. Weitere denkbare Rollen sind die einer RSU oder auch, abhängig von dem Overlaynetzwerk, die eines Routers oder Clusterkopfes.

**Identität:**

Wie bereits zuvor ausgeführt, sind digitale Identitäten ein zentrales Konzept, um Kommunikation zu ermöglichen und ebenso wichtig, um die Vertrauenswürdigkeit von Informationen, die von diesen kommuniziert werden, zu bestimmen. Für die Verwaltung von diesen wird eine eigene Komponente in Form eines Identitätsmanagers eingesetzt. Dieser muss jede Identität einer Kategorie zuordnen können, z. B. starke, mittlere und schwache Identität. Eine starke Identität könnte das digitale Nummernschild sein, das eine technische und juristische Identität ist, die fälschungssicher oder zumindest enormen technischen und finanziellen Aufwand benötigt, um diese manipulieren oder fälschen zu können. Anforderungen an diese Identität wären u. a. die Nichtabstreitbarkeit von Kommunikation oder Handlungen. Eine schwache Identität hingegen könnte eine IP-Adresse oder eine Identität aus einem Drittanbieterdienst wie Facebook sein, wobei die Stärke oder Schwäche auch von dem entsprechenden Kontext abhängt. Eine Facebook-Identität kann im Kontext von Facebook als stark gesehen werden, aber schwach im Zusammenhang mit sicherheitskritischen Vorgängen. Denn diese können auch rechtliche Konsequenzen haben und bei einer Facebook-Identität ist, mangels Überprüfung, nicht sichergestellt, ob eine echte juristische Person hinter der Identität steht. Eine exakte Kategorisierung der Identitäten benötigt weitere Untersuchungen der Kriterien und unterschiedlichen Ziele, die damit verbunden sind.

**Erfahrung:**

Bei der Bildung von Vertrauen basierend auf zurückliegenden Interaktionen oder beobachtetem Verhalten, allgemein Erfahrungen, muss darauf geachtet werden, dass einmal gewonnenes Vertrauen nicht ausgenutzt wird. Ein Angreifer könnte zunächst Vertrauen aufbauen und dann falsche oder manipulierte Informationen verbreiten oder schädliche Verhaltensweisen ausführen. Tran *et al.*[72] formulieren daher den Grundsatz, dass Vertrauen nur über einen längeren Zeitraum und durch viele positive Interaktionen und positivem Verhalten aufgebaut werden kann, aber dass dieses durch einige wenige negative Handlungen ruiniert werden kann. Daraus folgt, dass Vertrauen schwer aufzubauen, aber sehr leicht zu verlieren ist. Die Verwendung eines Vergesslichkeitsfaktors [39], [78] ermöglicht, dass kürzer zurückliegende Erfahrungen bevorzugt werden.

**Information:**

Der eigentliche Wert einer Kommunikation beruht auf den ausgetauschten Informationen. Eine Verkehrswarnung oder Kollisionswarnung sind möglicherweise wertvoller als eine Information über den nächsten Rastplatz oder auch Werbung von Lokalitäten. Wenn die Validität einer Information nicht gegeben ist, so verliert diese jeglichen Wert. Dieses Problem, bzgl. des Wahrheitsgehaltes von Informationen, können auch traditionelle Ansätze zur Nachrichtenintegrität oder -authentizität nicht vollständig lösen. Unabhängig ob eine zentrale, dezentrale oder eine hybride Vertrauensarchitektur genutzt wird, denn es besteht immer die Möglichkeit, dass ein Teilnehmer durch eine Fehlfunktion betroffen ist. Auch die Übernahme eines Teilnehmers durch einen

Angreifer, der diesen in Folge nutzt, um falsche Informationen zu verbreiten, ist denkbar. Um diese falschen Informationen oder Fehler zu erkennen, bietet sich die Verifikation der Daten an. Mögliche Ansätze sind im Kapitel 3.4 aufgeführt.

### 5.2.3 Simulation des Reputationssystems

Als Simulationsplattform wurde die MIT lizenzierte SimPy Bibliothek [49] in der Version 3.0.8 verwendet. Mit dieser lassen sich Simulationen, auf diskreten Ereignissen basierend, realisieren. SimPy beinhaltet keine vorgefertigten Abläufe oder Muster für Ad-hoc-Kommunikation. Da diese Simulation die

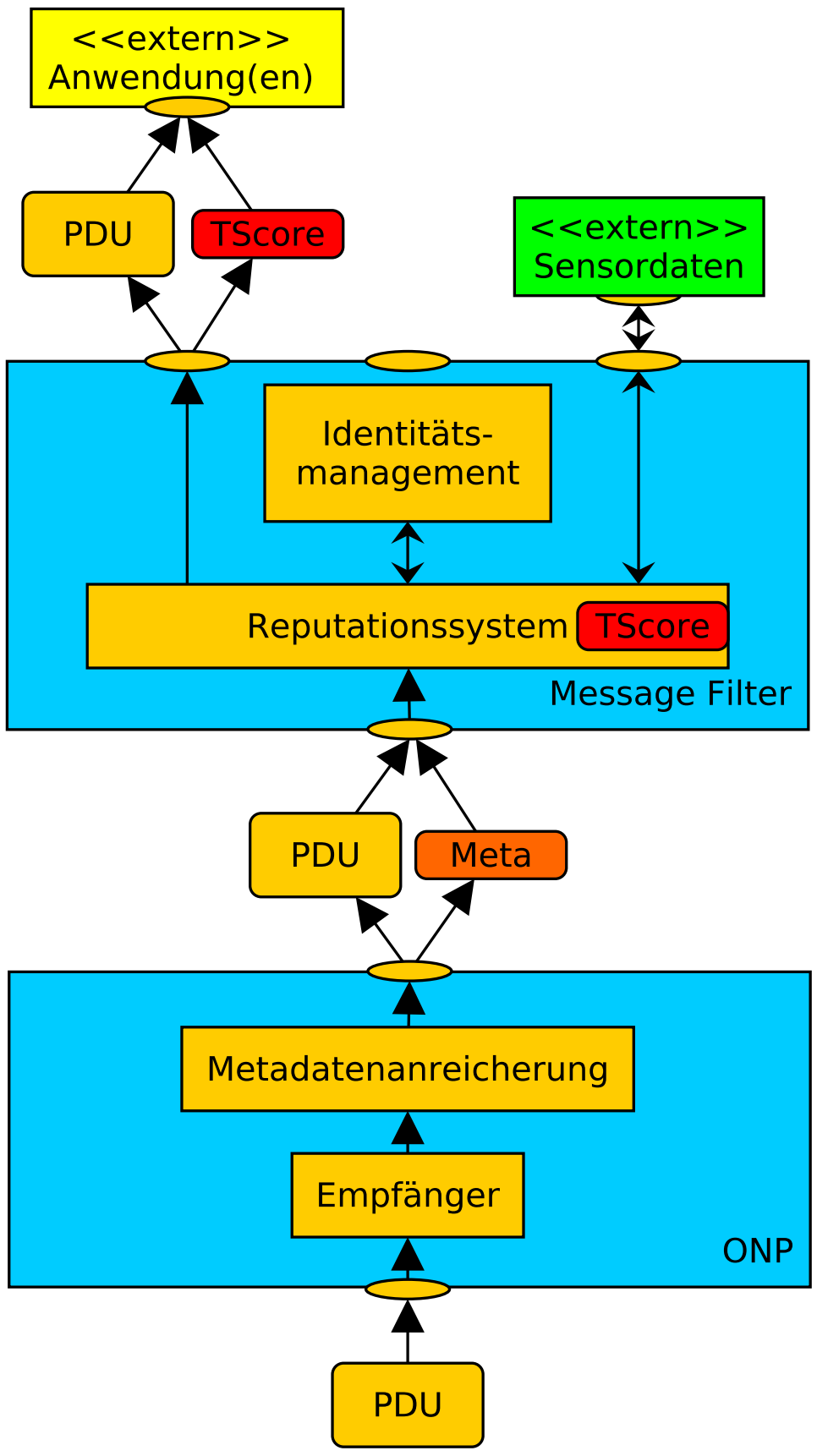

**Abbildung 5.4:** Detailansicht der Schichtarchitektur mit Kommunikationswegen bei Nachrichtenverarbeitung

Funktionsweise und Leistung im Sinne der Vertrauensbewertung durch ein Reputationssystem untersucht und dies zwar im Kontext eines VANET aber eben nicht ausschließlich darauf bezogen ist, eignet sich SimPy sehr gut für die hier implementierte Simulation. Darüber hinaus bietet die SimPy Bibliothek eine niedrige Einstiegshürde, da lediglich Programmierkenntnisse in Python und ein Grundwissen im Bereich von Simulationen basierend auf diskreten Ereignissen nötig sind. Ebenso bietet diese eine angenehme Lernkurve mit unterstützender Dokumentation, die ausreichend vorhanden ist. Wie bereits angesprochen, fokussiert die Simulation den Einsatz des Systems in einem VANET, unabhängig davon steht ein Einsatz in einem MANET oder generell in dem Gebiet der CPS nichts entgegen. Lediglich die spezifischen Regeln und Annahmen über Kontext, Rollen, Identitäten, usw., die für ein VANET getroffen wurden, müssen überprüft, angepasst oder erweitert werden, um allgemeingültiger auf ein Netzwerk aus Cyber-physikalischen Systemen anwendbar zu sein. Aus diesem Grund wird auch auf den Einsatz eines Simulators verzichtet, der die Besonderheiten von verschiedenen zellularen und Ad-hoc-Kommunikationstechnologien berücksichtigt, da diese die generelle Anwendbarkeit des Ansatzes in Frage stellen würden. Der vollständige Quellcode ist dem Anhang C zu entnehmen. Die Simulation bildet den, in der Abbildung 5.4 dargestellten, Ablauf einer Nachrichtenverarbeitung ab. Hierbei wird die Funktionalität der ONP- Schicht durch Nachrichtengeneratoren ersetzt. Diese erzeugen Nachrichten inklusive den dazugehörigen *Meta*-Informationen. Das Ergebnis eines Simulationsschrittes, nach der Verarbeitung durch das Reputationssystem, sind alle eingegangenen Nachrichten inklusive dem berechneten Reputationswert *TScore*.

Die Berechnung des *TScore* wird in der Simulation durch eine Mischung von diskreten Look-Up Tabellen, Summierung der Werte aus Teilschritten und Gewichtung von einzelnen Gesichtspunkten, wie bei der Identitäts- und Informationsüberprüfung, realisiert. Die verwendeten diskreten Werte und ihre umgangssprachliche Bedeutung sind wie folgt:

**1** vollstes Vertrauen
**0,75** Tendenz eher vertrauenswürdig
**0,5** keine klare Vertrauensaussage möglich
**0,25** Tendenz eher nicht vertrauenswürdig
**0** gar kein Vertrauen

Der berechnete *TScore* liegt, wie bereits im Kapitel 2 beschrieben, zwischen 0 und 1. Je näher dieser Wert an der Eins liegt, umso vertrauenswürdiger ist die Nachricht. Um eine exaktere Aussage über einen bestimmten Wert treffen zu können, wird der Wertebereich dreigeteilt. Im Folgenden sind die Bereiche und ihre Bedeutung definiert:

$0.0 \leqq$ **TScore** $< 0.5$ gar kein Vertrauen
$0.5 \leqq$ **TScore** $\leqq 0.65$ keine klare Vertrauensaussage möglich
$0.65 <$ **TScore** $\leqq 1.0$ vollstes Vertrauen

Hinter der Idee der Einteilung des *TScore* in Bereiche steht folgende Annah-

me: ein ideales Reputationssystem, das denjenigen maximal entlastet, der auf die Reputationsbewertung angewiesen ist, liefert eine binäre Bewertung, gut oder böse, vertrauenswürdig oder eben nicht. Um diesem Ideal so nahe wie möglich zukommen ohne sich von der Realität zu weit zu entfernen, wird auf die zuvor aufgeführte Dreiteilung zurückgegriffen. Welche einzelnen Bewertungsfunktionalitäten im Rahmen der Simulation umgesetzt werden, wird nun im Folgenden beschrieben.

**Kontext:**

Da der Schwerpunkt der Simulation bei der Kommunikation zwischen Teilnehmern, hier Fahrzeugen, liegt, werden als einfachste Definition eines Kontextes die Eigenschaften der Bitübertragungsschicht gewählt. Es wird unterschieden, ob die PDU über die zellulare Schnittstelle oder über die Ad-hoc-Schnittstelle, im Detail ob es über einen Dienst- oder Betriebssicherheitskanal, empfangen wurde. In diesem Abschnitt werden die Eingenschaften von verschiedenen Kommunikationstechnologien gegeneinander abgewogen.

**Rolle:**

Für die Simulation werden zwischen den Rollen eines zentralen Dienstes, einer RSU, einem Einsatzfahrzeug, einem generischen Fahrzeug und einem mobilen Endgerät unterschieden. Dies ist keine vollumfängliche Unterscheidung, da noch weitere Rollen denkbar sind, wie Fahrzeuge des öffentlichen Dienstes. Die Unterteilung der Rollen beruht im Wesentlichen darauf, wie sehr die Privatsphäre einer Rolle zu schützen ist. Dies findet Anwendung in der Berechnung über die Vertrauenswürdigkeit von Identitäten. Dafür wird aus dem nächsten Schritt vorgegriffen, um eine unterschiedliche Wertigkeit von Identitäten im Zusammenhang mit der Rolle darstellen zu können. Dies bedeutet, dass eine RSU weniger vertrauenswürdig ist, wenn sie schwache Identitäten verwendet. Denn RSU sind autonome Systeme, die keinen Schutz der Privatsphäre benötigen, aber sich durch Zuverlässigkeit und Vertrauenswürdigkeit hervortun sollten. Ein Polizeifahrzeug außer Dienst so wie im Dienst würde gemäß dieser Einteilung als normales Fahrzeug gewertet werden. Ist es hingegen im Einsatz, sind Blaulicht und Sirene oder eines von beiden aktiviert, so wird es zum Einsatzfahrzeug. Eine Nachricht, die ein vorausfahrendes Fahrzeug auffordert den Weg freizumachen, sollte nur von einem Einsatzwagen akzeptiert und vertraut werden, der sich ohne Zweifel als ein solcher identifizieren lässt. Einer solchen Nachricht mit schwacher Identität von einem Einsatzwagen oder einer anderen Rolle, sollte kein Vertrauen geschenkt werden. In diesem Abschnitt werden Rollen, sofern sie in einem Kommunikationsnetzwerk vorhanden sind, anhand von verschiedenen Kriterien, hier Privatsphäre, gewichtet.

**Identität:**

In einem ersten Schritt werden Identitäten überprüft, die aus technischen Gegebenheiten vorhanden sein müssen, wie MAC- oder IP-Adressen. In einem zweiten Schritt wird basierend auf den zuvor beschriebenen Kategorien von starken, mittleren und schwachen Identitäten zurückgegriffen, um eine qualitative Einschätzung über die Vertrauenswürdigkeit abgeben zu können. Dieser Abschnitt bewertet qualitativ technische und zusätzlich vorhandene Identitäten.

**Erfahrung:**

Bis zu diesem Schritt wird in der Simulation oder einer möglichen Implementierung kein Speicher benötigt, um Zustände und Zusammenhänge zwischen einzelnen Datenpaketen zu erhalten. Jedes Paket kann für sich abgeschlossen betrachtet werden. Das Bilden von Erfahrung über das Vertrauen bzgl. der Zuverlässigkeit oder in Entscheidungen, würde eine Zustandserhaltung allerdings nötig machen. Da diese auch den Gegebenheiten und Charakteristika der Kommunikation in VANETs unterliegen und diese in der eingesetzten Simulationsumgebung nur unzureichend abgebildet werden können, muss dieser Schritt in einer weiterführenden Arbeit untersucht werden und wird hier nicht betrachtet.

**Information:**

In der Simulation werden zwei lokale Verifikationen durchgeführt. Einmal wird überprüft, ob die angegebene Position eines Teilnehmers mit der gemessenen Sendeleistung zusammenpassen. Als Zweites wird verglichen, ob der angegebene Typ der Nachricht mit dem technisch genutzten Typ übereinstimmt. Hier wird ausgenutzt, dass sicherheitskritische Informationen, im Gegensatz zu Dienstleistungen, über andere Kanäle in der VANET Kommunikation übertragen werden. Dies müsste entfallen wenn, in einem CPS diese Unterscheidung nicht getroffen werden kann, da nur eine Kommunikationsschnittstelle vorhanden ist. Lokale oder entfernte Verifikation der Informationen wird in diesem Abschnitt durchgeführt, in der Simulation werden die zuvor beschriebenen lokalen Verifikationen eingesetzt.

### 5.2.4 Interpretation der Simulationsergebnisse

Der Pseudozufallszahlengenerator von Python verhält sich deterministisch, wenn er mit einem Startwert, auch Seed genannt, initialisiert wird. Über diesen Mechanismus können Simulationen immer wiederholt werden und liefern stets dieselben Ergebnisse. Aus diesem Grund wird auch darauf verzichtet die mitunter großen Datenmengen an diese Arbeit anzufügen, da der Quellcode C und der Seedwert der Simulation ausreichend sind, um alle Ergebnisse zu verifizieren. Der Startwert wird bei jeder Grafik in der Überschrift, mit der Bezeichnung *RND=* beginnend, angegeben. Alle hier präsentierten Grafiken verwenden die numerische Darstellung von $\pi$, aus der Python NumPy Bibliothek. $\pi$ wurde gewählt, da die Abbildungen alle zu besprechenden Phänomene enthalten. Das erste Diagramm 5.5 besteht aus zwei Teilgrafiken. Die obere Grafik stellt die Verteilung der *TScore*-Werte in den zuvor beschrieben Bereichen dar. Bei der Erzeugung der Nachrichten steht bereits fest, ob diese vertrauenswürdig sind oder nicht, daher können diese eindeutig gemäß dieser Beschaffenheit sortiert werden. Die Tabelle unterhalb der Legende gibt einen schnellen Überblick über die exakte Anzahl von Nachrichten und deren Bereichszuteilung. In der Grafik darunter sind die *TScore*-Werte jeder einzelnen Nachricht, von der pro Zeitschritt exakt eine verarbeitet wird, angetragen. In beiden Grafiken ist bei genauer Betrachtung lediglich eine einzige Nachricht, von 100, zu entdecken, die erzeugt wurde und vertrauenswürdig ist, alle anderen sind dies nicht. Der naive Generator scheint eine Präferenz zu nicht vertrauenswürdigen Nachrichten zu haben. Dass dies nicht an dem Generator liegt, sondern an den kombinatorischen Möglichkeiten des Nachrichtenmodells, zeigt eine weitere Analyse des Simulationsmodels. Hierfür werden die Möglichkeiten einer Nachricht je Nachrichteneigenschaft in Tabelle 5.1 aufgelistet. Einmal in der Spalte *positiv* für alle Möglichkeiten eine vertrauenswürdige Nachricht zu erzeugen und ein zweites Mal für *alle* möglichen Kombinationen einer Nachricht. Das Ergebnis ist, dass lediglich 2880 Kombinationsmöglichkeiten existieren, um eine vertrauenswürdige Nachricht zu erzeugen. Im Gegensatz dazu stehen 308160 Kombinationen nicht vertrauenswürdige Nachrichten zu generieren. In Prozent gerechnet bedeutet dies, dass nur 0.95% aller Kombinationen vertrauenswürdig sind. Es genügt, dass eine einzige der zehn überprüften Eigenschaften einen nicht vertrauenswürdigen Wert hat, damit die gesamte Nachricht als nicht vertrauenswürdig eingestuft wird. Bei nicht vertrauenswürdigen Nachrichten hingegen können beliebig viele nicht vertrauenswürdige Eigenschaften enthalten sein.

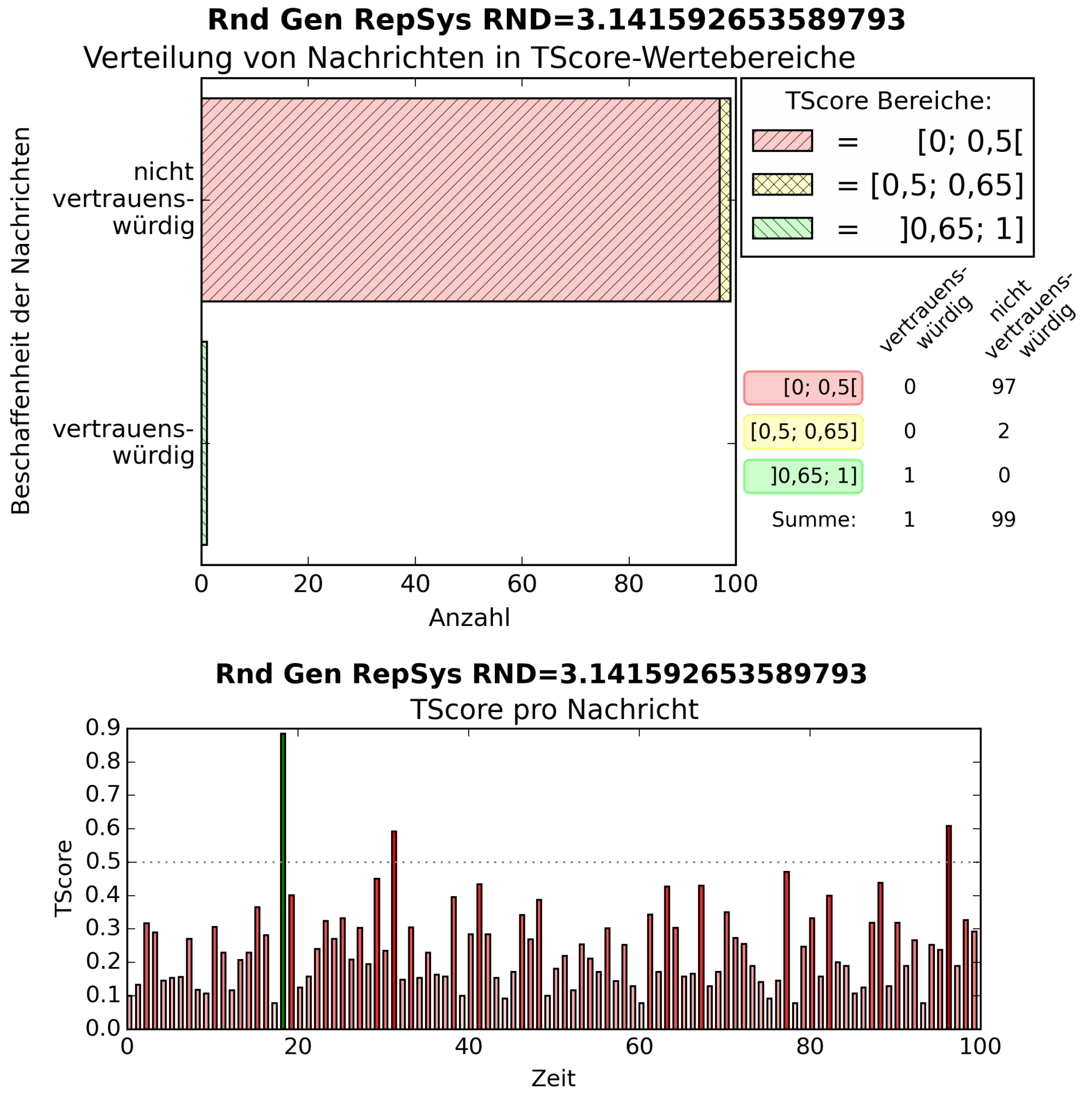

**Abbildung 5.5:** Standard Generator, 100er Stichprobe

| Eigenschaften | positiv | alle |
|---|---|---|
| Meta L1: | 2 | 2 |
| Meta L1 Signal: | 4 | 4 |
| Meta L2 Type: | 3 | 3 |
| Meta L2 MAC: | 2 | 3 |
| Meta L3 Addr: | 2 | 4 |
| Info Role: | 5 | 5 |
| Info ID: | 3 | 6 |
| Info Pos: | X | 4 |
| Info Type: | X | 3 |
| Info State: | 2 | 3 |
| Permutationen: | 2880 | 311040 |
| (Alle - Positiv): | 308160 | |

**Tabelle 5.1:** Kombinationsmöglichkeiten der Nachrichteneigenschaften

Um die Berechnungen zu verifizieren, kann ein Datensatz erzeugt werden, der alle Nachrichtenkombinationen enthält, dargestellt in der Abbildung 5.6. Hierfür muss die Simulationszeit (SIM_TIME) auf 1 gesetzt werden und der auskommentierte Generator, der alle Nachrichtenkombinationen in einem Simulationsschritt erzeugt, aktiviert und alle anderen Generatoren deaktiviert werden. Das Ergebnis ist ein Datensatz, der alle möglichen Nachrichtenkombinationen, inkl. Vertrauensbewertung enthält. Dieser eignet sich hervorragend, um die Leistung des Reputationssystems zu untersuchen.

Von 2880 vertrauenswürdigen Nachrichten werden 2008, fast 70% (genau: 69.72%), korrekt als solche eingestuft. Im Gegensatz dazu wird keine ein-

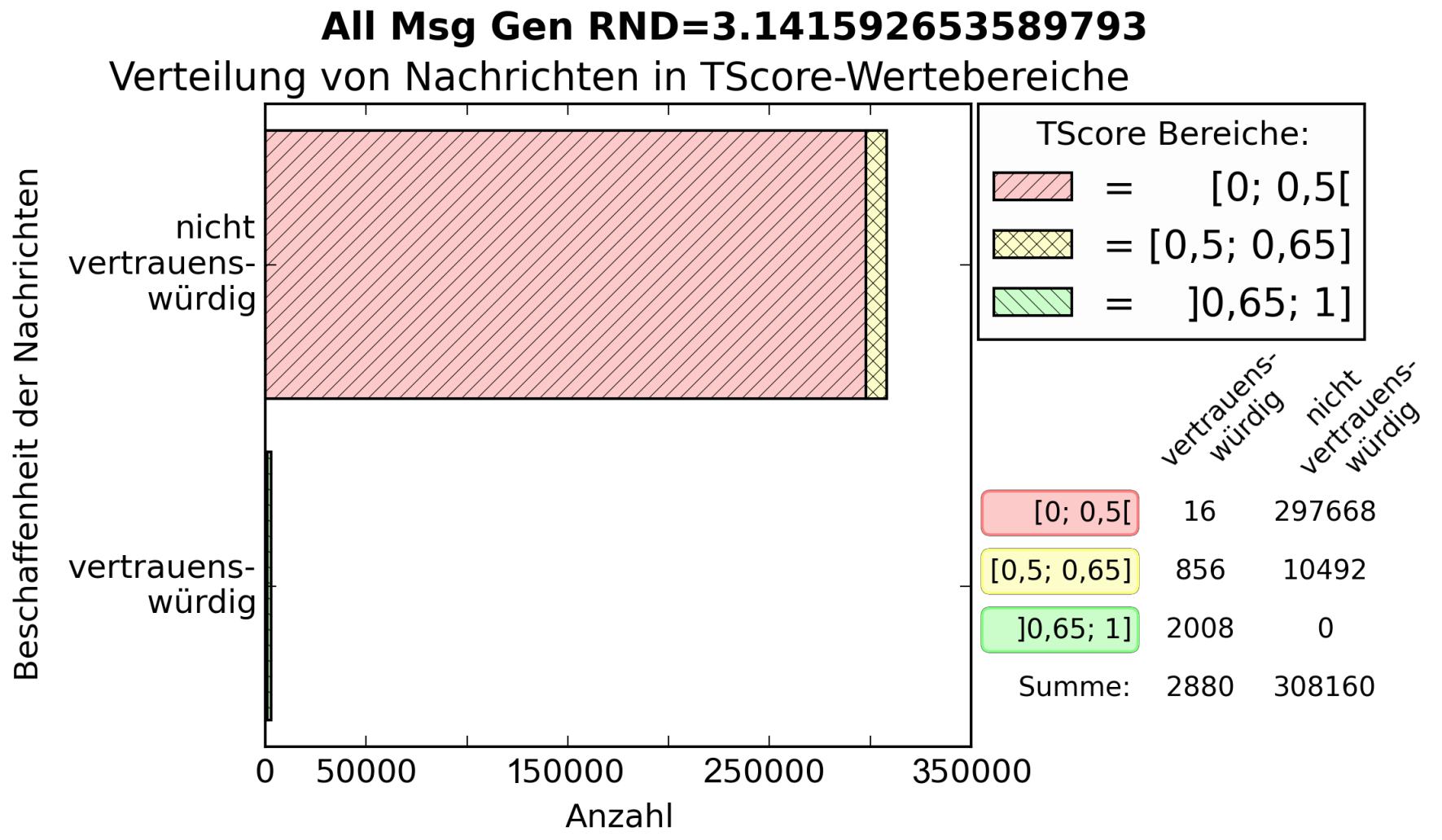

**Abbildung 5.6:** Alle möglichen Nachrichten durch Kombination aller Eigenschaften

zige nicht vertrauenswürdige Nachricht als vertrauenswürdig bewertet. Das Reputationssystem erzeugt somit keine falschen positiven Ergebnisse. Zunächst auf der anderen Seite des Spektrums, werden 16 von 2880 Nachrichten, 0.56%, als nicht vertrauenswürdig eingestuft, obwohl diese vertrauenswürdig wären. Dies liegt an einer fehlenden Überprüfung bei der Nachrichtenkategorisierung, nach vertrauenswürdig oder nicht. Denn alle betroffenen Nachrichten haben als Gemeinsamkeit, dass sie von den Rollen *RSU* oder Einsatzfahrzeug aus gesendet werden und dies mit einer schwachen Identität. Durch die implizite Annahme, wie in 5.2.3 beschrieben, dass ein Einsatzwagen oder eine RSU in ihrer Funktion nur mit starken Identitäten auftreten, wären diese Kombinationen ebenfalls als nicht vertrauenswürdig einzuordnen und können somit ignoriert werden. Ein Grund für diese Annahme, vor allem für Einsatzwägen, liegt in dem Datenschutz der Einsatzwägen und deren Insassen. Ein Einsatzwagen der keinen Einsatz hat, z. B. kein Martinshorn und Blaulicht aktiviert hat, taucht in der Kommunikation der anderen Verkehrsteilnehmer unter, indem er die Rolle von Einsatzwagen auf einfaches Fahrzeug wechselt und sich auch wie ein solches verhält. Es werden 297668 von 308160, 96.60%, korrekt als nicht vertrauenswürdig erkannt. Der mittlere Bereich wird von vertrauenswürdigen Nachrichten, 856 von 2880 ca. 29.72%, ausgefüllt, von denen ein Großteil lediglich über schwache Identitäten kommuniziert und der Rest zwar über mittelstarke Identitäten verfügt, allerdings die kommunizierten Informationen nicht verifizierbar sind. Deshalb ist das Vertrauen in diese Nachrichten nicht besonders hoch und spiegelt sich im *TScore* der Nachrichten wieder. Alle Nachrichten die nicht vertrauenswürdig sind, aber trotzdem einen *TScore* Wert größer als 0.5 erreicht haben, wurden entweder mit einer starken oder mittelstarken Identität erzeugt, 10492 von 308160, entspricht 3.40%. Diese machen es schwerer für das Reputationssystem ohne gezielte Anpassungen diese als eindeutig nicht vertrauenswürdig einzusortieren. Dieser Bereich setzt sich aus drei Teilgruppen zusammen. 50.10% der Nachrichten, 5256 von 10492, enthalten valide, verifizierbare Informationen, weitere 42.36%, 4444 von 10492 enthalten nicht verifizierbare Informationen und die restlichen 792 Nachrichten, 7.55%, enthalten falsche Informationen. Das Reputationssystem erkennt eine Fehlfunktion oder einen Angriff, kann aber keine klare Aussage zu diesen Nachrichten geben, außer, dass diese nicht voll vertrauenswürdig sind. Dies kann für eine sicherheitskritische Anwendung bereits Information genug sein, um Fehlverhalten zu verhindern. Um eine Gleichverteilung der Nachrichtenbeschaffenheit zu errei-

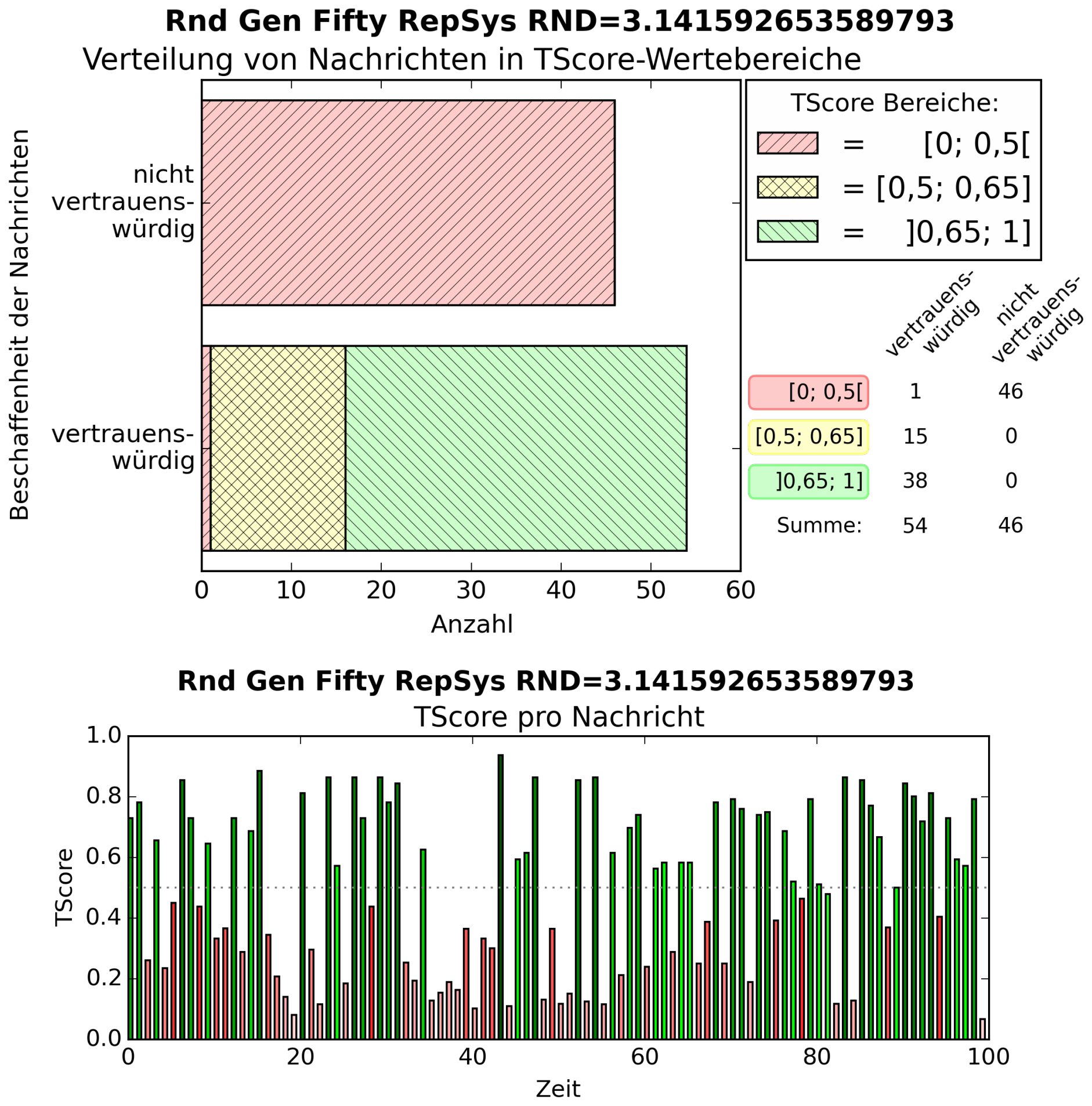

**Abbildung 5.7:** Gleichverteilte, 100er Stichprobe

chen, wurde ein Generator erstellt, der mit 50 prozentiger Wahrscheinlichkeit eine vertrauenswürdige Nachricht und mit der gleichen Wahrscheinlichkeit eine nicht vertrauenswürdige Nachricht erzeugt. Die Ergebnisse mit einer kleinen Stichprobe von 100 Nachrichten, eine pro Zeiteinheit, wird in der Abbildung 5.7 dargestellt. Das Verhältnis der Bereiche von vertrauenswürdigen Nachrichten bleiben im Rahmen von ±5 Prozentpunkten. Mit demselben

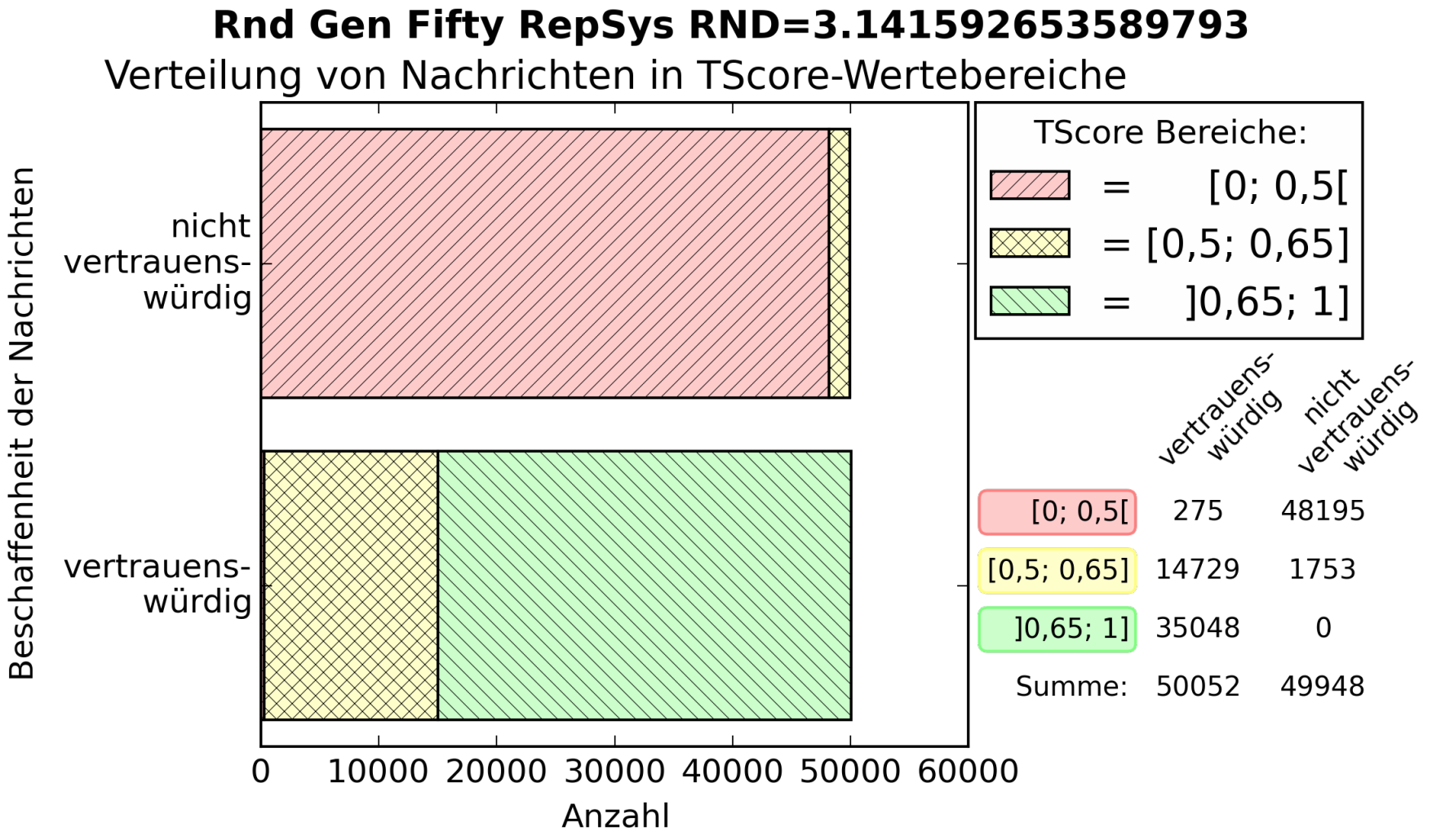

**Abbildung 5.8:** Gleichverteile, 100.000er Stichprobe

Generator und Startwert allerdings mit einer größeren Stichprobe von 100000 Nachrichten, liegen die Verhältnisse nur noch ±1 Prozentpunkt um die zuvor errechneten Werte, siehe Grafik 5.8.

Da fast 97% der nicht vertrauenswürdigen und fast 70% der vertrauenswürdigen Nachrichten korrekt erkannt werden, scheint das konzipierte Meta-Reputationssystem die gestellten Anforderungen bereits zufriedenstellend zu erfüllen, jedoch besteht in gewissen Bereichen noch Verbesserungspotenzial, siehe Abschnitt 5.2.5.

### 5.2.5 Abschließende Betrachtung des Reputationssystems

Ein Kritikpunkt an der Simulation und dem konzipierten Meta-Reputationssystem ist, dass zu wenig Komplexität vorhanden ist, sodass, nur auf Grund der Einfachheit und Abstraktion, gute Ergebnisse erzielt werden. Dies ist ein valider Einwand und daher ist einer der nächsten Schritte, für eine zukünftige Forschungsarbeit, eine Implementierung der Simulation in einer Simulationsumgebung, in der komplexere Umgebungen und Parameter realisiert werden können. Wie Eingangs beschrieben wurde diese Art der Simulation verwendet, da diese zwar konkret am Beispiel eines Systems für ein VANET angelehnt ist, aber trotzdem auch für ein MANET oder allgemeiner einer CPS-Umgebung angepasst werden könnte.

Ein weiterer Kritikpunkt ist die einfach gestrickte Generierung der Nachrichten, denn dadurch werden Kombinationen erzeugt, die realitätsfern sind. Diese einfache Generierung führt zu solchen Kombinationen, allerdings ist dies mit einer Art Fuzzing für das Meta-Reputationssystem zu vergleichen. Beim Software Fuzzing werden gerade auch solche Kombinationen gezielt erzeugt, um Schwachstellen in den Grenzbereichen von Software zu finden. Dieser Fuzzingeffekt ist hier somit erwünscht. Unabhängig davon sollte in Zukunft eine Simulation konzipiert werden, die bekannte Angriffe, wie beschrieben im Abschnitt 3.1, nachstellen kann.

In den nun folgenden Abschnitten werden kurze Anmerkungen zu den jeweiligen Teilschritten des Meta-Reputationssystem gegeben.

**Kontext:**

Der Kontext könnte an eine Verbindung gebunden sein und wie zuvor in Abschnitt 5.2.2 beschrieben und in der Simulation umgesetzt, einen Durchstich durch alle Schichten beinhalten, inkl. nötige Identitäten. Allerdings kann Kommunikation auch verbindungslos sein, daher sollte durch weitere Untersuchungen die Nützlichkeit dieser Maßnahme überprüft werden.

**Rolle:**

Die in der Simulation betrachteten Rollen stellen nur einen Ausschnitt der denkbaren Rollen dar. Zusätzlich wären weitere Rollen, wie Identitätsanbieter oder spezielle Rollen in einem Overlay-Netzwerk, möglich und die Verwaltung dieser könnte zu einer Herausforderung werden. Gerade wenn, wie in der Simulation durchgeführt, die Rolle den Kontext für die Identitäten bietet. So ist eine schwache Identität bei einer Infrastrukturdientsleistung in der Rolle der RSU weniger vertrauenswürdig als eine starke. Es sollte untersucht werden, inwiefern Rollen und Identitäten von dem Identitätsmanagementsystem sinnvoll gepflegt und bereitgestellt werden können.

**Identität:**

Anhand der Ausarbeitung über Anforderungen und Kriterien von Identitäten sollte überprüft werden, ob Privatsphäre schützende Kommunikation als nicht vertrauenswürdig eingestuft wird, da dies bis zu einem gewissen Grad abhängig von der Einstufung der verwendeten Identitäten als stark, mittel oder schwach ist.

**Erfahrung:**

Ob ein Gedächtnis, das Erfahrungen über das Vertrauen in andere Teilnehmer in einem Kommunikationsnetz speichert, wie durch Huang *et al.*[31] beschrieben, keinen Vorteil für ein Reputationssystem in einem VANET bringt, sollte überprüft werden. Hierfür ist allerdings eine Simulation nötigt, die verschiedene Szenarien (z. B. Stadt, Schnellstraße, viel oder wenig Verkehr) von VANETs abbilden kann. Gerade für Pendler, die regelmäßig um dieselbe Zeit verkehren, könnte sich durchaus ein Wiedererkennen von Teilnehmern und ein Aufbau von mittel- bis langfristigen Vertrauensverhältnissen lohnen. Dies sollte ebenfalls durch Simulation oder durch gezielte Verkehrsbeobachtung erhoben werden. Auch komplexere Ansätze wie *VEBAS* [65] schlagen vor längerfristige Vertrauensbeziehungen aufzubauen. Eine weitere Optimierungsmöglichkeit wäre auch, um für eine kurzfristige Erfahrungsbildung zu sorgen ohne ungültige oder irrelevanten Informationen aufzubewahren, über eine Vergesslichkeit des Systems zu realisieren. Dieses Vorgehen wurde bereits durch [39] oder [78] vorgeschlagen.

**Information:**

Lokale Verifikation bietet den Vorteil, dass jeder Teilnehmer unabhängiger ist. Es wird keine Kooperation und Kommunikation zwischen den verschiedenen Teilnehmern, über die zu verifizierende Nachricht hinaus, im VANET benötigt. Eine Grundvoraussetzung für eine Verifikation, unabhängig ob entfernt oder lokal, ist das die Informationen in einer Form vorliegen, sodass eine Komponente, hier das Reputationssystem, diese verstehen kann. Wenn Anwendungen ihre Informationen in einem anwendungsspezifischen Format nutzen, dann stellt dies, wegen der Vielfalt von Applikationen, eine sehr große Hürde für eine Realisierung des Reputationssystems dar. Eine Lösungsmöglichkeit für dieses Problem wäre eine Schnittstelle, über die eine Anwendung dem Reputationssystem eine Spezifikation des Datenformates vorgeben kann, so könnte das Reputationssystem auch unterschiedliche Informationsdarstellungen unterstützen.

# 6. Zusammenfassung und Fazit

Eine immer bessere Vernetzung bietet viele Chancen aber auch Risiken. Je eigenständiger ein System für seine eigene Sicherheit sorgen kann, desto besser. Dies gilt besonders im Automobilbereich, in dem das autonome oder teil-autonome Fahren eine am Technologiehorizont aufsteigende Errungenschaft darstellt. Dabei stellt die Vertrauensfrage – Wie können sich Systeme in einem ITS gegenseitig oder den Informationen vertrauen? – eine fundamentale Fragestellung dar. Hierzu wurde in dieser Arbeit ein Lösungsansatz entwickelt.

Am Beispiel des inHMotion Forschungsprojektes wurden Anwendungsfälle erarbeitet, auf deren Grundlage eine Bedrohungsmodellierung durchgeführt wurde. Zur Unterstützung des Erhebungsprozesses wurden Angriffsbäume eingesetzt. Die Bewertung der Bedrohung erfolgte durch das OWASP Risk Rating. Für die so identifizierten Bedrohungen, die als besonders hohes Schadensrisiko eingestuft wurden, wurden verschiedene Lösungsansätze angeboten. Ein Ansatz kann gleich mehrere Anforderungen erfüllen und stellt gleichzeitig einen Lösungsweg dar, um die Erfassung von Vertrauen oder Vertrauenswürdigkeit zu bewerkstelligen. Die Lösung dafür ist ein Reputationssystem, im Speziellen wurde ein Meta-Reputationssystem konzipiert und dessen Leistungsfähigkeit wurde anhand einer Simulation untersucht.

Das vorgeschlagene Modell liefert bereits gute Ergebnisse, dennoch muss in zukünftigen Arbeiten die Komplexität der Simulation erhöht werden. Dies könnte beispielsweise im Bereich der kontext-und rollenbasierten Verarbeitung geschehen. Zudem sind bei der Kategorisierung von Identitäten und beim Einfluss von Erfahrungsbildung auf die Leistung des Meta-Reputationssystems weitere Untersuchungen nötig. Die Informationsverarbeitung, im Sinne der Beschaffenheit von Schnittstellen und der Definitionen von Regeln zur Verarbeitung, stellt eine Implementierungsherausforderung dar.

Durch diese Arbeit wurde das Konzept eines Meta-Reputationssystems beschrieben und durch Simulation ein erster Beleg für dessen Nützlichkeit erbracht. Zusammenfassend lässt sich sagen, dass eine solide Grundlage geschaffen wurde auf deren Basis zukünftige Ansätze erarbeitet werden können.

# A. Tabellen

Im Tabellenanhang sind alle Tabellen und Listen aufgeführt, die der Vollständigkeit halber zu dieser Arbeit gehören oder im Laufe der Arbeit als Ergebnisse entstanden sind. Um nicht den Lesefluß des Lesers durch große Ansammlungen von Tabellen zu stören, wird an den nötigen Stellen nur auf den jeweiligen Abschnitt in diesem Anhang verwiesen.

## A.1 Anwendungsfälle

In den nun folgenden Abschnitten sind alle entworfenen Anwendungsfälle aufgeführt. Diese bilden die Grundlage für die grobe Systemarchitektur 2.1.

### A.1.1 Anwendungsfall 01: Pünktlich ankommen

| Anwendungsfall | Pünktlich ankommen |
|---|---|
| Ziel: | Pünktlich am Reiseziel ankommen |
| Level: | Systemüberblick |
| Vorbedingung: | Nutzer hat ein Smartphone mit ONP-Software und ein Profil beim Empfehlungsdienst |
| Akteur: | Empfehlungssystemnutzer |
| Ereignis: | Ein Stau hat sich auf der gewöhnlichen Route des Nutzers gebildet |
| Ergebnis: | Nutzer steht früher auf, kommt rechtzeitig an sein Reiseziel |
| Beschreibung: | Der Nutzer wird durch das System früher geweckt als durch den regulären Wecker, da das System einen Stau auf der gewöhnlichen Reiseroute des Nutzers (z.B.: in die Arbeit) erkannt hat. Er kann entscheiden, ob er der Empfehlung folgt oder nicht. Möglich wird die Benachrichtigung, da das System lokale Verkehrsinformationen erhält und die tägliche Routine eines Nutzers erlernt hat oder diese vorgegeben wurde. |
| Fehlerfall: | Falsche Verkehrsinformationen, es existiert kein Stau mehr. |

**Tabelle A.1:** Anwendungsfall 01: Pünktlich ankommen

### A.1.2 Anwendungsfall 02: Erweiterung für ACC

| Anwendungsfall | Erweiterung für ACC |
|---|---|
| Ziel: | Optimierung des Verkehrsflusses und Verbesserung der Fahrerassistenzsysteme |
| Level: | Systemüberblick |
| Vorbedingung: | Nutzer hat ein Fahrzeug mit ONP-Software |
| Akteur: | Systemnutzer |
| Ereignis: | - |
| Ergebnis: | Vorausschauende automatische Anpassung der Geschwindigkeit und entspanntes (autonomes) Fahren |
| Beschreibung: | Die ONP-Software sendet und empfängt stetig Telemetriedaten (z. B. Position, Geschwindigkeit, Bremsverhalten, Steigung, Reibungskoeffizienten), dadurch hat die Software ein aktuelles Bild der Verkehrslage in seiner direkten Umgebung und kann diese nutzen um das ACC vorausschauender zu steuern, nicht nur bis zur Stoßstange des Vorausfahrenden, sondern so weit wie es die drahtlose Kommunikation erlaubt. |
| Fehlerfall: | Falsche Verkehrsinformationen, defekte Sensoren, keine Kommunikationspartner. |

**Tabelle A.2:** Anwendungsfall 02: Erweiterung für ACC

### A.1.3 Anwendungsfall 03: Lokale Informationsverbreitung durch Infrastruktur

| Anwendungsfall | Lokale Informationsverbreitung durch Infrastruktur |
|---|---|
| Ziel: | Optimierung des Verkehrsflusses und Unterstützung von vorausschauendem Fahren |
| Level: | Systemüberblick |
| Vorbedingung: | Nutzer hat ein Fahrzeug mit ONP-Software |
| Akteur: | Systemnutzer |
| Ereignis: | Verkehrsflussveränderungen, Aufstellen einer RSU |
| Ergebnis: | Vorausschauende Fahrweise, Aufrechterhalten des Verkehrsflusses |
| Beschreibung: | Bei Veränderungen der Verkehrsführung, z. B. durch eine Baustelle oder einen Unfall, wird eine RSU aufgestellt, die als Verkehrsleitsystem fungiert und über unterschiedliche Kommunikationstechnologien, z. B. adhoc, über die neue Verkehrsführung informiert. Über den Ad-hoc-Kommunikationspfad werden diese Informationen durch das ONP weiter propagiert. |
| Fehlerfall: | Falsche Verkehrsinformationen (Abbau der RSU wurde übersehen). |

**Tabelle A.3:** Anwendungsfall 03: Lokale Informationsverbreitung durch Infrastruktur

### A.1.4 Anwendungsfall 04: Lokale Verfügbarkeitsinformationen

| Anwendungsfall | Lokale Verfügbarkeitsinformationen |
|---|---|
| Ziel: | Verfügbarkeitsinformationen zu unterschiedlichen Dienstleistungen oder Angeboten lokal verbreiten |
| Level: | Systemüberblick |
| Vorbedingung: | Nutzer hat ein Fahrzeug mit ONP-Software |
| Akteur: | Systemnutzer |
| Ereignis: | - |
| Ergebnis: | Nutzer findet zielstrebig ein geeignetes Angebot, Dienstleitung oder Objekt, das seine Interessen befriedigt |
| Beschreibung: | Am Ende oder als Zwischenstopp einer Reise benötigt der Nutzer einen Parkplatz, eine Ladestation für das Elektroauto, eine Tankstelle oder Ähnliches. Parkhäuser, Ladestationen oder Tankstellen verbreiten ihr Angebot, bzw. Verfügbarkeit, über das ONP-Netzwerk. So kann dem Nutzer die Verfügbarkeit von lokalen Ressourcen angezeigt werden und eine Ziel- oder Entscheidungsfindung unterstützen. |
| Fehlerfall: | Falsche Verfügbarketisinformationen, keine mit ONP ausgestatteten Ressourcen verfügbar. |

**Tabelle A.4:** Anwendungsfall 04: Lokale Verfügbarkeitsinformationen

### A.1.5 Anwendungsfall 05: Fahrzeuginformationen

| Anwendungsfall | Fahrzeuginformationen |
|---|---|
| Ziel: | Statusinformation an den Nutzer übermitteln |
| Level: | Systemüberblick |
| Vorbedingung: | Nutzer hat ein Fahrzeug mit ONP-Software und ein Smartphone mit ONP-Software |
| Akteur: | Systemnutzer |
| Ereignis: | - |
| Ergebnis: | Nutzer erhält aktuelle Statusinformationen zu seinem Fahrzeug |
| Beschreibung: | Ohne sich im Fahrzeug zu befinden, kann der Nutzer aktuelle Zustände des Fahrzeuges über ONP abfragen. Das Fahrzeug informiert direkt über Fahrersitzeinstellungen, Standheizung, Kindersicherung, Beifahrerairbag oder Position. Es ist keine zentrale Kommunikation nötig, dennoch als Ausweichlösung realisiert. |
| Fehlerfall: | ONP-Empfang nicht ausreichend, zentrales Ausweichsystem auch nicht erreichbar. |

**Tabelle A.5:** Anwendungsfall 05: Fahrzeuginformationen

### A.1.6 Anwendungsfall 06: Notfallinformationen

| Anwendungsfall | Notfallinformationen |
|---|---|
| Ziel: | Notfallinformationen an den Nutzer übermitteln, Verbesserung des Nutzerverhaltens, vorausschauendes Verhalten provozieren, Reduzierung von Unfällen und Vermeidung von Verzögerungen in Notfallsituationen |
| Level: | Systemüberblick |
| Vorbedingung: | Nutzer hat ein Fahrzeug mit ONP-Software |
| Akteur: | Systemnutzer |
| Ereignis: | Notfalleinsatz Rettungswagen- oder Polizeieinsatz |
| Ergebnis: | Nutzer erhält eine Warnung und verhält sich rücksichtsvoll, um den Einsatz nicht zu behindern |
| Beschreibung: | Der Einsatzwagen sendet eine Einsatzmeldung an alle, auf seiner Route liegenden, Fahrzeuge über ONP mit der Bitte den Weg frei zu machen. Ein Fahrzeug mit ONP-Software kann aus den enthaltenen Informationen, wie Teilstreckeninformationen und Geschwindigkeit des Einsatzfahrzeuges, den Zeitpunkt abschätzen ab wann das eigene Fahrzeug ein Hindernis darstellt und den Nutzer rechtzeitig auf den Notfall hinweisen. Dies geschieht noch bevor der Fahrer Blaulicht sehen kann oder eine Sirene hört. |
| Fehlerfall: | Falsche Routeninformationen oder falsches Kartenmaterial. |

**Tabelle A.6:** Anwendungsfall 06: Notfallinformationen

### A.1.7 Anwendungsfall 07: Geschwindigkeitsinformationen

| Anwendungsfall | Geschwindigkeitsinformationen |
|---|---|
| Ziel: | Einbindung des Fahrzeuges und anderer Verkehrsteilnehmer in Verkehrsleitsysteme |
| Level: | Systemüberblick |
| Vorbedingung: | Nutzer hat ein Fahrzeug mit ONP-Software |
| Akteur: | Systemnutzer |
| Ereignis: | Geschwindigkeitsanpassung |
| Ergebnis: | Dem Nutzer wird die zugelassene Geschwindigkeit in seinem Fahrzeug angezeigt |
| Beschreibung: | Zusätzlich zu den Geschwindigkeitsschildern senden RSUs oder andere Verkehrsteilnehmer die zugelasse Höchstgeschwindigkeit über ONP aus. Durch eigene Sensoren wie Verkehrsschilderkennungssysteme können diese Informationen verifiziert werden. Durch das Weiterleiten werden Verkehrsteilnehmer frühzeitig auf Änderungen aufmerksam gemacht und wissen jederzeit was die zugelassene Höchstgeschwindigkeit ist. |
| Fehlerfall: | Falsche Geschwindigkeitsinformationen. |

**Tabelle A.7:** Anwendungsfall 07: Geschwindigkeitsinformationen

### A.1.8 Anwendungsfall 08: Warnung bei hohem Verkehrsaufkommen

| Anwendungsfall | Warnung bei hohem Verkehrsaufkommen |
|---|---|
| Ziel: | Optimierung des Verkehrsflusses und Unterstützung von vorausschauendem Fahren, Reduzierung des Unfallrisikos |
| Level: | Systemüberblick |
| Vorbedingung: | Nutzer hat ein Fahrzeug mit ONP-Software |
| Akteur: | Systemnutzer |
| Ereignis: | hohes Verkehrsaufkommen |
| Ergebnis: | Anpassung des Fahrverhaltens an Verkehrslage |
| Beschreibung: | Die ONP-Software erkennt durch die ständige kooperative Bewusstseinskommunikation mit anderen Teilnehmern, dass eine hohe Verkehrsdichte vorliegt und unterstützt den Fahrer dahingehend, dass an Kreuzungen heranfahrende Fahrzeuge dem Fahrer angekündigt werden und gegebenenfalls je nach Vorfahrtregelung eine Empfehlung zur Geschwindigkeitsanpassung gegeben wird. |
| Fehlerfall: | Falsche Geschwindigkeitsinformationen, falsche Vorfahrtregelungsinformationen. |

**Tabelle A.8:** Anwendungsfall 08: Warnung bei hohem Verkehrsaufkommen

### A.1.9 Anwendungsfall 09: Lernende Navigation

| Anwendungsfall | Lernende Navigation |
|---|---|
| Ziel: | Optimierung des Verkehrsflusses und Berücksichtigung der Vorlieben der Nutzer |
| Level: | Systemüberblick |
| Vorbedingung: | Nutzer hat ein Fahrzeug mit ONP-Software |
| Akteur: | Systemnutzer |
| Ereignis: | - |
| Ergebnis: | Navigation berücksichtigt Vorlieben der Nutzer |
| Beschreibung: | Werden bestimmte Ziele und Routen durch den Nutzer öfter angefahren , so bietet sich die Möglichkeit für das Navigationssystem Vorlieben zu erlernen. Ein Nutzer könnte immer einen bestimmten Streckenabschnitt nehmen, z. B. weil die Landschaft schön ist, obwohl die Strecke stark befahren ist und zu Verzögerungen führt. Am Anfang würde es dem Nutzer nur eine Ausweichroute empfehlen, nach dem Lernprozess plant es zusätzlich die Verzögerung mit ein und benachrichtigt den Nutzer rechtzeitig aufzubrechen. (siehe A.1.1) Darüber hinaus nutzt das Navigationssystem die lokalen ONP-Informationen, um die Routenplanung nach den Vorlieben des Nutzers zu optimieren, z. B. schnell, pünktlich oder stressfrei ankommen. |
| Fehlerfall: | Falsche Verkehrsinformationen, erlernte Nutzervorlieben sind fehlerhaft bzw. identifizierte Vorlieben sind falsch. |

**Tabelle A.9:** Anwendungsfall 09: Lernende Navigation

### A.1.10 Anwendungsfall 10: Gefahrenmeldungen bei geringem Verkehrsaufkommen

| Anwendungsfall | Gefahrenmeldungen bei geringem Verkehrsaufkommen |
|---|---|
| Ziel: | Unterstützung von vorausschauendem Fahren, Reduzierung des Unfallrisikos |
| Level: | Systemüberblick |
| Vorbedingung: | Nutzer hat ein Fahrzeug mit ONP-Software |
| Akteur: | Systemnutzer |
| Ereignis: | - |
| Ergebnis: | Nutzer erhält einen Hinweis zur Anpassung des Fahrverhaltens |
| Beschreibung: | Die Sensoren eines Nutzers stellen eine Änderung der Straßenbeschaffenheit fest, z. B. Glatteis. Diese Information wird über das ONP verbreitet, falls keine Kommunikationspartner vorhanden sind, wird diese Information an ein externes System weitergeleitet. Dieses speichert die Daten und sendet sie gezielt an Teilnehmer in der betroffenen geographischen Region. |
| Fehlerfall: | Falsche Straßenzustandsmeldungen, keine Kommunikationspartner vorhanden, auch keine externen Systeme erreichbar. |

**Tabelle A.10:** Anwendungsfall 10: Gefahrenmeldungen bei geringem Verkehrsaufkommen

### A.1.11 Anwendungsfall 11: Intermodaler Transport

| Anwendungsfall | Intermodaler Transport |
|---|---|
| Ziel: | Verbinden von verschiedenen Transportmitteln |
| Level: | Systemüberblick |
| Vorbedingung: | Nutzer hat ein Smartphone mit ONP-Software |
| Akteur: | Empfehlungssystemnutzer |
| Ereignis: | - |
| Ergebnis: | Nutzer erhält Reiseplanempfehlungen mit verschiedenen Transportmitteln |
| Beschreibung: | Der Nutzer hat ein Reiseziel und verwendet die ONP-Software dazu einen Reiseplan zu erzeugen. Er gibt sein Start- und Zielort an und die Software fragt an einem Backend-System, das Fahrpläne von Bus, Bahn, Tram aggregiert, nach Reiseempfehlungen und bietet diese dem Nutzer an. |
| Fehlerfall: | Falsche Verkehrsinformationen, falsche externe Quellen, z. B. Fahrplaninformationen |

**Tabelle A.11:** Anwendungsfall 11: Intermodaler Transport

### A.1.12 Anwendungsfall 12: Optimierte Reiseempfehlungen

| Anwendungsfall | Optimierte Reiseempfehlungen |
|---|---|
| Ziel: | Optimierung der Reiseempfehlung nach den Wünschen des Nutzers |
| Level: | Systemüberblick |
| Vorbedingung: | Nutzer hat ein Smartphone mit ONP-Software |
| Akteur: | Empfehlungssystemnutzer |
| Ereignis: | - |
| Ergebnis: | Nutzer erhält nach seinen Wünschen optimierte Reiseplanempfehlungen |
| Beschreibung: | Zusätzlich zu dem Vorgehen in A.1.11 kann der Nutzer bestimmte Optimierungskriterien gemäß seinen Wünschen anpassen. Kriterien sind der Preis der Reise, die Bequemlichkeit, die Pünktlichkeit oder die Schonung der Umwelt. |
| Fehlerfall: | Kriterien sind widersprüchlich, Optimierung nicht möglich. |

**Tabelle A.12:** Anwendungsfall 12: Optimierte Reiseempfehlungen

### A.1.13 Anwendungsfall 13: Buchung der Reiseempfehlung

| Anwendungsfall | Buchung der Reiseempfehlung |
|---|---|
| Ziel: | Einfache Buchung der nötigen Transportmittel |
| Level: | Systemüberblick |
| Vorbedingung: | Nutzer hat ein Smartphone mit ONP-Software, Nutzer hat mindestens eine Reiseempfehlung erhalten |
| Akteur: | Empfehlungssystemnutzer |
| Ereignis: | - |
| Ergebnis: | Nutzer erhält die nötigen Tickets und Buchungsbestätigungen auf sein Smartphone |
| Beschreibung: | Nachdem der Nutzer mindestens eine Reiseempfehlung erhalten hat, durch A.1.11 oder A.1.12, möchte er diese Reise antreten und dazu alle Verkehrsmittel buchen. Dies ermöglicht ihm die ONP-Software, indem sie an ein Buchungssystem die Gesamtreiseanfrage stellt, dieses System liefert in Folge die nötigen Reservierungen. Tickets und Buchungen zurück an den Nutzer. |
| Fehlerfall: | Falsche Buchungsinformationen, externe Buchungssysteme nicht verfügbar. |

**Tabelle A.13:** Anwendungsfall 13: Buchung der Reiseempfehlung

### A.1.14 Anwendungsfall 14: Reiseplanänderungen

| Anwendungsfall | Reiseplanänderungen |
|---|---|
| Ziel: | Reaktion auf Verkehrslageänderungen in Bezug auf den Reiseplan |
| Level: | Systemüberblick |
| Vorbedingung: | Nutzer hat ein Smartphone mit ONP-Software, hat einen aktiven Reiseplan |
| Akteur: | Empfehlungssystemnutzer |
| Ereignis: | Änderung der Verkehrslage, z. B. Unfall auf Bus- oder Tram-Route |
| Ergebnis: | Nutzer erhält Hinweis und Empfehlung zur Anpassung des Reiseplanes |
| Beschreibung: | Das Empfehlungssystem beobachtet für den Nutzer die Verkehrslage auf den, durch den Reiseplan betroffenen, Strecken. Zusätzlich sammelt das System lokale Verkehrsinformationen durch das ONP. Wenn Veränderung auftreten, die eine Planänderung hervorrufen, z. B. weil ein Unfall eine Straße vollständig blockiert oder es zu großen Verzögerungen kommt, dann berechnet das Empfehlungssystem gemäß den Wünschen des Nutzers einen neuen Reiseplan und übermittelt diesen an die ONP-Software. |
| Fehlerfall: | Falsche Verkehrsinformationen, keine Ausweichmöglichkeiten wegen beschränkter Verkehrsmittelverfügbarkeit, z. B. Fähre. |

**Tabelle A.14:** Anwendungsfall 14: Reiseplanänderungen

### A.1.15 Anwendungsfall 15: Intermodaler Transport und Fußgängernavigation

| Anwendungsfall | Intermodaler Transport und Fußgängernavigation |
|---|---|
| Ziel: | Zeitnahe Nutzung der Verkehrsinformationen speziell auch für Fußgänger |
| Level: | Systemüberblick |
| Vorbedingung: | Nutzer hat ein Smartphone mit ONP-Software, hat einen aktiven Reiseplan |
| Akteur: | Empfehlungssystemnutzer |
| Ereignis: | Änderung der Verkehrslage, z. B. großer Andrang an öffentlichen Plätzen und Veranstaltungen |
| Ergebnis: | Nutzer erhält eine auf Fußgänger angepasste Navigation |
| Beschreibung: | Bei Änderungen der Verkehrslage, wie in A.1.14 beschrieben, empfiehlt das System aufgrund von hohem Verkehrsaufkommen das Fortsetzen der Reise zu Fuß. Bei Veranstaltungen sind meistens auch die Fußwege um die Lokalität herum stark frequentiert. Der Verkehrsinformationsaustausch funktioniert über eine situationsbedingte Kombination aus Ad-hoc- und Mobilfunkkommunikation. Diese Informationen berücksichtigt das Empfehlungssystem für alle Teilnehmer, die sich noch nicht am Ziel befinden. Eine Grundlage der Empfehlungen des Empfehlungssystems bilden die Ad-hoc-Informationen. Diese werden von allen aktiven Verkehrsteilnehmern, inklusive den Fußgängern erfasst. So kann der Nutzer eine hochaktuelle Fußgänger optimierte Navigation erhalten. |
| Fehlerfall: | Falsche Verkehrsinformationen |

**Tabelle A.15:** Anwendungsfall 15: Intermodaler Transport und Fußgängernavigation

### A.1.16 Anwendungsfall 16: Automatisches Fahrtenbuch

| Anwendungsfall | Automatisches Fahrtenbuch |
|---|---|
| Ziel: | Führung eines Automatischen Fahrtenbuchens |
| Level: | Systemüberblick |
| Vorbedingung: | Nutzer hat ein Fahrzeug mit ONP-Software, Aktivierung und Konfigurierung der Fahrtenbuchfunktion |
| Akteur: | Empfehlungssystemnutzer |
| Ereignis: | - |
| Ergebnis: | Nutzer kann, wann immer er möchte, sein automatisch geführtes Fahrtenbuch abrufen |
| Beschreibung: | Jede Fahrt des Nutzers kann aufgezeichnet werden, je nach Einstellungen des Nutzers kann dies deaktiviert sein oder lokal, zentral oder kombiniert erfolgen. Zu jedem Zeitpunkt kann der Nutzer über den Empfehlungsdienst sein automatisch geführtes Fahrtenbuch abrufen. |
| Fehlerfall: | Unvollständige Erfassung der Fahrten |

**Tabelle A.16:** Anwendungsfall 16: Automatisches Fahrtenbuch

### A.1.17 Anwendungsfall 17: Kfz-Versicherung

| Anwendungsfall | Kfz-Versicherung |
|---|---|
| Ziel: | Gerechtere Versicherungstarife und Disziplinierung der Verkehrsteilnehmer zur Unfallrisiko-Reduktion |
| Level: | Systemüberblick |
| Vorbedingung: | Nutzer hat ein Fahrzeug mit ONP-Software, Aktivierung Fahrerprofilfunktion |
| Akteur: | Empfehlungssystemnutzer |
| Ereignis: | - |
| Ergebnis: | Nutzer kann ein Fahrerprofil anlegen, um günstigere Versicherungstarife zu erhalten |
| Beschreibung: | Über die Erstellung eines Fahrerprofils, das die Fahrweise, z. B. Strecken, Uhrzeit, Brems- und Beschleunigungsverhalten, beinhaltet, kann ermittelt werden, welcher Typ von Fahrer man ist. Dies kann bedeuten, dass er auf dem Fahrstil basierend ein geringes, normales oder ein hohes Unfallrisiko hat. Bei vorbildlichem Verhalten können Rabatte und Boni gewährt werden. Denkbar ist auch das Nutzen von mehreren Autos auf demselben Profil, z. B. bei mehreren eigenen Fahrzeugen, in einer Familie, bei Fahrzeugvermietung. Im Falle der Vermietung könnten bei vorbildlichem Fahrerprofil Rabatte gegeben werden. |
| Fehlerfall: | Unvollständige Erfassung der Fahrten |

**Tabelle A.17:** Anwendungsfall 17: KFZ Versicherung

### A.1.18 Anwendungsfall 18: Streckenabhängige Maut

| Anwendungsfall | Streckenabhängige Maut |
|---|---|
| Ziel: | Realisierung eines streckenabhängigen Mautsystems |
| Level: | Systemüberblick |
| Vorbedingung: | Nutzer hat ein Fahrzeug mit ONP-Software |
| Akteur: | Empfehlungssystemnutzer |
| Ereignis: | - |
| Ergebnis: | Automatische Erfassung von mautpflichtigen Strecken und Bezahlung |
| Beschreibung: | Durch die Teilnahme am streckenabhängigen Mautsystem wird jede Fahrt des Nutzers aufgezeichnet. Durch die Erfassung der gefahrenen Strecken kann exakt und auf Wunsch auch automatisch die fällige Maut beglichen werden. Dazu müssen Zahlungsinformationen beim Empfehlungsdienst hinterlegt sein. Das ONP-System unterstützt die korrekte Erfassung der Strecken u. a. durch verifizierte Positionsbestimmung mit Hilfe von RSU-Informationen. Eine RSU kann auch als virtuelle Mautschranke genutzt werden, so dass diese das Passieren an das Empfehlungssystem meldet. |
| Fehlerfall: | Unvollständige Erfassung der Fahrten |

**Tabelle A.18:** Anwendungsfall 18: Strecken abhängige Maut

## A.2 Auflistung aller konstruierten Angriffsbäume

Im Folgenden sind alle konstruierten Angriffsbäume aufgeführt. Jeder Angriffsbaum wird als Auflistung seiner einzelnen Bedrohungen und einer Übersicht der betroffenen Akteure und Werte dargestellt.

### A.2.1 Angriffsbaum: Monetäre Vorteile

**Ziel:** *Kostenpflichtige Ressourcen (Anwendungen, Dienste) kostenlos nutzen*

1. Ausnutzen von schwacher Authentifizierung bei Diensten B.32
   (a) Klonen der PHY-MAC B.3
2. Ausnutzen von fehlenden Kontrollen in der Geschäftslogik eines Dienstes B.33
3. Anwendungen aus dem AppStore stehlen B.49
   (a) Einmal installieren, dann von Gerät zu Gerät kopieren
   (b) Anwendung als Installationspaket aus dem AppStore herunterladen und verteilen

Die folgenden Werte sind durch dieses Ziel und die einzelnen Bedrohungen betroffen:

- Hersteller:
  - hohe Marktdurchdringung
  - vielseitiges Angebot von Diensten

### A.2.2 Angriffsbaum: Leibliches Wohlergehen

**Ziel:** *Schaden an Leib und Leben der Nutzer anrichten*

1. Komponenten dazu bringen manipulierte Informationen zu nutzen
   (a) Manipulierte Informationen aussenden, um Teilnehmer zu beeinflussen und möglicherweise schädliche Reaktionen der Nutzer oder sicherheitskritischer Anwendungen provozieren B.47
   (b) Software-Manipulation der Sensoren oder Steuereinheiten B.41
      i. Einspielen manipulierter Firmware
   (a) Hardware-Manipulation, Kurzschließen von Kontakten, Beschädigen des Sensors, z.B. GPS-Antenne B.40
2. Ausfälle von sicherheitskritischen Komponenten provozieren
   (a) Software-Manipulation der Sensoren oder Steuereinheiten B.41
   (b) Hardware-Manipulation, Kurzschließen von Kontakten, Beschädigen des Sensors, z.B. GPS-Antenne B.40

Die folgenden Werte sind durch dieses Ziel und die einzelnen Bedrohungen betroffen:

- spezieller Dienstnutzer:
  - Schutz von Leib und Leben bei Nutzung
- Overlaynutzer:
  - Leib und Leben der Nutzer, auch aus Gewährleistungsgründen und Schutz vor Schadensersatzansprüchen
  - Nichtabstreitbarkeit bei sicherheitskritischen Aktionen
- Hersteller:
  - Leib und Leben der Nutzer
  - Vertrauen und Ansehen
  - Hohes Maß an Betriebssicherheit
  - Fehlerfreiheit der Software
  - sichere Fehlerbehandlung

### A.2.3 Angriffsbaum: Denial of Service

**Ziel:** *Einen ordnungsgemäßen Betrieb behindern: Denial of Service Angriffe (DoS)*

1. Überfluten mit Datenpaketen im zellularen Netzwerk B.7
2. Überfluten mit Datenpaketen im Ad-hoc-Netzwerk B.6
3. Störsender im zellularen Frequenzbereich B.5
4. Störsender im Ad-hoc-Frequenzbereich B.4
5. Dienste mit hohem Ressourcenverbrauch anbieten B.46
6. Gezieltes Unterdrücken von einzelnen Informationen
   (a) Gezieltes Unterdrücken von einzelnen Informationen via Ad-hoc-Kommunikation B.8
   (b) Gezieltes Unterdrücken von einzelnen Informationen via zellularer Kommunikation B.9
7. Beeinflussung von Softwarekomponenten durch manipulierte andere Komponenten B.48

Die folgenden Werte sind durch dieses Ziel und die einzelnen Bedrohungen betroffen:

- allgemeiner Dienstnutzer:
  - Verfügbarkeit des Dienstes
  - Vertrauen in den Dienst
- Overlaynutzer:
  - hohe Effizienz

- Priorisierung der Nachrichten
  - Rufschädigung des Dienstanbieters
  - Keine ungewollte gegenseitige Beeinflussung von verschiedenen Diensten

- Hersteller:
  - wenig Infrastruktur
  - Minimierung der Betriebskosten
  - Vertrauen und Ansehen des Herstellers

### A.2.4 Angriffsbaum: Manipulation von Monetären Werten

**Ziel:** *Manipulation von Daten um, den Wert einer Ressource zu erhöhen (Kilometerzähler zurückdrehen, Chiptuning)*

1. Hardware-Manipulation, Kurzschließen von Kontakten, Beschädigen des Sensors, z.B. GPS-Antenne B.40
2. Software-Manipulation der Sensoren oder Steuereinheiten B.41
   (a) Einspielen manipulierter Firmware
3. Manipulation von Archivdaten B.42
   (a) Einfügen von manipulierten Einträgen (Log Injection)
   (b) Löschen der Archivdaten
4. Manipulierte Identitäten (als mehrere/eine bestimmte ausgeben)
   (a) Manipulierte Identitäten via Software B.1
   (b) Manipulierte Identitäten via Hardware B.2

Die folgenden Werte sind durch dieses Ziel und die einzelnen Bedrohungen betroffen:

- Overlaynutzer:
  - unterschiedliche Niveaus von Vertraulichkeit, Integrität, Authentizität, Verfügbarkeit und Datenschutz
  - Nichtabstreitbarkeit bei sicherheitskritischen Aktionen
  - Qualität der Informationen

### A.2.5 Angriffsbaum: Manipulation von Diensten

**Ziel:** *Manipulation von Daten, um den Wert der Daten bzw. den Wert eines Dienstes zu verringern (Empfehlungssystem)*

1. Injection-Angriff auf die Anwendungsdatenhaltung B.25
2. Injection-Angriff auf die Dienst-Datenhaltung B.31
3. Injection-Angriff auf die ONP-Datenhaltung B.27
4. Daten während der Übertragung manipulieren

(a) Daten über das Ad-hoc-Netzwerk manipulieren B.13
   (b) Daten über das zellulare Netzwerk manipulieren B.14
   (c) Daten über das Backbone-Netzwerk manipulieren (Internet) B.15
5. Aggregation von Daten via ad-hoc manipulieren B.45
6. Manipulierte Fahrpläne durch DNS-Hijacking (backbone) unterschieben B.44
7. Manipulierte Positionsdaten via ad-hoc senden B.43
   (a) Manipulation von Sensoren durch Umwelteinflüsse
      i. Opfer/Ziel muss sich in einem Tunnel befinden (AND)
      ii. Manipuliertes GPS-Signal aussenden (AND)
   (a) Software-Manipulation der Sensoren oder Steuereinheiten B.41
      i. Einspielen manipulierter Firmware
   (b) Hardware-Manipulation, Kurzschließen von Kontakten, Beschädigen des Sensors, z.B. GPS-Antenne B.40
8. Manipulierte Positionsdaten via zelullar senden
   (a) Manipulation von Sensoren durch Umwelteinflüsse
      i. Opfer/Ziel muss sich in einem Tunnel befinden (AND)
      ii. Manipuliertes GPS-Signal aussenden (AND)
   (a) Software-Manipulation der Sensoren oder Steuereinheiten B.41
      i. Einspielen manipulierter Firmware
   (b) Hardware-Manipulation, Kurzschließen von Kontakten, Beschädigen des Sensors, z.B. GPS-Antenne B.40

Die folgenden Werte sind durch dieses Ziel und die einzelnen Bedrohungen betroffen:

- spezieller Dienstnutzer:
  - zuverlässige Handlungsempfehlungen
  - Verfügbarkeit der Ortsinformationen
- allgemeiner Dienstnutzer:
  - Vertrauen in den Dienst
  - Keine negative Beeinflussung durch Overlay
- Overlaynutzer:
  - Rufschädigung des Dienstanbieters
  - unterschiedliche Niveaus von Vertraulichkeit, Integrität, Authentizität, Verfügbarkeit und Datenschutz
  - Qualität der Informationen

- Hersteller:
  - Vertrauen und Ansehen des Herstellers
  - funktionierendes, intelligentes Empfehlungssystem

### A.2.6 Angriffsbaum: Diebstahl von vertraulichen Informationen

**Ziel:** *Stehlen von vertraulichen Daten: Positionsdaten (aktuell/vergangen/ zukünftig), personenbezogene Daten (Name, Adresse, etc.)*

1. Daten durch einen Dienst stehlen
   (a) Daten von dem Dienstanbieter stehlen B.34
      i. Injection-Angriff auf die Dienst-Datenhaltung B.31
      ii. Einen Dienst eines Anbieters übernehmen
         A. Durch Schwachstellen in Service-Software
            Bufferoverflow o. Ä. resultierend in Remote Code Execution (RCE)
         B. Durch Schwachstellen im Betriebssystem des Services
   (b) Einen Dienst imitieren
      i. Einen eigenen Dienst bereitstellen (AND)
      ii. Durch DNS-Hijacking (AND)
2. Daten am Endgerät des Teilnehmers stehlen
   (a) Übernehmen von Anwendungen auf einem Endgerät B.28
      i. Durch Schwachstellen in Anwendungssoftware
         A. Bufferoverflow o. Ä. resultierend in Remote Code Execution (RCE)
      ii. Durch manipulierte Sensoren (Hardware) B.25
   (b) Spionage-Anwendung auf ein Endgerät anbringen B.29
      i. Benutzer dazu bringen die Spionage App zu installieren und zu autorisieren (Phishing)
      ii. Sicherheitsüberprüfung des AppStores umgehen
         A. Injection-Angriff auf die AppStore-Datenhaltung B.30
      iii. Übernehmen eines AppStores
         A. Durch DNS-Hijacking B.19
         B. Durch Schwachstellen in AppStore-Software
            - Bufferoverflow o. Ä. resultierend in Remote Code Execution (RCE)

C. Durch Schwachstellen im Betriebssystem des Services

(c) Die Kontrolle über das Endgerät übernehmen

i. Durch publizierte Schwachstellen im Betriebssystem des Endgerätes B.17

ii. Durch unpublizierte Schwachstellen im Betriebssystem des Endgerätes B.18

iii. Durch Kompromittierung eines verbauten Coprozessors (z.B.: Modem) B.16

iv. Durch Schwachstellen in der ONP-Software

A. Durch Kompromittierung des ONP-Message-Filter B.26

- Bufferoverflow o. Ä. resultierend in Remote Code Execution (RCE)

B. Durch Kompromittierung des ONP-Recievers

- Als schädlicher AppStore B.20
  - Bufferoverflow o. Ä. resultierend in Remote Code Execution (RCE)
- Als schädlicher Dienst B.21
  - Bufferoverflow o. Ä. resultierend in Remote Code Execution (RCE)
- Als schädlicher Teilnehmer B.22
  - Bufferoverflow o. Ä. resultierend in Remote Code Execution (RCE)

C. Durch Kompromittierung des ONP-Senders

- Als schädliche Anwendung B.23

D. Durch manipulierte Sensoren (Hardware) B.24

3. Daten über das drahtlose Medium mitlesen

(a) Daten über das Ad-hoc-Netzwerk mitlesen B.10

i. Manipulation der Netzwerk Topologie, (z.B.: Wormhole Angriff) Man in the Middle (MitM) B.37

ii. Daten müssen unverschlüsselt sein Man on the Side (MotS)

iii. Verschlüsselung muss gebrochen/umgangen werden

A. MitM B.13

B. Weak Cipher / Protocol Downgrade

C. Brute Force

(b) Daten über das zellulare Netzwerk mitlesen B.11

i. Verschlüsselung muss gebrochen/umgangen werden

A. MitM B.14
      B. Weak Cipher / Protocol Downgrade
      C. Brute Force
    ii. Stehlen der symmetrischen Schlüssel
      A. aus dem Authentication Center (AuC) eines Telekommunikationsanbieters
      B. Bei Hersteller der SIM-Karten (NSA,GCHQ vs Gemalto) stehlen
      C. Seitenkanalangriffe auf Hardware Security Module (HSM) (SIM-Karten)
    iii. Manipulation der Netzwerk Topologie, (z.B.: „Stingrays“ Cell Site Simulators, IMSI-Catcher) Man in the Middle (MitM) B.36
4. Daten über das Backbone-Netzwerk mitlesen (Internet) B.12
  (a) Daten müssen unverschlüsselt sein
  (b) Verschlüsselung muss gebrochen/umgangen werden
    i. MitM B.15
    ii. Weak Cipher / Protocol Downgrade
    iii. Brute Force

Die folgenden Werte sind durch dieses Ziel und die einzelnen Bedrohungen betroffen:

- spezieller Dienstnutzer:
  - Vertraulichkeit der Ortsinformationen, der Historie dieser und die Vorhersagen über zukünftige Ortsinformationen
- allgemeiner Dienstnutzer:
  - Vertrauen in den Dienst
  - Vertraulichkeit der Kommunikation
  - Vertraulichkeit der personenbezogenen Daten
  - Datenverarbeitung gemäß dem BDSG
- Overlaynutzer:
  - Rufschädigung des Dienstanbieters
  - unterschiedliche Niveaus von Vertraulichkeit, Integrität, Authentizität, Verfügbarkeit und Datenschutz
- Hersteller:
  - Vertrauen und Ansehen des Herstellers

# B. Auflistung aller betrachteten Bedrohungen

Im folgenden sind alle betrachteten und bewerteten Bedrohungen in tabellarischer Form aufgeführt.

| | | | |
|---|---|---|---|
| *Nummer* | 1 | | |
| *Beschreibung* | Falsche Identitäten (Fahrzeug gibt sich als viele Fahrzeuge aus.) (Software) | | |
| *Bedrohung* | Verlust der Nachvollziehbarkeit durch Impersonierung | | |
| *Kategorie* | Impersonation-Attacks | | |
| *Angreifer* | Softwareentwickler im Bereich Kommunikationstechnologien | | |
| *Schnittstelle* | ONP-Message-Filter | | |
| *Entwickler, Methode* | CP, Brainstorming | | |
| *Threat Agent* | *Value* | *Comment* | |
| *Skill Level* | 3 | Softwareentwickler, bei Verkehrsdelikten Identitäten (ver-)tauschen, vor Ort sein, eigene Software nötig, wenige | |
| *Motive* | 9 | | |
| *Opportunity* | 5 | | |
| *Size* | 2 | | |
| *Vulnerability* | *Value* | *Comment* | |
| *Ease of discovery* | 7 | einfach zu entdecken, schwer ausnutzbar, offensichtliche Schwachstelle, dokumentiert aber ohne Kontrolle | |
| *Ease of exploit* | 3 | | |
| *Awareness* | 6 | | |
| *Intrusion* | 8 | | |
| *Likelihood* | 5.38 | MEDIUM | |
| *Technical Impact* | *Value* | *Comment* | |
| *Loss of Confidentiality* | 1 | keine sensiblen Daten enthüllen, wenige Daten leicht beschädigt, wenige sekundäre Dienste unterbrochen, möglicherweise nachvollziehbar | |
| *Loss of Integrity* | 1 | | |
| *Loss of Availability* | 1 | | |
| *Loss of Accountability* | 7 | | |
| *Technical Impact* | 2.50 | LOW | |
| *Business Impact* | *Value* | *Comment* | |
| *Financial damage* | 7 | signifikante Auswirkungen auf jährlichen Gewinn, Beschädigung des Markennamens, hochkarätige Verletzung, keine sensiblen Daten | |
| *Reputation damage* | 9 | | |
| *Non-compliance* | 7 | | |
| *Privacy Violation* | 1 | | |
| *Business Impact* | 6.00 | MEDIUM | |
| *Gesamtrisiko* | *technisch* | Low | |
| | *geschäftlich* | Medium | |

**Tabelle B.1:** Risk Rating für Bedrohung Nr. 1

| | | | |
|---|---|---|---|
| *Nummer* | 2 | | |
| *Beschreibung* | Falsche Identitäten (Fahrzeug gibt sich als viele Fahrzeuge aus.) (Hardware, z.B. mehrere OBUs oder spezielle Hardware [Simulatoren]) | | |
| *Bedrohung* | Verlust der Nachvollziehbarkeit durch Impersonierung | | |
| *Kategorie* | Impersonation-Attacks | | |
| *Angreifer* | Spezialisten im Bereich Hardwareentwicklung für Hochfrequenztechnik | | |
| *Schnittstelle* | ONP-Message-Filter | | |
| *Entwickler, Methode* | CP, Brainstorming | | |
| *Threat Agent* | *Value* | *Comment* | |
| *Skill Level* | 2 | Spezialisten, bei Verkehrsdelikten Identitäten (ver-)tauschen, vor Ort teure spezielle Hardware nötig, kleine Gruppe | |
| *Motive* | 9 | | |
| *Opportunity* | 4 | | |
| *Size* | 1 | | |
| *Vulnerability* | *Value* | *Comment* | |
| *Ease of discovery* | 7 | einfach zu entdecken, theoretisch ausnutzbar, offensichtliche Schwachstelle, keine Erkennung | |
| *Ease of exploit* | 1 | | |
| *Awareness* | 6 | | |
| *Intrusion* | 9 | | |
| *Likelihood* | 4.88 | MEDIUM | |
| *Technical Impact* | *Value* | *Comment* | |
| *Loss of Confidentiality* | 1 | keine sensiblen Daten enthüllen, wenige Daten leicht beschädigt, wenige sekundäre Dienste unterbrochen, möglicherweise nachvollziehbar | |
| *Loss of Integrity* | 1 | | |
| *Loss of Availability* | 1 | | |
| *Loss of Accountability* | 7 | | |
| *Technical Impact* | 2.50 | LOW | |
| *Business Impact* | *Value* | *Comment* | |
| *Financial damage* | 7 | signifikante Auswirkungen auf jährlichen Gewinn, Beschädigung des Markennamens, hochkarätige Verletzung, keine sensiblen Daten | |
| *Reputation damage* | 9 | | |
| *Non-compliance* | 7 | | |
| *Privacy Violation* | 1 | | |
| *Business Impact* | 6.00 | MEDIUM | |
| *Gesamtrisiko* | *technisch* | | Low |
| | *geschäftlich* | | Medium |

**Tabelle B.2:** Risk Rating für Bedrohung Nr. 2

| | | | |
|---|---|---|---|
| *Nummer* | 3 | | |
| *Beschreibung* | Falsche Identitäten Teilnehmeridentitätsvortäuschung / Teilnehmeridentitätsdiebstahl (z.B. Klonen der PHY-MAC) | | |
| *Bedrohung* | Verlust der Nachvollziehbarkeit durch Impersonierung | | |
| *Kategorie* | Impersonation-Attacks | | |
| *Angreifer* | Erfahrener Computer Nutzer mit Kenntnissen im Bereich Netzwerke | | |
| *Schnittstelle* | OTA (Over-the-Air) ONP-Receiver | | |
| *Entwickler, Methode* | CP, Brainstorming | | |
| *Threat Agent* | *Value* | *Comment* | |
| *Skill Level* | 4 | Erfahrener Computernutzer mit Kenntnissen im Bereich Netzwerke, bei Verkehrsdelikten Identitäten tauschen, einfache Soft- und Hardware nötig und vor Ort, sind nicht selten und bleiben anonym | |
| *Motive* | 9 | | |
| *Opportunity* | 7 | | |
| *Size* | 7 | | |
| *Vulnerability* | *Value* | *Comment* | |
| *Ease of discovery* | 7 | einfach zu entdecken, einfach ausnutzbar, allgemein bekannte Schwachstelle, dokumentiert aber ohne Kontrolle | |
| *Ease of exploit* | 5 | | |
| *Awareness* | 9 | | |
| *Intrusion* | 8 | | |
| *Likelihood* | 7.00 | HIGH | |
| *Technical Impact* | *Value* | *Comment* | |
| *Loss of Confidentiality* | 1 | keine sensiblen Daten enthüllen, wenige Daten leicht beschädigt, wenige sekundäre Dienste unterbrochen, möglicherweise nachvollziehbar | |
| *Loss of Integrity* | 1 | | |
| *Loss of Availability* | 1 | | |
| *Loss of Accountability* | 7 | | |
| *Technical Impact* | 2.50 | LOW | |
| *Business Impact* | *Value* | *Comment* | |
| *Financial damage* | 7 | signifikante Auswirkungen auf jährlichen Gewinn, Beschädigung des Markennamens, hochkarätige Verletzung, keine sensiblen Daten | |
| *Reputation damage* | 9 | | |
| *Non-compliance* | 7 | | |
| *Privacy Violation* | 1 | | |
| *Business Impact* | 6.00 | MEDIUM | |
| *Gesamtrisiko* | *technisch* | | Medium |
| | *geschäftlich* | | High |

**Tabelle B.3:** Risk Rating für Bedrohung Nr. 3

| | | |
|---|---|---|
| *Nummer* | 4 | |
| *Beschreibung* | Störsender im Ad-hoc-Frequenzbereich | |
| *Bedrohung* | Dienstleistungsausfall durch Blockieren von Kommuikation | |
| *Kategorie* | Denial of Service (DoS) | |
| *Angreifer* | Sicherheitsbehörden/Geheimdienste/private Sicherheitsdienste/organisierte Kriminalität/anonymer Erpresser/Hacktivist/böswilliger Nutzer | |
| *Schnittstelle* | OTA (Over-the-Air) ONP-Receiver | |
| *Entwickler, Methode* | CP, Brainstorming | |
| *Threat Agent* | *Value* | *Comment* |
| *Skill Level* | 9 | Keine Fähigkeiten; geringer Nutzen (Erpressungsversuch); Störsender vor Ort nötig; jeder, der einen Störsender bauen oder kaufen kann |
| *Motive* | 1 | |
| *Opportunity* | 8 | |
| *Size* | 9 | |
| *Vulnerability* | *Value* | *Comment* |
| *Ease of discovery* | 7 | einfach zu entdecken, einfach ausnutzbar, allgemein bekannte Schwachstelle, keine Erkennung |
| *Ease of exploit* | 5 | |
| *Awareness* | 9 | |
| *Intrusion* | 9 | |
| *Likelihood* | 7.12 | HIGH |
| *Technical Impact* | *Value* | *Comment* |
| *Loss of Confidentiality* | 1 | keine sensiblen Daten enthüllen, wenige Daten leicht beschädigt, alle Dienste vollständig unterbrochen, vollständig anonym |
| *Loss of Integrity* | 1 | |
| *Loss of Availability* | 9 | |
| *Loss of Accountability* | 9 | |
| *Technical Impact* | 5.00 | MEDIUM |
| *Business Impact* | *Value* | *Comment* |
| *Financial damage* | 1 | weniger als die Kosten zur Behebung, geringer Schaden, keine Verletzung, keine sensiblen Daten |
| *Reputation damage* | 1 | |
| *Non-compliance* | 1 | |
| *Privacy Violation* | 1 | |
| *Business Impact* | 1.00 | LOW |
| *Gesamtrisiko* | *technisch* | High |
| | *geschäftlich* | Medium |

**Tabelle B.4:** Risk Rating für Bedrohung Nr. 4

| | | |
|---|---|---|
| *Nummer* | 5 | |
| *Beschreibung* | Störsender im zellularen Frequenzbereich | |
| *Bedrohung* | Dienstleistungsausfall durch Blockieren von Kommuikation | |
| *Kategorie* | Denial of Service (DoS) | |
| *Angreifer* | Sicherheitsbehörden/Geheimdienste/private Sicherheitsdienste/organisierte Kriminalität/anonymer Erpresser/Hacktivist/böswilliger Nutzer | |
| *Schnittstelle* | OTA (Over-the-Air) ONP-Receiver | |
| *Entwickler, Methode* | CP, Brainstorming | |
| *Threat Agent* | *Value* | *Comment* |
| *Skill Level* | 9 | Keine Fähigkeiten; geringer Nutzen (Erpressungsversuch); Störsender vor Ort nötig; jeder, der einen Störsender bauen oder kaufen kann |
| *Motive* | 1 | |
| *Opportunity* | 9 | |
| *Size* | 9 | |
| *Vulnerability* | *Value* | *Comment* |
| *Ease of discovery* | 7 | einfach zu entdecken, einfach ausnutzbar, allgemein bekannte Schwachstelle, keine Erkennung |
| *Ease of exploit* | 5 | |
| *Awareness* | 9 | |
| *Intrusion* | 9 | |
| *Likelihood* | 7.25 | HIGH |
| *Technical Impact* | *Value* | *Comment* |
| *Loss of Confidentiality* | 1 | keine sensiblen Daten enthüllen, wenige Daten leicht beschädigt, alle Dienste vollständig unterbrochen, vollständig anonym |
| *Loss of Integrity* | 1 | |
| *Loss of Availability* | 9 | |
| *Loss of Accountability* | 9 | |
| *Technical Impact* | 5.00 | MEDIUM |
| *Business Impact* | *Value* | *Comment* |
| *Financial damage* | 1 | weniger als die Kosten zur Behebung, geringer Schaden, keine Verletzung, keine sensiblen Daten |
| *Reputation damage* | 1 | |
| *Non-compliance* | 1 | |
| *Privacy Violation* | 1 | |
| *Business Impact* | 1.00 | LOW |
| *Gesamtrisiko* | *technisch* | High |
| | *geschäftlich* | Medium |

**Tabelle B.5:** Risk Rating für Bedrohung Nr. 5

| | | |
|---|---|---|
| *Nummer* | 6 | |
| *Beschreibung* | Überfluten mit Datenpaketen im Ad-hoc-Netzwerk | |
| *Bedrohung* | Dienstleistungsausfall durch Verbrauchen von Ressourcen | |
| *Kategorie* | Denial of Service (DoS) | |
| *Angreifer* | Sicherheitsbehörden/Geheimdienste/private Sicherheitsdienste/organisierte Kriminalität/anonymer Erpresser/Hacktivist/böswilliger Nutzer | |
| *Schnittstelle* | OTA (Over-the-Air) ONP-Receiver | |
| *Entwickler, Methode* | CP, Brainstorming | |
| *Threat Agent* | *Value* | *Comment* |
| *Skill Level* | 9 | Keine Fähigkeiten, geringer Nutzen (Erpressungsversuch), Antenne vor Ort nötig und geeignete Software (Anpassung von existierender) |
| *Motive* | 1 | |
| *Opportunity* | 6 | |
| *Size* | 9 | |
| *Vulnerability* | *Value* | *Comment* |
| *Ease of discovery* | 7 | einfach zu entdecken, einfach ausnutzbar, allgemein bekannte Schwachstelle, keine Erkennung |
| *Ease of exploit* | 5 | |
| *Awareness* | 9 | |
| *Intrusion* | 9 | |
| *Likelihood* | 6.88 | HIGH |
| *Technical Impact* | *Value* | *Comment* |
| *Loss of Confidentiality* | 1 | keine sensiblen Daten enthüllen, wenige Daten leicht beschädigt, alle Dienste vollständig unterbrochen, vollständig anonym |
| *Loss of Integrity* | 1 | |
| *Loss of Availability* | 9 | |
| *Loss of Accountability* | 9 | |
| *Technical Impact* | 5.00 | MEDIUM |
| *Business Impact* | *Value* | *Comment* |
| *Financial damage* | 5 | mittlere Auswirkungen auf jährlichen Gewinn, Beschädigung des Markennamens, keine Verletzung, keine sensiblen Daten |
| *Reputation damage* | 9 | |
| *Non-compliance* | 1 | |
| *Privacy Violation* | 1 | |
| *Business Impact* | 4.00 | MEDIUM |
| *Gesamtrisiko* | *technisch*<br>*geschäftlich* | High<br>High |

**Tabelle B.6:** Risk Rating für Bedrohung Nr. 6

| | | | |
|---|---|---|---|
| *Nummer* | 7 | | |
| *Beschreibung* | Überfluten mit Datenpaketen im zellularen Netzwerk | | |
| *Bedrohung* | Dienstleistungsausfall durch Verbrauchen von Ressourcen | | |
| *Kategorie* | Denial of Service (DoS) | | |
| *Angreifer* | Sicherheitsbehörden/Geheimdienste/private Sicherheitsdienste/organisierte Kriminalität/anonymer Erpresser/Hacktivist/böswilliger Nutzer | | |
| *Schnittstelle* | OTA (Over-the-Air) ONP-Receiver | | |
| *Entwickler, Methode* | CP, Brainstorming | | |
| *Threat Agent* | *Value* | *Comment* | |
| *Skill Level* | 9 | Keine Fähigkeiten, geringer Nutzen (Erpressungsversuch), Antenne vor Ort nötig und geeignete Software (Anpassung von existierender) | |
| *Motive* | 1 | | |
| *Opportunity* | 7 | | |
| *Size* | 9 | | |
| *Vulnerability* | *Value* | *Comment* | |
| *Ease of discovery* | 7 | einfach zu entdecken, einfach ausnutzbar, allgemein bekannte Schwachstelle, keine Erkennung | |
| *Ease of exploit* | 5 | | |
| *Awareness* | 9 | | |
| *Intrusion* | 9 | | |
| *Likelihood* | 7.00 | HIGH | |
| *Technical Impact* | *Value* | *Comment* | |
| *Loss of Confidentiality* | 1 | keine sensiblen Daten enthüllen, wenige Daten leicht beschädigt, alle Dienste vollständig unterbrochen, vollständig anonym | |
| *Loss of Integrity* | 1 | | |
| *Loss of Availability* | 9 | | |
| *Loss of Accountability* | 9 | | |
| *Technical Impact* | 5.00 | MEDIUM | |
| *Business Impact* | *Value* | *Comment* | |
| *Financial damage* | 5 | mittlere Auswirkungen auf jährlichen Gewinn, Beschädigung des Markennamens, keine Verletzung, keine sensiblen Daten | |
| *Reputation damage* | 9 | | |
| *Non-compliance* | 1 | | |
| *Privacy Violation* | 1 | | |
| *Business Impact* | 4.00 | MEDIUM | |
| *Gesamtrisiko* | *technisch* | | High |
| | *geschäftlich* | | High |

**Tabelle B.7:** Risk Rating für Bedrohung Nr. 7

| | | |
|---|---|---|
| *Nummer* | 8 | |
| *Beschreibung* | Gezieltes Unterdrücken (durch Störung von ad-hoc) von einzelnen Informationen | |
| *Bedrohung* | Dienstleistungsausfall durch Blockieren von Kommuikation | |
| *Kategorie* | Man on the Side (MotS) | |
| *Angreifer* | IT-Security-Spezialisten im Bereich Ad-hoc-Kommunikation | |
| *Schnittstelle* | OTA (Over-the-Air) ONP-Receiver | |
| *Entwickler, Methode* | CP, Brainstorming | |
| *Threat Agent* | *Value* | *Comment* |
| *Skill Level* | 2 | Spezialisten im Bereich Ad-hoc-Kommunikation, Vorbereitung für weitere Angriffe (gezielte Angriffe), Antenne vor Ort und spezielle Hard- o. Software nötig |
| *Motive* | 2 | |
| *Opportunity* | 1 | |
| *Size* | 2 | |
| *Vulnerability* | *Value* | *Comment* |
| *Ease of discovery* | 5 | mäßig schwierig zu entdecken, schwer ausnutzbar, versteckte Schwachstelle, dokumentiert aber ohne Kontrolle |
| *Ease of exploit* | 3 | |
| *Awareness* | 4 | |
| *Intrusion* | 8 | |
| *Likelihood* | 3.38 | MEDIUM |
| *Technical Impact* | *Value* | *Comment* |
| *Loss of Confidentiality* | 1 | keine sensiblen Daten enthüllen, alle Daten vollständig beschädigt, wenige sekundäre Dienste unterbrochen, vollständig anonym |
| *Loss of Integrity* | 9 | |
| *Loss of Availability* | 1 | |
| *Loss of Accountability* | 9 | |
| *Technical Impact* | 5.00 | MEDIUM |
| *Business Impact* | *Value* | *Comment* |
| *Financial damage* | 3 | geringe Auswirkungen auf jährlichen Gewinn, Verlust von Kunden mittlerer Größe, geringe Verletzung, keine sensiblen Daten |
| *Reputation damage* | 3 | |
| *Non-compliance* | 2 | |
| *Privacy Violation* | 1 | |
| *Business Impact* | 2.25 | LOW |
| *Gesamtrisiko* | *technisch* | Medium |
| | *geschäftlich* | Low |

**Tabelle B.8:** Risk Rating für Bedrohung Nr. 8

| | | | |
|---|---|---|---|
| *Nummer* | 9 | | |
| *Beschreibung* | Gezieltes Unterdrücken (durch Störung von zellular) von einzelnen Informationen | | |
| *Bedrohung* | Dienstleistungsausfall durch Blockieren von Kommuikation | | |
| *Kategorie* | Man on the Side (MotS) | | |
| *Angreifer* | IT-Security-Spezialisten im Bereich zellularer Kommunikation | | |
| *Schnittstelle* | OTA (Over-the-Air) ONP-Receiver | | |
| *Entwickler, Methode* | CP, Brainstorming | | |
| *Threat Agent* | *Value* | *Comment* | |
| *Skill Level* | 2 | Spezialisten im Bereich zellularer Kommunikation, Vorbereitung für weitere Angriffe (gezielte Angriffe), Antenne vor Ort und spezielle Hard- o. Software nötig | |
| *Motive* | 2 | | |
| *Opportunity* | 2 | | |
| *Size* | 2 | | |
| *Vulnerability* | *Value* | *Comment* | |
| *Ease of discovery* | 3 | schwer zu entdecken, schwer ausnutzbar, versteckte Schwachstelle, dokumentiert aber ohne Kontrolle | |
| *Ease of exploit* | 3 | | |
| *Awareness* | 4 | | |
| *Intrusion* | 8 | | |
| *Likelihood* | 3.25 | MEDIUM | |
| *Technical Impact* | *Value* | *Comment* | |
| *Loss of Confidentiality* | 1 | keine sensiblen Daten enthüllen, alle Daten vollständig beschädigt, wenige sekundäre Dienste unterbrochen, vollständig anonym | |
| *Loss of Integrity* | 9 | | |
| *Loss of Availability* | 1 | | |
| *Loss of Accountability* | 9 | | |
| *Technical Impact* | 5.00 | MEDIUM | |
| *Business Impact* | *Value* | *Comment* | |
| *Financial damage* | 3 | geringe Auswirkungen auf jährlichen Gewinn, Verlust von Kunden mittlerer Größe, geringe Verletzung, keine sensiblen Daten | |
| *Reputation damage* | 3 | | |
| *Non-compliance* | 2 | | |
| *Privacy Violation* | 1 | | |
| *Business Impact* | 2.25 | LOW | |
| *Gesamtrisiko* | *technisch* | | Medium |
| | *geschäftlich* | | Low |

**Tabelle B.9:** Risk Rating für Bedrohung Nr. 9

| | | | |
|---|---|---|---|
| *Nummer* | 10 | | |
| *Beschreibung* | Mitlesen der Kommunikation (ad-hoc) | | |
| *Bedrohung* | Verlust von vertraulichen Daten | | |
| *Kategorie* | Man on the Side (MotS) | | |
| *Angreifer* | Erfahrener Computernutzer mit Kenntnissen im Bereich Netzwerke | | |
| *Schnittstelle* | OTA (Over-the-Air) ONP-Receiver | | |
| *Entwickler, Methode* | CP, Brainstorming | | |
| *Threat Agent* | *Value* | *Comment* | |
| *Skill Level* | 4 | Erfahrener Computernutzer mit Kenntnissen im Bereich Netzwerke, Stehlen von privaten Daten und der Verkauf dieser, Antenne vor Ort | |
| *Motive* | 7 | | |
| *Opportunity* | 8 | | |
| *Size* | 7 | | |
| *Vulnerability* | *Value* | *Comment* | |
| *Ease of discovery* | 7 | einfach zu entdecken, einfach ausnutzbar, allgemein bekannte Schwachstelle, keine Erkennung | |
| *Ease of exploit* | 5 | | |
| *Awareness* | 9 | | |
| *Intrusion* | 9 | | |
| *Likelihood* | 7.00 | HIGH | |
| *Technical Impact* | *Value* | *Comment* | |
| *Loss of Confidentiality* | 9 | alle Daten enthüllen, wenige Daten leicht beschädigt, wenige sekundäre Dienste unterbrochen, vollständig anonym | |
| *Loss of Integrity* | 1 | | |
| *Loss of Availability* | 1 | | |
| *Loss of Accountability* | 9 | | |
| *Technical Impact* | 5.00 | MEDIUM | |
| *Business Impact* | *Value* | *Comment* | |
| *Financial damage* | 7 | signifikante Auswirkungen auf jährlichen Gewinn, Unterstellung von böswilligem Handeln, geringe Verletzung, millionen von betroffenen Personen | |
| *Reputation damage* | 5 | | |
| *Non-compliance* | 2 | | |
| *Privacy Violation* | 9 | | |
| *Business Impact* | 5.75 | MEDIUM | |
| *Gesamtrisiko* | *technisch* | | High |
| | *geschäftlich* | | High |

**Tabelle B.10:** Risk Rating für Bedrohung Nr. 10

| | | | |
|---|---|---|---|
| *Nummer* | 11 | | |
| *Beschreibung* | Mitlesen der Kommunikation (zellular) | | |
| *Bedrohung* | Verlust von vertraulichen Daten | | |
| *Kategorie* | Man on the Side (MotS) | | |
| *Angreifer* | Erfahrener Computernutzer mit Kenntnissen im Bereich Netzwerke | | |
| *Schnittstelle* | OTA (Over-the-Air) ONP-Receiver | | |
| *Entwickler, Methode* | CP, Brainstorming | | |
| *Threat Agent* | *Value* | *Comment* | |
| *Skill Level* | 4 | Erfahrener Computernutzer mit Kenntnissen im Bereich Netzwerke, Handel mit privaten Daten, Antenne vor Ort, je nach eingesetztem zellularen Standard muss die Verschlüsselung gebrochen werden, z.B. LTE | |
| *Motive* | 7 | | |
| *Opportunity* | 3 | | |
| *Size* | 7 | | |
| *Vulnerability* | *Value* | *Comment* | |
| *Ease of discovery* | 7 | einfach zu entdecken, theoretisch ausnutzbar, allgemein bekannte Schwachstelle, keine Erkennung | |
| *Ease of exploit* | 1 | | |
| *Awareness* | 9 | | |
| *Intrusion* | 9 | | |
| *Likelihood* | 5.88 | MEDIUM | |
| *Technical Impact* | *Value* | *Comment* | |
| *Loss of Confidentiality* | 9 | alle Daten enthüllen, wenige Daten leicht beschädigt, wenige sekundäre Dienste unterbrochen, vollständig anonym | |
| *Loss of Integrity* | 1 | | |
| *Loss of Availability* | 1 | | |
| *Loss of Accountability* | 9 | | |
| *Technical Impact* | 5.00 | MEDIUM | |
| *Business Impact* | *Value* | *Comment* | |
| *Financial damage* | 7 | signifikante Auswirkungen auf jährlichen Gewinn, Unterstellung von böswilligem Handeln, geringe Verletzung, millionen von betroffenen Personen | |
| *Reputation damage* | 5 | | |
| *Non-compliance* | 2 | | |
| *Privacy Violation* | 9 | | |
| *Business Impact* | 5.75 | MEDIUM | |
| *Gesamtrisiko* | *technisch* | | Medium |
| | *geschäftlich* | | Medium |

**Tabelle B.11:** Risk Rating für Bedrohung Nr. 11

| | | |
|---|---|---|
| *Nummer* | 12 | |
| *Beschreibung* | Mitlesen der Kommunikation (backbone) | |
| *Bedrohung* | Verlust von vertraulichen Daten | |
| *Kategorie* | Man on the Side (MotS) | |
| *Angreifer* | Sicherheitsbehörden/Geheimdienste/private Sicherheitsdienste/organisierte Kriminalität (LI=Lawful Intercept) | |
| *Schnittstelle* | OTA (Over-the-Air) ONP-Receiver | |
| *Entwickler, Methode* | CP, Brainstorming | |
| *Threat Agent* | *Value* | *Comment* |
| *Skill Level* | 4 | Spezialisten im Bereich Netzwerkkommunikation, Handel mit privaten Daten, Angreifer muss an privilegierter Stelle im Netzwerk sitzen (Router) und spezielle Software nötig, allerdings macht es LI einfacher |
| *Motive* | 7 | |
| *Opportunity* | 4 | |
| *Size* | 7 | |
| *Vulnerability* | *Value* | *Comment* |
| *Ease of discovery* | 9 | automatisierte Tools verfügbar, einfach ausnutzbar, allgemein bekannte Schwachstelle, keine Erkennung |
| *Ease of exploit* | 5 | |
| *Awareness* | 9 | |
| *Intrusion* | 9 | |
| *Likelihood* | 6.75 | HIGH |
| *Technical Impact* | *Value* | *Comment* |
| *Loss of Confidentiality* | 9 | alle Daten enthüllen, wenige Daten leicht beschädigt, wenige sekundäre Dienste unterbrochen, vollständig anonym |
| *Loss of Integrity* | 1 | |
| *Loss of Availability* | 1 | |
| *Loss of Accountability* | 9 | |
| *Technical Impact* | 5.00 | MEDIUM |
| *Business Impact* | *Value* | *Comment* |
| *Financial damage* | 7 | signifikante Auswirkungen auf jährlichen Gewinn, Unterstellung von böswilligem Handeln, geringe Verletzung, millionen von betroffenen Personen |
| *Reputation damage* | 5 | |
| *Non-compliance* | 2 | |
| *Privacy Violation* | 9 | |
| *Business Impact* | 5.75 | MEDIUM |
| *Gesamtrisiko* | *technisch*<br>*geschäftlich* | High<br>High |

**Tabelle B.12:** Risk Rating für Bedrohung Nr. 12

| | | | |
|---|---|---|---|
| *Nummer* | 13 | | |
| *Beschreibung* | Einfügen/Verändern/Verwerfen/Mitlesen von Datenpaketen in einer Kommunikation (ad-hoc) | | |
| *Bedrohung* | Verlust der Vertraulichkeit, Integrität, Verfügbarkeit und Authentizität | | |
| *Kategorie* | Man in the Middle (MitM) | | |
| *Angreifer* | IT-Security-Spezialisten im Bereich Ad-hoc-Kommunikation | | |
| *Schnittstelle* | OTA (Over-the-Air) ONP-Receiver | | |
| *Entwickler, Methode* | CP, Brainstorming | | |
| *Threat Agent* | *Value* | *Comment* | |
| *Skill Level* | 2 | Spezialisten im Bereich Ad-hoc-Kommunikation, Vorbereitung für weitere Angriffe (gezielte Angriffe), Antenne vor Ort und spezielle Hard- o. Software nötig | |
| *Motive* | 4 | | |
| *Opportunity* | 6 | | |
| *Size* | 2 | | |
| *Vulnerability* | *Value* | *Comment* | |
| *Ease of discovery* | 3 | schwer zu entdecken, schwer ausnutzbar, versteckte Schwachstelle, keine Erkennung | |
| *Ease of exploit* | 3 | | |
| *Awareness* | 4 | | |
| *Intrusion* | 9 | | |
| *Likelihood* | 4.12 | MEDIUM | |
| *Technical Impact* | *Value* | *Comment* | |
| *Loss of Confidentiality* | 9 | alle Daten enthüllen, alle Daten vollständig beschädigt, wenige sekundäre Dienste unterbrochen, vollständig anonym | |
| *Loss of Integrity* | 9 | | |
| *Loss of Availability* | 1 | | |
| *Loss of Accountability* | 9 | | |
| *Technical Impact* | 7.00 | HIGH | |
| *Business Impact* | *Value* | *Comment* | |
| *Financial damage* | 7 | signifikante Auswirkungen auf jährlichen Gewinn, Beschädigung des Markennamens, geringe Verletzung, millionen von betroffenen Personen | |
| *Reputation damage* | 9 | | |
| *Non-compliance* | 2 | | |
| *Privacy Violation* | 9 | | |
| *Business Impact* | 6.75 | HIGH | |
| *Gesamtrisiko* | *technisch* | | High |
| | *geschäftlich* | | High |

**Tabelle B.13:** Risk Rating für Bedrohung Nr. 13

| | | |
|---|---|---|
| *Nummer* | 14 | |
| *Beschreibung* | Einfügen/Verändern/Verwerfen/Mitlesen von Datenpaketen in einer Kommunikation (zellular) | |
| *Bedrohung* | Verlust der Vertraulichkeit, Integrität, Verfügbarkeit und Authentizität | |
| *Kategorie* | Man in the Middle (MitM) | |
| *Angreifer* | IT-Security-Spezialisten im Bereich zellularer Kommunikation | |
| *Schnittstelle* | OTA (Over-the-Air) ONP-Receiver | |
| *Entwickler, Methode* | CP, Brainstorming | |
| *Threat Agent* | *Value* | *Comment* |
| *Skill Level* | 2 | Spezialisten im Bereich zellularer Kommunikation, Vorbereitung für weitere Angriffe (gezielte Angriffe), Antenne vor Ort und spezielle Hard- o. Software nötig |
| *Motive* | 4 | |
| *Opportunity* | 7 | |
| *Size* | 2 | |
| *Vulnerability* | *Value* | *Comment* |
| *Ease of discovery* | 3 | schwer zu entdecken, schwer ausnutzbar, versteckte Schwachstelle, keine Erkennung |
| *Ease of exploit* | 3 | |
| *Awareness* | 4 | |
| *Intrusion* | 9 | |
| *Likelihood* | 4.25 | MEDIUM |
| *Technical Impact* | *Value* | *Comment* |
| *Loss of Confidentiality* | 9 | alle Daten enthüllen, alle Daten vollständig beschädigt, wenige sekundäre Dienste unterbrochen, vollständig anonym |
| *Loss of Integrity* | 9 | |
| *Loss of Availability* | 1 | |
| *Loss of Accountability* | 9 | |
| *Technical Impact* | 7.00 | HIGH |
| *Business Impact* | *Value* | *Comment* |
| *Financial damage* | 7 | signifikante Auswirkungen auf jährlichen Gewinn, Beschädigung des Markennamens, geringe Verletzung, millionen von betroffenen Personen |
| *Reputation damage* | 9 | |
| *Non-compliance* | 2 | |
| *Privacy Violation* | 9 | |
| *Business Impact* | 6.75 | HIGH |
| *Gesamtrisiko* | *technisch* | High |
| | *geschäftlich* | High |

**Tabelle B.14:** Risk Rating für Bedrohung Nr. 14

| | | | |
|---|---|---|---|
| *Nummer* | 15 | | |
| *Beschreibung* | Einfügen/Verändern/Verwerfen/Mitlesen von Datenpaketen in einer Kommunikation (backbone) | | |
| *Bedrohung* | Verlust der Vertraulichkeit, Integrität, Verfügbarkeit und Authentizität | | |
| *Kategorie* | Man in the Middle (MitM) | | |
| *Angreifer* | Sicherheitsbehörden/Geheimdienste/private Sicherheitsdienste/organisierte Kriminalität | | |
| *Schnittstelle* | OTA (Over-the-Air) ONP-Receiver | | |
| *Entwickler, Methode* | CP, Brainstorming | | |
| *Threat Agent* | *Value* | *Comment* | |
| *Skill Level* | 2 | Spezialisten im Bereich Netzwerkkommunikation, Vorbereitung für weitere Angriffe (gezielte Angriffe), Angreifer muss an privilegierter Stelle im Netzwerk sitzen und spezielle Software nötig | |
| *Motive* | 4 | | |
| *Opportunity* | 4 | | |
| *Size* | 7 | | |
| *Vulnerability* | *Value* | *Comment* | |
| *Ease of discovery* | 3 | schwer zu entdecken, schwer ausnutzbar, versteckte Schwachstelle, keine Erkennung | |
| *Ease of exploit* | 3 | | |
| *Awareness* | 4 | | |
| *Intrusion* | 9 | | |
| *Likelihood* | 4.50 | MEDIUM | |
| *Technical Impact* | *Value* | *Comment* | |
| *Loss of Confidentiality* | 9 | alle Daten enthüllen, alle Daten vollständig beschädigt, wenige sekundäre Dienste unterbrochen, vollständig anonym | |
| *Loss of Integrity* | 9 | | |
| *Loss of Availability* | 1 | | |
| *Loss of Accountability* | 9 | | |
| *Technical Impact* | 7.00 | HIGH | |
| *Business Impact* | *Value* | *Comment* | |
| *Financial damage* | 7 | signifikante Auswirkungen auf jährlichen Gewinn, Beschädigung des Markennamens, geringe Verletzung, millionen von betroffenen Personen | |
| *Reputation damage* | 9 | | |
| *Non-compliance* | 2 | | |
| *Privacy Violation* | 9 | | |
| *Business Impact* | 6.75 | HIGH | |
| *Gesamtrisiko* | *technisch* | | High |
| | *geschäftlich* | | High |

**Tabelle B.15:** Risk Rating für Bedrohung Nr. 15

| | | | |
|---|---|---|---|
| *Nummer* | 16 | | |
| *Beschreibung* | Kompromittierung eines verbauten Coprocessors (z.B. Modem), um Hauptsystem (OS) oder Systemkomponenten (ONP) zu beeinflussen (z.B. Schreiben in gemeinsamen Speicherbereich o.Ä.) | | |
| *Bedrohung* | Verlust der Vertraulichkeit, Integrität, Verfügbarkeit und Authentizität | | |
| *Kategorie* | Hardware based Attacks | | |
| *Angreifer* | Schadsoftwareautor vergleichbar mit IT-Security-Spezialisten im Bereich eingebetteter Systeme, Hardware nahe und Betriebssystemprogrammierung | | |
| *Schnittstelle* | Alle innerhalb des ONP fähigen Gerätes | | |
| *Entwickler, Methode* | CP, Brainstorming | | |
| *Threat Agent* | *Value* | *Comment* | |
| *Skill Level* | 2 | Spezialisten im Bereich eingebetteter Systeme; Betriebssystemprogrammierung; hohe Persistenz, da unberührt durch Software Updates; Vorbereitung für weitere Angriffe; detailliertes Wissen über Coprozessoren nötig | |
| *Motive* | 7 | | |
| *Opportunity* | 3 | | |
| *Size* | 2 | | |
| *Vulnerability* | *Value* | *Comment* | |
| *Ease of discovery* | 3 | schwer zu entdecken, schwer ausnutzbar, versteckte Schwachstelle, keine Erkennung | |
| *Ease of exploit* | 3 | | |
| *Awareness* | 4 | | |
| *Intrusion* | 9 | | |
| *Likelihood* | 4.12 | MEDIUM | |
| *Technical Impact* | *Value* | *Comment* | |
| *Loss of Confidentiality* | 9 | alle Daten enthüllen, alle Daten vollständig beschädigt, alle Dienste vollständig unterbrochen, vollständig anonym | |
| *Loss of Integrity* | 9 | | |
| *Loss of Availability* | 9 | | |
| *Loss of Accountability* | 9 | | |
| *Technical Impact* | 9.00 | HIGH | |
| *Business Impact* | *Value* | *Comment* | |
| *Financial damage* | 1 | weniger als die Kosten zur Behebung, geringer Schaden, geringe Verletzung, hunderttausende von betroffenen Personen | |
| *Reputation damage* | 1 | | |
| *Non-compliance* | 2 | | |
| *Privacy Violation* | 8 | | |
| *Business Impact* | 3.00 | LOW | |
| *Gesamtrisiko* | *technisch* | | High |
| | *geschäftlich* | | Low |

**Tabelle B.16:** Risk Rating für Bedrohung Nr. 16

| | | | |
|---|---|---|---|
| *Nummer* | 17 | | |
| *Beschreibung* | Kompromittierung des Betriebssystems durch publizierte Schwachstellen | | |
| *Bedrohung* | Verlust der Vertraulichkeit, Integrität, Verfügbarkeit und Authentizität | | |
| *Kategorie* | privilegierte Remote Code Execution (RCE) | | |
| *Angreifer* | Schadsoftwarenutzer mit geringen technischen Fähigkeiten | | |
| *Schnittstelle* | Alle innerhalb des ONP fähigen Gerätes | | |
| *Entwickler, Methode* | CP, Brainstorming | | |
| *Threat Agent* | *Value* | *Comment* | |
| *Skill Level* | 6 | Schadsoftwarenutzer mit technischen Fähigkeiten, Erbeuten von privaten Daten zum Verkauf dieser, Nutzung der Rechenleistung oder kostenpflichtigen Diensten, viele | |
| *Motive* | 9 | | |
| *Opportunity* | 9 | | |
| *Size* | 7 | | |
| *Vulnerability* | *Value* | *Comment* | |
| *Ease of discovery* | 9 | automatisierte Tools verfügbar, vollautomatische Tools verfügbar, allgemein bekannte Schwachstelle, keine Erkennung | |
| *Ease of exploit* | 9 | | |
| *Awareness* | 9 | | |
| *Intrusion* | 9 | | |
| *Likelihood* | 8.38 | HIGH | |
| *Technical Impact* | *Value* | *Comment* | |
| *Loss of Confidentiality* | 9 | alle Daten enthüllen, alle Daten vollständig beschädigt, alle Dienste vollständig unterbrochen, vollständig anonym | |
| *Loss of Integrity* | 9 | | |
| *Loss of Availability* | 9 | | |
| *Loss of Accountability* | 9 | | |
| *Technical Impact* | 9.00 | HIGH | |
| *Business Impact* | *Value* | *Comment* | |
| *Financial damage* | 7 | signifikante Auswirkungen auf jährlichen Gewinn, Verlust von größeren Kunden, eindeutige Verletzung, millionen von betroffenen Personen | |
| *Reputation damage* | 4 | | |
| *Non-compliance* | 5 | | |
| *Privacy Violation* | 9 | | |
| *Business Impact* | 6.25 | HIGH | |
| *Gesamtrisiko* | *technisch* | | Critical |
| | *geschäftlich* | | Critical |

**Tabelle B.17:** Risk Rating für Bedrohung Nr. 17

| | | |
|---|---|---|
| *Nummer* | 18 | |
| *Beschreibung* | Kompromittierung des Betriebssystems durch unpulizierte Schwachstellen | |
| *Bedrohung* | Verlust der Vertraulichkeit, Integrität, Verfügbarkeit und Authentizität | |
| *Kategorie* | privilegierte Remote Code Execution (RCE) | |
| *Angreifer* | Schadsoftwareautor vergleichbar mit IT-Security-Spezialisten im Bereich Betriebssystemprogrammierung | |
| *Schnittstelle* | Alle innerhalb des ONP fähigen Gerätes | |
| *Entwickler, Methode* | CP, Brainstorming | |
| *Threat Agent* | *Value* | *Comment* |
| *Skill Level* | 2 | Schadsoftwareautor vgl. IT-Security-Spezialisten im Bereich Betriebssystemprogrammierung, finanzieller Wert der Schwachstelle, Zugang zu Hard- und Software der Zielplattform nötig, wenige |
| *Motive* | 9 | |
| *Opportunity* | 7 | |
| *Size* | 2 | |
| *Vulnerability* | *Value* | *Comment* |
| *Ease of discovery* | 3 | schwer zu entdecken, schwer ausnutzbar, unbekannte Schwachstelle, keine Erkennung |
| *Ease of exploit* | 3 | |
| *Awareness* | 1 | |
| *Intrusion* | 9 | |
| *Likelihood* | 4.50 | MEDIUM |
| *Technical Impact* | *Value* | *Comment* |
| *Loss of Confidentiality* | 9 | alle Daten enthüllen, alle Daten vollständig beschädigt, alle Dienste vollständig unterbrochen, vollständig anonym |
| *Loss of Integrity* | 9 | |
| *Loss of Availability* | 9 | |
| *Loss of Accountability* | 9 | |
| *Technical Impact* | 9.00 | HIGH |
| *Business Impact* | *Value* | *Comment* |
| *Financial damage* | 7 | signifikante Auswirkungen auf jährlichen Gewinn, Verlust von größeren Kunden, eindeutige Verletzung, millionen von betroffenen Personen |
| *Reputation damage* | 4 | |
| *Non-compliance* | 5 | |
| *Privacy Violation* | 9 | |
| *Business Impact* | 6.25 | HIGH |
| *Gesamtrisiko* | *technisch* | High |
| | *geschäftlich* | High |

**Tabelle B.18:** Risk Rating für Bedrohung Nr. 18

| | | |
|---|---|---|
| *Nummer* | 19 | |
| *Beschreibung* | DNS-Hijacking, ONP dazu bringen einen schädlichen AppStore zu verwenden | |
| *Bedrohung* | Verlust der Integrität durch nicht vertrauenswürdige Anwendungen | |
| *Kategorie* | unprivilegierte Remote Code Execution (RCE) | |
| *Angreifer* | Schadsoftwareautor vergleichbar mit IT-Security-Spezialisten im Bereich Webservice/Webapplikations-Entwicklung | |
| *Schnittstelle* | AS-WS2OR | |
| *Entwickler, Methode* | CP, Brainstorming | |
| *Threat Agent* | *Value* | *Comment* |
| *Skill Level* | 2 | Spezialisten im Bereich Netzwerkkommunikation, Kompromittierung einer kleinen Gruppe/einzelner Teilnehmer, geeignete Position im Netzwerk erforderlich, wenige |
| *Motive* | 7 | |
| *Opportunity* | 4 | |
| *Size* | 2 | |
| *Vulnerability* | *Value* | *Comment* |
| *Ease of discovery* | 5 | mäßig schwierig zu entdecken, mittelschwer ausnutzbar, offensichtliche Schwachstelle, dokumentiert aber ohne Kontrolle |
| *Ease of exploit* | 4 | |
| *Awareness* | 6 | |
| *Intrusion* | 8 | |
| *Likelihood* | 4.75 | MEDIUM |
| *Technical Impact* | *Value* | *Comment* |
| *Loss of Confidentiality* | 7 | umfassend kritische Daten enthüllen, viele Daten stark beschädigt, umfassend sekundäre Dienste unterbrochen, möglicherweise nachvollziehbar |
| *Loss of Integrity* | 7 | |
| *Loss of Availability* | 5 | |
| *Loss of Accountability* | 7 | |
| *Technical Impact* | 6.50 | HIGH |
| *Business Impact* | *Value* | *Comment* |
| *Financial damage* | 3 | geringe Auswirkungen auf jährlichen Gewinn, Beschädigung des Markennamens, eindeutige Verletzung, millionen von betroffenen Personen |
| *Reputation damage* | 9 | |
| *Non-compliance* | 5 | |
| *Privacy Violation* | 9 | |
| *Business Impact* | 6.50 | HIGH |
| *Gesamtrisiko* | *technisch*<br>*geschäftlich* | High<br>High |

**Tabelle B.19:** Risk Rating für Bedrohung Nr. 19

| | | |
|---|---|---|
| *Nummer* | 20 | |
| *Beschreibung* | Ausnutzen von Schwachstellen im Anwendungscode von ONP (Schädlicher/falscher AppStore) | |
| *Bedrohung* | Verlust der Integrität durch die Ausführung von nicht vertrauenswürdigem Code, Verlust der Verfügbarkeit durch Dientsausfall | |
| *Kategorie* | privilegierte Remote Code Execution (RCE) | |
| *Angreifer* | Schadsoftwareautor vergleichbar mit IT-Security-Spezialisten im Bereich Webservice/Webapplikations-Entwicklung | |
| *Schnittstelle* | AS-WS2OR | |
| *Entwickler, Methode* | CP, Brainstorming | |
| *Threat Agent* | *Value* | *Comment* |
| *Skill Level* | 2 | Spezialisten im Bereich Webservice/Webapplikations-Entwicklung, Kompromittierung einer großen Nutzergemeinde, kompromittieren eines AppStores oder betreiben eines eigenen nötig, wenige |
| *Motive* | 7 | |
| *Opportunity* | 4 | |
| *Size* | 2 | |
| *Vulnerability* | *Value* | *Comment* |
| *Ease of discovery* | 5 | mäßig schwierig zu entdecken, mittelschwer ausnutzbar, offensichtliche Schwachstelle, dokumentiert aber ohne Kontrolle |
| *Ease of exploit* | 4 | |
| *Awareness* | 6 | |
| *Intrusion* | 8 | |
| *Likelihood* | 4.75 | MEDIUM |
| *Technical Impact* | *Value* | *Comment* |
| *Loss of Confidentiality* | 7 | umfassend kritische Daten enthüllen, viele Daten stark beschädigt, umfassend sekundäre Dienste unterbrochen, möglicherweise nachvollziehbar |
| *Loss of Integrity* | 7 | |
| *Loss of Availability* | 5 | |
| *Loss of Accountability* | 7 | |
| *Technical Impact* | 6.50 | HIGH |
| *Business Impact* | *Value* | *Comment* |
| *Financial damage* | 3 | geringe Auswirkungen auf jährlichen Gewinn, Beschädigung des Markennamens, eindeutige Verletzung, millionen von betroffenen Personen |
| *Reputation damage* | 9 | |
| *Non-compliance* | 5 | |
| *Privacy Violation* | 9 | |
| *Business Impact* | 6.50 | HIGH |
| *Gesamtrisiko* | *technisch* | High |
| | *geschäftlich* | High |

**Tabelle B.20:** Risk Rating für Bedrohung Nr. 20

| | | | |
|---|---|---|---|
| *Nummer* | 21 | | |
| *Beschreibung* | Ausnutzen von Schwachstellen im Anwendungscode von ONP (Schädlicher/falscher Service) | | |
| *Bedrohung* | Verlust der Integrität durch die Ausführung von nicht vertrauenswürdigem Code, Verlust der Verfügbarkeit durch Dienstausfall | | |
| *Kategorie* | privilegierte Remote Code Execution (RCE) | | |
| *Angreifer* | Schadsoftwareautor vergleichbar mit IT-Security-Spezialisten im Bereich Webservice/Webapplikations Entwicklung | | |
| *Schnittstelle* | SP-WS2OR | | |
| *Entwickler, Methode* | CP, Brainstorming | | |
| *Threat Agent* | *Value* | *Comment* | |
| *Skill Level* | 2 | Spezialisten im Bereich Webservice/Webapplikations-Entwicklung, Kompromittierung einer großen Nutzergemeinde, kompromittieren eines Diensteanbieters oder betreiben eines eigenen nötig, wenige | |
| *Motive* | 6 | | |
| *Opportunity* | 5 | | |
| *Size* | 2 | | |
| *Vulnerability* | *Value* | *Comment* | |
| *Ease of discovery* | 5 | mäßig schwierig zu entdecken, mittelschwer ausnutzbar, offensichtliche Schwachstelle, dokumentiert aber ohne Kontrolle | |
| *Ease of exploit* | 4 | | |
| *Awareness* | 6 | | |
| *Intrusion* | 8 | | |
| *Likelihood* | 4.75 | MEDIUM | |
| *Technical Impact* | *Value* | *Comment* | |
| *Loss of Confidentiality* | 7 | umfassend kritische Daten enthüllen, viele Daten stark beschädigt, umfassend sekundäre Dienste unterbrochen, möglicherweise nachvollziehbar | |
| *Loss of Integrity* | 7 | | |
| *Loss of Availability* | 5 | | |
| *Loss of Accountability* | 7 | | |
| *Technical Impact* | 6.50 | HIGH | |
| *Business Impact* | *Value* | *Comment* | |
| *Financial damage* | 3 | geringe Auswirkungen auf jährlichen Gewinn, Beschädigung des Markennamens, eindeutige Verletzung, millionen von betroffenen Personen | |
| *Reputation damage* | 9 | | |
| *Non-compliance* | 5 | | |
| *Privacy Violation* | 9 | | |
| *Business Impact* | 6.50 | HIGH | |
| *Gesamtrisiko* | *technisch* | High | |
| | *geschäftlich* | High | |

**Tabelle B.21:** Risk Rating für Bedrohung Nr. 21

| | | |
|---|---|---|
| *Nummer* | 22 | |
| *Beschreibung* | Ausnutzen von Schwachstellen im Anwendungscode von ONP (Schädliche ONP-Teilnehmer) | |
| *Bedrohung* | Verlust der Integrität durch die Ausführung von nicht vertrauenswürdigem Code, Verlust der Verfügbarkeit durch Dienstausfall | |
| *Kategorie* | privilegierte Remote Code Execution (RCE) | |
| *Angreifer* | Schadsoftwareautor vergleichbar mit IT-Security-Spezialisten im Bereich Netzwerk und Systemprogrammierung | |
| *Schnittstelle* | OS2OR | |
| *Entwickler, Methode* | CP, Brainstorming | |
| *Threat Agent* | *Value* | *Comment* |
| *Skill Level* | 2 | Spezialisten im Bereich Netzwerk und Systemprogrammierung, Kompromittierung von vielen ONP-Nutzer, modifizierter ONP-Sender nötig und lokale Begrenzung durch Ad-hoc-Kommunkation, wenige |
| *Motive* | 5 | |
| *Opportunity* | 6 | |
| *Size* | 2 | |
| *Vulnerability* | *Value* | *Comment* |
| *Ease of discovery* | 5 | mäßig schwierig zu entdecken, mittelschwer ausnutzbar, offensichtliche Schwachstelle, dokumentiert aber ohne Kontrolle |
| *Ease of exploit* | 4 | |
| *Awareness* | 6 | |
| *Intrusion* | 8 | |
| *Likelihood* | 4.75 | MEDIUM |
| *Technical Impact* | *Value* | *Comment* |
| *Loss of Confidentiality* | 7 | umfassend kritische Daten enthüllen, viele Daten stark beschädigt, umfassend sekundäre Dienste unterbrochen, möglicherweise nachvollziehbar |
| *Loss of Integrity* | 7 | |
| *Loss of Availability* | 5 | |
| *Loss of Accountability* | 7 | |
| *Technical Impact* | 6.50 | HIGH |
| *Business Impact* | *Value* | *Comment* |
| *Financial damage* | 3 | geringe Auswirkungen auf jährlichen Gewinn, Beschädigung des Markennamens, eindeutige Verletzung, millionen von betroffenen Personen |
| *Reputation damage* | 9 | |
| *Non-compliance* | 5 | |
| *Privacy Violation* | 9 | |
| *Business Impact* | 6.50 | HIGH |
| *Gesamtrisiko* | *technisch* | High |
| | *geschäftlich* | High |

**Tabelle B.22:** Risk Rating für Bedrohung Nr. 22

| | | | |
|---|---|---|---|
| *Nummer* | 23 | | |
| *Beschreibung* | Ausnutzen von Schwachstellen im Anwendungscode von ONP (Schädliche Anwendung) | | |
| *Bedrohung* | Verlust der Integrität durch die Ausführung von nicht vertrauenswürdigem Code, Verlust der Verfügbarkeit durch Dienstausfall | | |
| *Kategorie* | privilegierte Local Code Execution (LCE) | | |
| *Angreifer* | Schadsoftwareautor zu vergleichen mit Softwareentwickler im Bereich Mobileanwendungen | | |
| *Schnittstelle* | AP2OS | | |
| *Entwickler, Methode* | CP, Brainstorming | | |
| *Threat Agent* | *Value* | *Comment* | |
| *Skill Level* | 3 | Softwareentwickler im Bereich Mobileanwendungen, Kompromittierung von vielen ONP-Nutzern, schädliche Anwendung muss auf die ONP-Systeme gelangen, registrierter Entwickler | |
| *Motive* | 5 | | |
| *Opportunity* | 5 | | |
| *Size* | 3 | | |
| *Vulnerability* | *Value* | *Comment* | |
| *Ease of discovery* | 5 | mäßig schwierig zu entdecken, mittelschwer ausnutzbar, offensichtliche Schwachstelle, dokumentiert aber ohne Kontrolle | |
| *Ease of exploit* | 4 | | |
| *Awareness* | 6 | | |
| *Intrusion* | 8 | | |
| *Likelihood* | 4.88 | MEDIUM | |
| *Technical Impact* | *Value* | *Comment* | |
| *Loss of Confidentiality* | 7 | umfassend kritische Daten enthüllen, viele Daten stark beschädigt, umfassend sekundäre Dienste unterbrochen, möglicherweise nachvollziehbar | |
| *Loss of Integrity* | 7 | | |
| *Loss of Availability* | 5 | | |
| *Loss of Accountability* | 7 | | |
| *Technical Impact* | 6.50 | HIGH | |
| *Business Impact* | *Value* | *Comment* | |
| *Financial damage* | 3 | geringe Auswirkungen auf jährlichen Gewinn, Beschädigung des Markennamens, eindeutige Verletzung, millionen von betroffenen Personen | |
| *Reputation damage* | 9 | | |
| *Non-compliance* | 5 | | |
| *Privacy Violation* | 9 | | |
| *Business Impact* | 6.50 | HIGH | |
| *Gesamtrisiko* | *technisch* | | High |
| | *geschäftlich* | | High |

**Tabelle B.23:** Risk Rating für Bedrohung Nr. 23

| | | |
|---|---|---|
| *Nummer* | 24 | |
| *Beschreibung* | Ausnutzen von Schwachstellen im Anwendungscode von ONP (Manipulierte Sensoren) | |
| *Bedrohung* | Verlust der Integrität durch die Ausführung von nicht vertrauenswürdigem Code, Verlust der Verfügbarkeit durch Dienstausfall | |
| *Kategorie* | privilegierte Local Code Execution (LCE) | |
| *Angreifer* | Geheimdienste/private Sicherheitsdienste/organisierte Kriminalität | |
| *Schnittstelle* | SD2OM | |
| *Entwickler, Methode* | CP, Brainstorming | |
| *Threat Agent* | *Value* | *Comment* |
| *Skill Level* | 1 | Hard- und Softwareentwickler mit fundierten Kenntnissen über die Applikationsschnittstelle, Kompromittierung von vielen ONP-Nutzern, Manipulation der Sensoren möglicherweise während der Herstellung nötig, wenige |
| *Motive* | 5 | |
| *Opportunity* | 1 | |
| *Size* | 2 | |
| *Vulnerability* | *Value* | *Comment* |
| *Ease of discovery* | 3 | schwer zu entdecken, schwer ausnutzbar, offensichtliche Schwachstelle, dokumentiert aber ohne Kontrolle |
| *Ease of exploit* | 3 | |
| *Awareness* | 6 | |
| *Intrusion* | 8 | |
| *Likelihood* | 3.62 | MEDIUM |
| *Technical Impact* | *Value* | *Comment* |
| *Loss of Confidentiality* | 7 | umfassend kritische Daten enthüllen, viele Daten stark beschädigt, umfassend sekundäre Dienste unterbrochen, möglicherweise nachvollziehbar |
| *Loss of Integrity* | 7 | |
| *Loss of Availability* | 5 | |
| *Loss of Accountability* | 7 | |
| *Technical Impact* | 6.50 | HIGH |
| *Business Impact* | *Value* | *Comment* |
| *Financial damage* | 3 | geringe Auswirkungen auf jährlichen Gewinn, Unterstellung von böswilligem Handeln, eindeutige Verletzung, millionen von betroffenen Personen |
| *Reputation damage* | 5 | |
| *Non-compliance* | 5 | |
| *Privacy Violation* | 9 | |
| *Business Impact* | 5.50 | MEDIUM |
| *Gesamtrisiko* | *technisch* | High |
| | *geschäftlich* | Medium |

**Tabelle B.24:** Risk Rating für Bedrohung Nr. 24

| | | | |
|---|---|---|---|
| *Nummer* | 25 | | |
| *Beschreibung* | Ausnutzen von Schwachstellen im Anwendungscode von APP (Manipulierte Sensoren) | | |
| *Bedrohung* | Verlust der Integrität durch die Ausführung von nicht vertrauenswürdigem Code, Verlust der Verfügbarkeit durch Dienstausfall | | |
| *Kategorie* | unprivilegierte Local Code Execution (LCE) | | |
| *Angreifer* | Geheimdienste/private Sicherheitsdienste/organisierte Kriminalität | | |
| *Schnittstelle* | SD2AP | | |
| *Entwickler, Methode* | CP, Brainstorming | | |
| *Threat Agent* | *Value* | *Comment* | |
| *Skill Level* | 1 | Hard- und Softwareentwickler mit fundierten Kenntnissen über die Applikationsschnittstelle, Kompromittierung von vielen ONP-Nutzern, Manipulation der Sensoren möglicherweise während der Herstellung nötig, wenige | |
| *Motive* | 5 | | |
| *Opportunity* | 1 | | |
| *Size* | 2 | | |
| *Vulnerability* | *Value* | *Comment* | |
| *Ease of discovery* | 3 | schwer zu entdecken, schwer ausnutzbar, offensichtliche Schwachstelle, dokumentiert aber ohne Kontrolle | |
| *Ease of exploit* | 3 | | |
| *Awareness* | 6 | | |
| *Intrusion* | 8 | | |
| *Likelihood* | 3.62 | MEDIUM | |
| *Technical Impact* | *Value* | *Comment* | |
| *Loss of Confidentiality* | 7 | umfassend kritische Daten enthüllen, viele Daten stark beschädigt, umfassend sekundäre Dienste unterbrochen, möglicherweise nachvollziehbar | |
| *Loss of Integrity* | 7 | | |
| *Loss of Availability* | 5 | | |
| *Loss of Accountability* | 7 | | |
| *Technical Impact* | 6.50 | HIGH | |
| *Business Impact* | *Value* | *Comment* | |
| *Financial damage* | 1 | weniger als die Kosten zur Behebung, geringer Schaden, eindeutige Verletzung, tausende von betroffenen Personen | |
| *Reputation damage* | 1 | | |
| *Non-compliance* | 5 | | |
| *Privacy Violation* | 7 | | |
| *Business Impact* | 3.50 | MEDIUM | |
| *Gesamtrisiko* | *technisch* | High | |
| | *geschäftlich* | Medium | |

**Tabelle B.25:** Risk Rating für Bedrohung Nr. 25

| | | | |
|---|---|---|---|
| *Nummer* | 26 | | |
| *Beschreibung* | Ausnutzen von Schwachstellen im ONP-Message-Filter (z.B. Pufferüberlauf) | | |
| *Bedrohung* | Verlust der Vertraulichkeit, Integrität, Verfügbarkeit und Authentizität | | |
| *Kategorie* | privilegierte Remote Code Execution (RCE) | | |
| *Angreifer* | Schadsoftwareautor vergleichbar mit IT-Security-Spezialisten im Bereich Betriebssystemprogrammierung | | |
| *Schnittstelle* | OR2OM | | |
| *Entwickler, Methode* | CP, Brainstorming | | |
| *Threat Agent* | *Value* | *Comment* | |
| *Skill Level* | 2 | Spezialisten im Bereich Betriebssystemprogrammierung, Ruf Schädigung, Kompromittierung von vielen ONP-Nutzern, wenige | |
| *Motive* | 6 | | |
| *Opportunity* | 9 | | |
| *Size* | 2 | | |
| *Vulnerability* | *Value* | *Comment* | |
| *Ease of discovery* | 5 | mäßig schwierig zu entdecken, mittelschwer ausnutzbar, offensichtliche Schwachstelle, dokumentiert aber ohne Kontrolle | |
| *Ease of exploit* | 4 | | |
| *Awareness* | 6 | | |
| *Intrusion* | 8 | | |
| *Likelihood* | 5.25 | MEDIUM | |
| *Technical Impact* | *Value* | *Comment* | |
| *Loss of Confidentiality* | 7 | umfassend kritische Daten enthüllen, viele Daten stark beschädigt, umfassend primäre Dienste unterbrochen, möglicherweise nachvollziehbar | |
| *Loss of Integrity* | 7 | | |
| *Loss of Availability* | 7 | | |
| *Loss of Accountability* | 7 | | |
| *Technical Impact* | 7.00 | HIGH | |
| *Business Impact* | *Value* | *Comment* | |
| *Financial damage* | 3 | geringe Auswirkungen auf jährlichen Gewinn, Beschädigung des Markennamens, eindeutige Verletzung, millionen von betroffenen Personen | |
| *Reputation damage* | 9 | | |
| *Non-compliance* | 5 | | |
| *Privacy Violation* | 9 | | |
| *Business Impact* | 6.50 | HIGH | |
| *Gesamtrisiko* | *technisch* | | High |
| | *geschäftlich* | | High |

**Tabelle B.26:** Risk Rating für Bedrohung Nr. 26

| | | | |
|---|---|---|---|
| *Nummer* | 27 | | |
| *Beschreibung* | Ausnutzen von Schwachstellen im ONP-Message-Filter (z.B. Injection Angriff auf die ONP-Datenhaltung, Nachricht mit Sensordaten, die interpretiert durch das Storagebackend, sich selber einen hohen Vertrauenswert gibt oder alle Vergleichsdaten löscht) | | |
| *Bedrohung* | Verlust der Integrität und Authentizität durch Manipulation der Daten, Verlust der Verfügbarkeit durch Löschen von Daten | | |
| *Kategorie* | Eingabevalidierung | | |
| *Angreifer* | Schadsoftwareautor vergleichbar mit IT-Security-Spezialisten im Bereich Datenbanken | | |
| *Schnittstelle* | OM2OD | | |
| *Entwickler, Methode* | CP, Brainstorming | | |
| *Threat Agent* | *Value* | *Comment* | |
| *Skill Level* | 2 | Spezialisten im Bereich Datenbanken, Persistenz und dauerhafte Kompromittierung des ONP-Systems (betrifft alle Anwendungen), wenige | |
| *Motive* | 6 | | |
| *Opportunity* | 9 | | |
| *Size* | 2 | | |
| *Vulnerability* | *Value* | *Comment* | |
| *Ease of discovery* | 5 | mäßig schwierig zu entdecken, mittelschwer ausnutzbar, offensichtliche Schwachstelle, dokumentiert aber ohne Kontrolle | |
| *Ease of exploit* | 4 | | |
| *Awareness* | 6 | | |
| *Intrusion* | 8 | | |
| *Likelihood* | 5.25 | MEDIUM | |
| *Technical Impact* | *Value* | *Comment* | |
| *Loss of Confidentiality* | 7 | umfassend kritische Daten enthüllen, viele Daten stark beschädigt, umfassend primäre Dienste unterbrochen, möglicherweise nachvollziehbar | |
| *Loss of Integrity* | 7 | | |
| *Loss of Availability* | 7 | | |
| *Loss of Accountability* | 7 | | |
| *Technical Impact* | 7.00 | HIGH | |
| *Business Impact* | *Value* | *Comment* | |
| *Financial damage* | 3 | geringe Auswirkungen auf jährlichen Gewinn, Beschädigung des Markennamens, eindeutige Verletzung, millionen von betroffenen Personen | |
| *Reputation damage* | 9 | | |
| *Non-compliance* | 5 | | |
| *Privacy Violation* | 9 | | |
| *Business Impact* | 6.50 | HIGH | |
| *Gesamtrisiko* | *technisch*<br>*geschäftlich* | | High<br>High |

**Tabelle B.27:** Risk Rating für Bedrohung Nr. 27

| | | | |
|---|---|---|---|
| *Nummer* | 28 | | |
| *Beschreibung* | Ausnutzen von Schwachstellen in Anwendungen (APPs) (z.B. Pufferüberlauf) | | |
| *Bedrohung* | Lokaler Dienstausfall und Integritätsverlust durch Ausführung von Schadcode | | |
| *Kategorie* | unprivilegierte Remote Code Execution (RCE) | | |
| *Angreifer* | Schadsoftwareautor vergleichbar mit IT-Security-Spezialisten im Bereich Mobileanwendungentwicklung | | |
| *Schnittstelle* | APPs | | |
| *Entwickler, Methode* | CP, Brainstorming | | |
| *Threat Agent* | *Value* | *Comment* | |
| *Skill Level* | 2 | Softwareentwickler im Bereich Mobileanwendungsentwicklung, Kompromittierung einer bestimmten Anwendung, Installation durch Nutzerinteraktion und im AppStore verfügbar sein, registrierter Entwickler | |
| *Motive* | 5 | | |
| *Opportunity* | 4 | | |
| *Size* | 2 | | |
| *Vulnerability* | *Value* | *Comment* | |
| *Ease of discovery* | 5 | mäßig schwierig zu entdecken, einfach ausnutzbar, offensichtliche Schwachstelle, dokumentiert aber ohne Kontrolle | |
| *Ease of exploit* | 5 | | |
| *Awareness* | 6 | | |
| *Intrusion* | 8 | | |
| *Likelihood* | 4.62 | MEDIUM | |
| *Technical Impact* | *Value* | *Comment* | |
| *Loss of Confidentiality* | 7 | umfassend kritische Daten enthüllen, wenige Daten stark beschädigt, wenige sekundäre Dienste unterbrochen, möglicherweise nachvollziehbar | |
| *Loss of Integrity* | 3 | | |
| *Loss of Availability* | 1 | | |
| *Loss of Accountability* | 7 | | |
| *Technical Impact* | 4.50 | MEDIUM | |
| *Business Impact* | *Value* | *Comment* | |
| *Financial damage* | 1 | weniger als die Kosten zur Behebung, geringer Schaden, geringe Verletzung, keine sensiblen Daten | |
| *Reputation damage* | 1 | | |
| *Non-compliance* | 2 | | |
| *Privacy Violation* | 1 | | |
| *Business Impact* | 1.25 | LOW | |
| *Gesamtrisiko* | *technisch* | | Medium |
| | *geschäftlich* | | Low |

**Tabelle B.28:** Risk Rating für Bedrohung Nr. 28

| | | | |
|---|---|---|---|
| *Nummer* | 29 | | |
| *Beschreibung* | Ausspähung von vertraulichen Informationen durch (z.B. Spionage App) | | |
| *Bedrohung* | Verlust von vertraulichen Daten | | |
| *Kategorie* | Sensitve Data Exposure | | |
| *Angreifer* | Schadsoftwareautor vergleichbar mit Softwareentwickler im Bereich Mobileanwendungsentwicklung | | |
| *Schnittstelle* | OM2AP | | |
| *Entwickler, Methode* | CP, Brainstorming | | |
| *Threat Agent* | *Value* | *Comment* | |
| *Skill Level* | 3 | Softwareentwickler im Bereich Mobileanwendungsentwicklung, Kompromittierung einer bestimmten Anwendung, Installation durch Nutzerinteraktion und im AppStore verfügbar sein, registrierter Entwickler | |
| *Motive* | 7 | | |
| *Opportunity* | 4 | | |
| *Size* | 3 | | |
| *Vulnerability* | *Value* | *Comment* | |
| *Ease of discovery* | 9 | automatisierte Tools verfügbar, vollautomatische Tools verfügbar, allgemein bekannte Schwachstelle, dokumentiert aber ohne Kontrolle | |
| *Ease of exploit* | 9 | | |
| *Awareness* | 9 | | |
| *Intrusion* | 8 | | |
| *Likelihood* | 6.50 | HIGH | |
| *Technical Impact* | *Value* | *Comment* | |
| *Loss of Confidentiality* | 9 | alle Daten enthüllen, wenige Daten leicht beschädigt, wenige sekundäre Dienste unterbrochen, vollständig nachvollziehbar | |
| *Loss of Integrity* | 1 | | |
| *Loss of Availability* | 1 | | |
| *Loss of Accountability* | 1 | | |
| *Technical Impact* | 3.00 | LOW | |
| *Business Impact* | *Value* | *Comment* | |
| *Financial damage* | 1 | weniger als die Kosten zur Behebung, Beschädigung des Markennamens, geringe Verletzung, millionen von betroffenen Personen | |
| *Reputation damage* | 9 | | |
| *Non-compliance* | 2 | | |
| *Privacy Violation* | 9 | | |
| *Business Impact* | 5.25 | MEDIUM | |
| *Gesamtrisiko* | *technisch* | | Medium |
| | *geschäftlich* | | High |

**Tabelle B.29:** Risk Rating für Bedrohung Nr. 29

| | | | |
|---|---|---|---|
| *Nummer* | 30 | | |
| *Beschreibung* | Ausnutzen von Schwachstellen im AppStore WebService (z.B. Injection Angriff auf die AppStore-Datenhaltung) | | |
| *Bedrohung* | Verlust der Integrität und Authentizität durch Ersetzen von Apps durch schädliche Kopien | | |
| *Kategorie* | Eingabevalidierung | | |
| *Angreifer* | Schadsoftwareautor vergleichbar mit IT-Security-Spezialisten im Bereich Datenbanken und Webservice/Webapplikations-Entwicklung | | |
| *Schnittstelle* | ASB-WS2D | | |
| *Entwickler, Methode* | CP, Brainstorming | | |
| *Threat Agent* | *Value* | *Comment* | |
| *Skill Level* | 2 | Spezialisten im Bereich Datenbanken und Webservice/Webapplikations- Entwicklung, Kompromittierung einer großen Nutzergemeinde durch Kompromittieren eines AppStores, authentifizierte Nutzer | |
| *Motive* | 7 | | |
| *Opportunity* | 9 | | |
| *Size* | 6 | | |
| *Vulnerability* | *Value* | *Comment* | |
| *Ease of discovery* | 9 | automatisierte Tools verfügbar, einfach ausnutzbar, offensichtliche Schwachstelle, dokumentiert aber ohne Kontrolle | |
| *Ease of exploit* | 5 | | |
| *Awareness* | 6 | | |
| *Intrusion* | 8 | | |
| *Likelihood* | 6.50 | HIGH | |
| *Technical Impact* | *Value* | *Comment* | |
| *Loss of Confidentiality* | 6 | wenige kritische Daten enthüllen, wenige Daten stark beschädigt, umfassend sekundäre Dienste unterbrochen, möglicherweise nachvollziehbar | |
| *Loss of Integrity* | 3 | | |
| *Loss of Availability* | 5 | | |
| *Loss of Accountability* | 7 | | |
| *Technical Impact* | 5.25 | MEDIUM | |
| *Business Impact* | *Value* | *Comment* | |
| *Financial damage* | 7 | signifikante Auswirkungen auf jährlichen Gewinn, Verlust von größeren Kunden, geringe Verletzung, millionen von betroffenen Personen | |
| *Reputation damage* | 4 | | |
| *Non-compliance* | 2 | | |
| *Privacy Violation* | 9 | | |
| *Business Impact* | 5.50 | MEDIUM | |
| *Gesamtrisiko* | *technisch* | | High |
| | *geschäftlich* | | High |

**Tabelle B.30:** Risk Rating für Bedrohung Nr. 30

| | | |
|---|---|---|
| *Nummer* | 31 | |
| *Beschreibung* | Ausnutzen von Schwachstellen im ServiceProvider WebService, um Nutzerprofile (personenbezogene Daten) zu stehlen (z.B. Injection Angriff auf die ServiceProvider-Datenhaltung) | |
| *Bedrohung* | Verlust von vertraulichen Daten, Verlust der Integrität und Authentizität durch Manipulation der Daten, Verlust der Verfügbarkeit durch Löschen von Daten | |
| *Kategorie* | Eingabevalidierung | |
| *Angreifer* | Schadsoftwareautor vergleichbar mit IT-Security-Spezialisten im Bereich Datenbanken und Webservice/Webapplikations-Entwicklung | |
| *Schnittstelle* | SPB-WS2D | |
| *Entwickler, Methode* | CP, Brainstorming | |
| *Threat Agent* | *Value* | *Comment* |
| *Skill Level* | 2 | Spezialisten im Bereich Datenbanken und Webservice/Webapplikations- Entwicklung, Kompromittierung einer großen Nutzergemeinde durch Kompromittieren eines Dienstanbieters, authentifizierte Nutzer |
| *Motive* | 7 | |
| *Opportunity* | 9 | |
| *Size* | 6 | |
| *Vulnerability* | *Value* | *Comment* |
| *Ease of discovery* | 9 | automatisierte Tools verfügbar, einfach ausnutzbar, offensichtliche Schwachstelle, dokumentiert aber ohne Kontrolle |
| *Ease of exploit* | 5 | |
| *Awareness* | 6 | |
| *Intrusion* | 8 | |
| *Likelihood* | 6.50 | HIGH |
| *Technical Impact* | *Value* | *Comment* |
| *Loss of Confidentiality* | 7 | umfassend kritische Daten enthüllen, wenige Daten stark beschädigt, wenige primäre Dienste unterbrochen, möglicherweise nachvollziehbar |
| *Loss of Integrity* | 3 | |
| *Loss of Availability* | 3 | |
| *Loss of Accountability* | 7 | |
| *Technical Impact* | 5.00 | MEDIUM |
| *Business Impact* | *Value* | *Comment* |
| *Financial damage* | 7 | signifikante Auswirkungen auf jährlichen Gewinn, Verlust von größeren Kunden, eindeutige Verletzung, millionen von betroffenen Personen |
| *Reputation damage* | 4 | |
| *Non-compliance* | 5 | |
| *Privacy Violation* | 9 | |
| *Business Impact* | 6.25 | HIGH |
| *Gesamtrisiko* | *technisch* | High |
| | *geschäftlich* | Critical |

**Tabelle B.31:** Risk Rating für Bedrohung Nr. 31

| | | | |
|---|---|---|---|
| *Nummer* | 32 | | |
| *Beschreibung* | Ausnutzen von Schwachstellen im ServiceProvider WebService zum Erschleichen von Leistungen (bei kostenpflichtigen Diensten) (z.B. schwache Authentifizierung und Autorisierung: Nummernschild) | | |
| *Bedrohung* | Kostenlose Nutzung von kostenpflichtigen Diensten | | |
| *Kategorie* | Berechtigungsmissbrauch | | |
| *Angreifer* | Böswilliger Nutzer mit geringen technischen Fähigkeiten | | |
| *Schnittstelle* | SP-WS2OR | | |
| *Entwickler, Methode* | CP, Brainstorming | | |
| *Threat Agent* | *Value* | *Comment* | |
| *Skill Level* | 7 | Geringe Fähigkeiten nötig, von kostenpflichtig zu kostenlos (Free riding), man muss registriert, also ein Nutzer sein | |
| *Motive* | 9 | | |
| *Opportunity* | 7 | | |
| *Size* | 6 | | |
| *Vulnerability* | *Value* | *Comment* | |
| *Ease of discovery* | 7 | einfach zu entdecken, einfach ausnutzbar, versteckte Schwachstelle, dokumentiert aber ohne Kontrolle | |
| *Ease of exploit* | 5 | | |
| *Awareness* | 4 | | |
| *Intrusion* | 8 | | |
| *Likelihood* | 6.62 | HIGH | |
| *Technical Impact* | *Value* | *Comment* | |
| *Loss of Confidentiality* | 1 | keine sensiblen Daten enthüllen, wenige Daten leicht beschädigt, wenige sekundäre Dienste unterbrochen, möglicherweise nachvollziehbar | |
| *Loss of Integrity* | 1 | | |
| *Loss of Availability* | 1 | | |
| *Loss of Accountability* | 7 | | |
| *Technical Impact* | 2.50 | LOW | |
| *Business Impact* | *Value* | *Comment* | |
| *Financial damage* | 7 | signifikante Auswirkungen auf jährlichen Gewinn, geringer Schaden, keine Verletzung, keine sensiblen Daten | |
| *Reputation damage* | 1 | | |
| *Non-compliance* | 1 | | |
| *Privacy Violation* | 1 | | |
| *Business Impact* | 2.50 | LOW | |
| *Gesamtrisiko* | *technisch* | | Medium |
| | *geschäftlich* | | Medium |

**Tabelle B.32:** Risk Rating für Bedrohung Nr. 32

| | | | |
|---|---|---|---|
| *Nummer* | 33 | | |
| *Beschreibung* | Ausnutzen von Schwachstellen im ServiceProvider WebService durch Ausnutzung von Business Logik (z.B. Mehrfach Registrierung, um Boni zu akkumulieren) | | |
| *Bedrohung* | Kostenlose Nutzung von kostenpflichtigen Diensten | | |
| *Kategorie* | Geschäftslogikfehler | | |
| *Angreifer* | Böswilliger Nutzer | | |
| *Schnittstelle* | SP-WS2OR | | |
| *Entwickler, Methode* | CP, Brainstorming | | |
| *Threat Agent* | *Value* | *Comment* | |
| *Skill Level* | 9 | Keine Fähigkeiten nötig, da Registrierungen meistens einfach gestaltet sind; Belohnung je nach Bonus; jeder, der einen Dienst nutzen können soll also Kunde werden soll | |
| *Motive* | 9 | | |
| *Opportunity* | 9 | | |
| *Size* | 9 | | |
| *Vulnerability* | *Value* | *Comment* | |
| *Ease of discovery* | 7 | einfach zu entdecken, einfach ausnutzbar, allgemein bekannte Schwachstelle, dokumentiert aber ohne Kontrolle | |
| *Ease of exploit* | 5 | | |
| *Awareness* | 9 | | |
| *Intrusion* | 8 | | |
| *Likelihood* | 8.12 | HIGH | |
| *Technical Impact* | *Value* | *Comment* | |
| *Loss of Confidentiality* | 1 | keine sensiblen Daten enthüllen, wenige Daten leicht beschädigt, wenige sekundäre Dienste unterbrochen, möglicherweise nachvollziehbar | |
| *Loss of Integrity* | 1 | | |
| *Loss of Availability* | 1 | | |
| *Loss of Accountability* | 7 | | |
| *Technical Impact* | 2.50 | LOW | |
| *Business Impact* | *Value* | *Comment* | |
| *Financial damage* | 3 | geringe Auswirkungen auf jährlichen Gewinn, geringer Schaden, keine Verletzung, keine sensiblen Daten | |
| *Reputation damage* | 1 | | |
| *Non-compliance* | 1 | | |
| *Privacy Violation* | 1 | | |
| *Business Impact* | 1.50 | LOW | |
| *Gesamtrisiko* | *technisch*<br>*geschäftlich* | | Medium<br>Medium |

**Tabelle B.33:** Risk Rating für Bedrohung Nr. 33

| | | | |
|---|---|---|---|
| *Nummer* | 34 | | |
| *Beschreibung* | Diebstahl von Nutzerprofilen, Orstinformationen/Bewegungsprofilen im Falle des Empfehlungssystems (ServiceProvider WebService) | | |
| *Bedrohung* | Verlust von vertraulichen/personenbezogenen Daten | | |
| *Kategorie* | Sensitve Data Exposure | | |
| *Angreifer* | Geheimdienste/private Sicherheitsdienste/organisierte Kriminalität | | |
| *Schnittstelle* | SP-WS2OR | | |
| *Entwickler, Methode* | CP, Brainstorming | | |
| *Threat Agent* | *Value* | *Comment* | |
| *Skill Level* | 2 | Spezialisten im Bereich Datenbanken und Webservice/Webapplikations-Entwicklung, Kompromittierung einer großen Nutzergemeinde durch einen Dienstanbieter, hoher Wert der Bewegungsprofile | |
| *Motive* | 9 | | |
| *Opportunity* | 9 | | |
| *Size* | 2 | | |
| *Vulnerability* | *Value* | *Comment* | |
| *Ease of discovery* | 7 | einfach zu entdecken, einfach ausnutzbar, offensichtliche Schwachstelle, dokumentiert aber ohne Kontrolle | |
| *Ease of exploit* | 5 | | |
| *Awareness* | 6 | | |
| *Intrusion* | 8 | | |
| *Likelihood* | 6.00 | MEDIUM | |
| *Technical Impact* | *Value* | *Comment* | |
| *Loss of Confidentiality* | 9 | alle Daten enthüllen, wenige Daten stark beschädigt, wenige primäre Dienste unterbrochen, möglicherweise nachvollziehbar | |
| *Loss of Integrity* | 3 | | |
| *Loss of Availability* | 3 | | |
| *Loss of Accountability* | 7 | | |
| *Technical Impact* | 5.50 | MEDIUM | |
| *Business Impact* | *Value* | *Comment* | |
| *Financial damage* | 7 | signifikante Auswirkungen auf jährlichen Gewinn, Verlust von größeren Kunden, eindeutige Verletzung, millionen von betroffenen Personen | |
| *Reputation damage* | 4 | | |
| *Non-compliance* | 5 | | |
| *Privacy Violation* | 9 | | |
| *Business Impact* | 6.25 | HIGH | |
| *Gesamtrisiko* | *technisch* | | Medium |
| | *geschäftlich* | | High |

**Tabelle B.34:** Risk Rating für Bedrohung Nr. 34

| | | |
|---|---|---|
| *Nummer* | 35 | |
| *Beschreibung* | Injection Angriff auf die APP-Datenhaltung | |
| *Bedrohung* | Verlust von vertraulichen App-Daten, Verlust der Integrität und Authentizität durch Manipulation der App-Daten, Verlust der Verfügbarkeit durch Löschen von App Daten | |
| *Kategorie* | Eingabevalidierung | |
| *Angreifer* | Schadsoftwareautor vergleichbar mit IT-Security-Spezialisten im Bereich Mobileanwendungsentwicklung | |
| *Schnittstelle* | AP2D | |
| *Entwickler, Methode* | CP, Brainstorming | |
| *Threat Agent* | *Value* | *Comment* |
| *Skill Level* | 2 | Spezialisten im Bereich Mobileanwendungsentwicklung, Persistenz dauerhafte Kompromittierung einer Applikation, wenige |
| *Motive* | 5 | |
| *Opportunity* | 7 | |
| *Size* | 2 | |
| *Vulnerability* | *Value* | *Comment* |
| *Ease of discovery* | 7 | einfach zu entdecken, einfach ausnutzbar, offensichtliche Schwachstelle, dokumentiert aber ohne Kontrolle |
| *Ease of exploit* | 5 | |
| *Awareness* | 6 | |
| *Intrusion* | 8 | |
| *Likelihood* | 5.25 | MEDIUM |
| *Technical Impact* | *Value* | *Comment* |
| *Loss of Confidentiality* | 6 | wenige kritische Daten enthüllen, wenige Daten stark beschädigt, wenige primäre Dienste unterbrochen, möglicherweise nachvollziehbar |
| *Loss of Integrity* | 3 | |
| *Loss of Availability* | 3 | |
| *Loss of Accountability* | 7 | |
| *Technical Impact* | 4.75 | MEDIUM |
| *Business Impact* | *Value* | *Comment* |
| *Financial damage* | 3 | geringe Auswirkungen auf jährlichen Gewinn, Verlust von Kunden mittlerer Größe, keine Verletzung, tausende von betroffenen Personen |
| *Reputation damage* | 3 | |
| *Non-compliance* | 1 | |
| *Privacy Violation* | 7 | |
| *Business Impact* | 3.50 | MEDIUM |
| *Gesamtrisiko* | *technisch*<br>*geschäftlich* | Medium<br>Medium |

**Tabelle B.35:** Risk Rating für Bedrohung Nr. 35

| | | |
|---|---|---|
| *Nummer* | 36 | |
| *Beschreibung* | Manipulation der Netzwerktopologie im zellularen Netzwerk (z.B. „Stingrays“ Cell Site Simulators für MitM Angriffe) | |
| *Bedrohung* | Mitlesen vertraulicher Kommunikation, Dienstausfall | |
| *Kategorie* | Man in the Middle (MitM) | |
| *Angreifer* | Sicherheitsbehörden/Geheimdienste/private Sicherheitsdienste/organisierte Kriminalität | |
| *Schnittstelle* | OTA (Over-the-Air) ONP-Receiver | |
| *Entwickler, Methode* | CP, Brainstorming | |
| *Threat Agent* | *Value* | *Comment* |
| *Skill Level* | 1 | Spezialist für Kommunikationstechnik, Massenüberwachung, spezielle Hard- und Software nötig, wenige |
| *Motive* | 9 | |
| *Opportunity* | 4 | |
| *Size* | 2 | |
| *Vulnerability* | *Value* | *Comment* |
| *Ease of discovery* | 7 | einfach zu entdecken, halb automatisierte Tools verfügbar, offensichtliche Schwachstelle, dokumentiert aber ohne Kontrolle |
| *Ease of exploit* | 7 | |
| *Awareness* | 6 | |
| *Intrusion* | 8 | |
| *Likelihood* | 5.50 | MEDIUM |
| *Technical Impact* | *Value* | *Comment* |
| *Loss of Confidentiality* | 9 | alle Daten enthüllen, alle Daten vollständig beschädigt, umfassend sekundäre Dienste unterbrochen, möglicherweise nachvollziehbar |
| *Loss of Integrity* | 9 | |
| *Loss of Availability* | 5 | |
| *Loss of Accountability* | 7 | |
| *Technical Impact* | 7.50 | HIGH |
| *Business Impact* | *Value* | *Comment* |
| *Financial damage* | 5 | mittlere Auswirkungen auf jährlichen Gewinn, Unterstellung von böswilligem Handeln, keine Verletzung, millionen von betroffenen Personen |
| *Reputation damage* | 5 | |
| *Non-compliance* | 1 | |
| *Privacy Violation* | 9 | |
| *Business Impact* | 5.00 | MEDIUM |
| *Gesamtrisiko* | *technisch* | High |
| | *geschäftlich* | Medium |

**Tabelle B.36:** Risk Rating für Bedrohung Nr. 36

| | | |
|---|---|---|
| *Nummer* | 37 | |
| *Beschreibung* | Manipulation der Netzwerktopologie im Ad-hoc-Netzwerk (z.B. Wormhole Angriff, Verbinden von entfernten Netzwerksegmenten über einen Tunnel, Verbreiten von falschen Routing-Informationen für MitM Angriffe) | |
| *Bedrohung* | Mitlesen vertraulicher Kommunikation, Dienstausfall | |
| *Kategorie* | Man in the Middle (MitM) | |
| *Angreifer* | Sicherheitsbehörden/Geheimdienste/private Sicherheitsdienste/organisierte Kriminalität | |
| *Schnittstelle* | OTA (Over-the-Air) ONP-Receiver | |
| *Entwickler, Methode* | CP, Brainstorming | |
| *Threat Agent* | *Value* | *Comment* |
| *Skill Level* | 1 | Spezialist für Kommunikationstechnik, Massenüberwachung, spezielle Hard- und Software nötig, wenige |
| *Motive* | 9 | |
| *Opportunity* | 4 | |
| *Size* | 2 | |
| *Vulnerability* | *Value* | *Comment* |
| *Ease of discovery* | 7 | einfach zu entdecken, einfach ausnutzbar, offensichtliche Schwachstelle, dokumentiert aber ohne Kontrolle |
| *Ease of exploit* | 5 | |
| *Awareness* | 6 | |
| *Intrusion* | 8 | |
| *Likelihood* | 5.25 | MEDIUM |
| *Technical Impact* | *Value* | *Comment* |
| *Loss of Confidentiality* | 9 | alle Daten enthüllen, alle Daten vollständig beschädigt, umfassend sekundäre Dienste unterbrochen, möglicherweise nachvollziehbar |
| *Loss of Integrity* | 9 | |
| *Loss of Availability* | 5 | |
| *Loss of Accountability* | 7 | |
| *Technical Impact* | 7.50 | HIGH |
| *Business Impact* | *Value* | *Comment* |
| *Financial damage* | 5 | mittlere Auswirkungen auf jährlichen Gewinn, Unterstellung von böswilligem Handeln, geringe Verletzung, millionen von betroffenen Personen |
| *Reputation damage* | 5 | |
| *Non-compliance* | 2 | |
| *Privacy Violation* | 9 | |
| *Business Impact* | 5.25 | MEDIUM |
| *Gesamtrisiko* | *technisch* | High |
| | *geschäftlich* | Medium |

**Tabelle B.37:** Risk Rating für Bedrohung Nr. 37

| | | | |
|---|---|---|---|
| *Nummer* | 38 | | |
| *Beschreibung* | Manipulation der Ausführung durch physikalische Einflüsse (Strahlung, Kurzschlüsse/Überbrücken von Kontakten, etc.) | | |
| *Bedrohung* | Verlust der Vertraulichkeit, Integrität, Verfügbarkeit und Authentizität | | |
| *Kategorie* | Physical Security | | |
| *Angreifer* | Geheimdienste/private Sicherheitsdienste/organisierte Kriminalität | | |
| *Schnittstelle* | Alle innerhalb des ONP fähigen Gerätes | | |
| *Entwickler, Methode* | CP, Brainstorming | | |
| *Threat Agent* | *Value* | *Comment* | |
| *Skill Level* | 1 | Spezialisten im Bereich Seitenkanäle von Hard- und Software, gezielte Manipulation (mysteriöse „Unfälle“), man muss an die Hardware kommen, sehr wenige | |
| *Motive* | 6 | | |
| *Opportunity* | 1 | | |
| *Size* | 1 | | |
| *Vulnerability* | *Value* | *Comment* | |
| *Ease of discovery* | 1 | praktisch unmöglich, theoretisch ausnutzbar, versteckte Schwachstelle, keine Erkennung | |
| *Ease of exploit* | 1 | | |
| *Awareness* | 4 | | |
| *Intrusion* | 9 | | |
| *Likelihood* | 3.00 | LOW | |
| *Technical Impact* | *Value* | *Comment* | |
| *Loss of Confidentiality* | 1 | keine sensiblen Daten enthüllen, wenige Daten stark beschädigt, wenige sekundäre Dienste unterbrochen, vollständig anonym | |
| *Loss of Integrity* | 3 | | |
| *Loss of Availability* | 1 | | |
| *Loss of Accountability* | 9 | | |
| *Technical Impact* | 3.50 | MEDIUM | |
| *Business Impact* | *Value* | *Comment* | |
| *Financial damage* | 1 | weniger als die Kosten zur Behebung, geringer Schaden, keine Verletzung, Daten eines einzigen Individiums | |
| *Reputation damage* | 1 | | |
| *Non-compliance* | 1 | | |
| *Privacy Violation* | 3 | | |
| *Business Impact* | 1.50 | LOW | |
| *Gesamtrisiko* | *technisch* | | Low |
| | *geschäftlich* | | Note |

**Tabelle B.38:** Risk Rating für Bedrohung Nr. 38

| | | | |
|---|---|---|---|
| *Nummer* | 39 | | |
| *Beschreibung* | Manipulation von Sensordaten (z.B. durch Blockieren von Sensoren (GPS im Tunnel) und das Übermitteln von falschen Daten an ein Opfer durch einen GPS-Signal-Simulator) | | |
| *Bedrohung* | Verlust der Integrität | | |
| *Kategorie* | Spoofing Attacks | | |
| *Angreifer* | Sicherheitsbehörden/Geheimdienste/private Sicherheitsdienste/organisierte Kriminalität | | |
| *Schnittstelle* | Sensoren | | |
| *Entwickler, Methode* | CP, Brainstorming (Arktile Prof Spanien, Luxus jacht entführt ) | | |
| *Threat Agent* | *Value* | *Comment* | |
| *Skill Level* | 1 | Spezialisten im Bereich Positionsbestimmung, gezielte Manipulation (mysteriöse „Unfälle“), man muss nah an das Opfer heran, spezielle die Hard- und Software nötig, sehr wenige | |
| *Motive* | 6 | | |
| *Opportunity* | 1 | | |
| *Size* | 1 | | |
| *Vulnerability* | *Value* | *Comment* | |
| *Ease of discovery* | 3 | schwer zu entdecken, schwer ausnutzbar, offensichtliche Schwachstelle, keine Erkennung | |
| *Ease of exploit* | 3 | | |
| *Awareness* | 6 | | |
| *Intrusion* | 9 | | |
| *Likelihood* | 3.75 | MEDIUM | |
| *Technical Impact* | *Value* | *Comment* | |
| *Loss of Confidentiality* | 1 | keine sensiblen Daten enthüllen, viele Daten leicht beschädigt, wenige sekundäre Dienste unterbrochen, vollständig anonym | |
| *Loss of Integrity* | 5 | | |
| *Loss of Availability* | 1 | | |
| *Loss of Accountability* | 9 | | |
| *Technical Impact* | 4.00 | MEDIUM | |
| *Business Impact* | *Value* | *Comment* | |
| *Financial damage* | 7 | signifikante Auswirkungen auf jährlichen Gewinn, geringer Schaden, keine Verletzung, Daten eines einzigen Individiums | |
| *Reputation damage* | 1 | | |
| *Non-compliance* | 1 | | |
| *Privacy Violation* | 3 | | |
| *Business Impact* | 3.00 | LOW | |
| *Gesamtrisiko* | *technisch* | | Medium |
| | *geschäftlich* | | Low |

**Tabelle B.39:** Risk Rating für Bedrohung Nr. 39

| | | | |
|---|---|---|---|
| *Nummer* | 40 | | |
| *Beschreibung* | Hardware-Manipulation der Sensoren/Steuereinheiten (z.B. Kurzschließen von Kontakten, um Zugriffsschutz auf die Software zu deaktivieren, Auflöten einer Debugging Schnittstelle oder Beschädigung der GPS-Antenne, um eine Positionsbestimmung zu verhindern) | | |
| *Bedrohung* | Verlust der Integrität, Nachvollziehbarkeit, Nichtabstreitbarkeit | | |
| *Kategorie* | Physical Security | | |
| *Angreifer* | Geheimdienste/private Sicherheitsdienste/organisierte Kriminalität zur Verhinderung von Spuren | | |
| *Schnittstelle* | Sensoren | | |
| *Entwickler, Methode* | CP, Brainstorming | | |
| *Threat Agent* | *Value* | *Comment* | |
| *Skill Level* | 1 | Person mit Wissen über die Lage der Sensoren und Erfahrungen im Bereich Hardware Modifikation, nötig um Spuren von einer anderen Straftat zu verhindern, Zugang zu Hardware nötig | |
| *Motive* | 6 | | |
| *Opportunity* | 5 | | |
| *Size* | 7 | | |
| *Vulnerability* | *Value* | *Comment* | |
| *Ease of discovery* | 3 | schwer zu entdecken, mittelschwer ausnutzbar, offensichtliche Schwachstelle, keine Erkennung | |
| *Ease of exploit* | 4 | | |
| *Awareness* | 6 | | |
| *Intrusion* | 9 | | |
| *Likelihood* | 5.12 | MEDIUM | |
| *Technical Impact* | *Value* | *Comment* | |
| *Loss of Confidentiality* | 1 | keine sensiblen Daten enthüllen, viele Daten leicht beschädigt, wenige sekundäre Dienste unterbrochen, vollständig anonym | |
| *Loss of Integrity* | 5 | | |
| *Loss of Availability* | 1 | | |
| *Loss of Accountability* | 9 | | |
| *Technical Impact* | 4.00 | MEDIUM | |
| *Business Impact* | *Value* | *Comment* | |
| *Financial damage* | 7 | signifikante Auswirkungen auf jährlichen Gewinn, Verlust von Kunden mittlerer Größe, eindeutige Verletzung, Daten eines einzigen Individuums | |
| *Reputation damage* | 3 | | |
| *Non-compliance* | 5 | | |
| *Privacy Violation* | 3 | | |
| *Business Impact* | 4.50 | MEDIUM | |
| *Gesamtrisiko* | *technisch* | | Medium |
| | *geschäftlich* | | Medium |

**Tabelle B.40:** Risk Rating für Bedrohung Nr. 40

| | | |
|---|---|---|
| *Nummer* | 41 | |
| *Beschreibung* | Software-Manipulation der Sensoren/Steuereinheiten (z.B. Einspielen von Firmware, die Gewährleistungsüberschreitungen nicht aufzeichnet, Betrieb an den Grenzen der Spezifikation, Chiptuning oder falsche Daten melden) | |
| *Bedrohung* | Verlust der Integrität, Nachvollziehbarkeit, Nichtabstreitbarkeit | |
| *Kategorie* | Spoofing Attacks | |
| *Angreifer* | Geheimdienste/private Sicherheitsdienste/organisierte Kriminalität zur Verhinderung von Spuren | |
| *Schnittstelle* | Sensoren | |
| *Entwickler, Methode* | CP, Brainstorming | |
| *Threat Agent* | *Value* | *Comment* |
| *Skill Level* | 1 | Spezialisten im Bereich eingebetteter Systeme, Hardware nahe und Betriebssystemprogrammierung, gezielte Manipulation (mysteriöse „Unfälle“), Nähe zu Opfer, außer Firmware via Netzwerk einspielbar, wenige |
| *Motive* | 6 | |
| *Opportunity* | 7 | |
| *Size* | 3 | |
| *Vulnerability* | *Value* | *Comment* |
| *Ease of discovery* | 3 | schwer zu entdecken, mittelschwer ausnutzbar, offensichtliche Schwachstelle, keine Erkennung |
| *Ease of exploit* | 4 | |
| *Awareness* | 6 | |
| *Intrusion* | 9 | |
| *Likelihood* | 4.88 | MEDIUM |
| *Technical Impact* | *Value* | *Comment* |
| *Loss of Confidentiality* | 1 | keine sensiblen Daten enthüllen, viele Daten leicht beschädigt, wenige sekundäre Dienste unterbrochen, vollständig anonym |
| *Loss of Integrity* | 5 | |
| *Loss of Availability* | 1 | |
| *Loss of Accountability* | 9 | |
| *Technical Impact* | 4.00 | MEDIUM |
| *Business Impact* | *Value* | *Comment* |
| *Financial damage* | 7 | signifikante Auswirkungen auf jährlichen Gewinn, Verlust von Kunden mittlerer Größe, eindeutige Verletzung, Daten eines einzigen Individuums |
| *Reputation damage* | 3 | |
| *Non-compliance* | 5 | |
| *Privacy Violation* | 3 | |
| *Business Impact* | 4.50 | MEDIUM |
| *Gesamtrisiko* | *technisch*<br>*geschäftlich* | Medium<br>Medium |

**Tabelle B.41:** Risk Rating für Bedrohung Nr. 41

| | | |
|---|---|---|
| *Nummer* | 42 | |
| *Beschreibung* | Manipulation von Archivdaten (z.B. vor der Archivierung oder nachträglich, Gewährleistungsüberschreitungen nicht archivieren, Tacho zurückdrehen bzw. Betrieb an den Grenzen der Spezifikation, Gewährleistungsangelegenheiten) | |
| *Bedrohung* | Verlust der Integrität, Nachvollziehbarkeit, Nichtabstreitbarkeit | |
| *Kategorie* | Spoofing Attacks | |
| *Angreifer* | Böswilliger Verkäufer/Händler, organisierte Kriminalität, Geheimdienste | |
| *Schnittstelle* | Log/Archivierung | |
| *Entwickler, Methode* | CP, Brainstorming | |
| *Threat Agent* | *Value* | *Comment* |
| *Skill Level* | 2 | Spezialisten in Log- und Archivierungssysteme, verwischen von Spuren, gute Kenntnis über System nötig, wenige |
| *Motive* | 9 | |
| *Opportunity* | 5 | |
| *Size* | 3 | |
| *Vulnerability* | *Value* | *Comment* |
| *Ease of discovery* | 5 | mäßig schwierig zu entdecken, mittelschwer ausnutzbar, offensichtliche Schwachstelle, dokumentiert aber ohne Kontrolle |
| *Ease of exploit* | 4 | |
| *Awareness* | 6 | |
| *Intrusion* | 8 | |
| *Likelihood* | 5.25 | MEDIUM |
| *Technical Impact* | *Value* | *Comment* |
| *Loss of Confidentiality* | 1 | keine sensiblen Daten enthüllen, viele Daten leicht beschädigt, wenige sekundäre Dienste unterbrochen, vollständig anonym |
| *Loss of Integrity* | 5 | |
| *Loss of Availability* | 1 | |
| *Loss of Accountability* | 9 | |
| *Technical Impact* | 4.00 | MEDIUM |
| *Business Impact* | *Value* | *Comment* |
| *Financial damage* | 3 | geringe Auswirkungen auf jährlichen Gewinn, Verlust von größeren Kunden, hochkarätige Verletzung, keine sensiblen Daten |
| *Reputation damage* | 4 | |
| *Non-compliance* | 7 | |
| *Privacy Violation* | 1 | |
| *Business Impact* | 3.75 | MEDIUM |
| *Gesamtrisiko* | *technisch*<br>*geschäftlich* | Medium<br>Medium |

**Tabelle B.42:** Risk Rating für Bedrohung Nr. 42

| | | |
|---|---|---|
| *Nummer* | 43 | |
| *Beschreibung* | Falsche Informationen dem Dienst (Empfehlungssystem) unterschieben (z.B. Senden von falschen Positionsinformationen) (ad-hoc) | |
| *Bedrohung* | Verlust der Integrität | |
| *Kategorie* | Spoofing Attacks | |
| *Angreifer* | Böswilliger Nutzer, der Softwareentwickler im Bereich Kommunikationstechnologien ist | |
| *Schnittstelle* | Dienste | |
| *Entwickler, Methode* | CP, Brainstorming | |
| *Threat Agent* | *Value* | *Comment* |
| *Skill Level* | 3 | Software Entwickler, eigene Software oder Manipulation vorhandener nötig, muss vor Ort sein, wenige, außer Schaden am Dienst, Beeinträchtigung anderer Teilnehmer, keinen Nutzen, wenige |
| *Motive* | 4 | |
| *Opportunity* | 5 | |
| *Size* | 2 | |
| *Vulnerability* | *Value* | *Comment* |
| *Ease of discovery* | 7 | einfach zu entdecken, schwer ausnutzbar, offensichtliche Schwachstelle, dokumentiert aber ohne Kontrolle |
| *Ease of exploit* | 3 | |
| *Awareness* | 6 | |
| *Intrusion* | 8 | |
| *Likelihood* | 4.75 | MEDIUM |
| *Technical Impact* | *Value* | *Comment* |
| *Loss of Confidentiality* | 1 | keine sensiblen Daten enthüllen, viele Daten leicht beschädigt, wenige sekundäre Dienste unterbrochen, möglicherweise nachvollziehbar |
| *Loss of Integrity* | 5 | |
| *Loss of Availability* | 1 | |
| *Loss of Accountability* | 7 | |
| *Technical Impact* | 3.50 | MEDIUM |
| *Business Impact* | *Value* | *Comment* |
| *Financial damage* | 7 | signifikante Auswirkungen auf jährlichen Gewinn, Beschädigung des Markennamens, keine Verletzung, keine sensiblen Daten |
| *Reputation damage* | 9 | |
| *Non-compliance* | 1 | |
| *Privacy Violation* | 1 | |
| *Business Impact* | 4.50 | MEDIUM |
| *Gesamtrisiko* | *technisch*<br>*geschäftlich* | Medium<br>Medium |

**Tabelle B.43:** Risk Rating für Bedrohung Nr. 43

| | | | |
|---|---|---|---|
| *Nummer* | 44 | | |
| *Beschreibung* | Falsche Informationen dem Dienst (Empfehlungssystem) unterschieben (z.B. falsche Fahrpläne via DNS-Hijacking des MVV/MVG-Portals) (backbone) | | |
| *Bedrohung* | Verlust der Integrität | | |
| *Kategorie* | Spoofing Attacks | | |
| *Angreifer* | Konkurrent des Dienstes | | |
| *Schnittstelle* | Dienste | | |
| *Entwickler, Methode* | CP, Brainstorming | | |
| *Threat Agent* | *Value* | *Comment* | |
| *Skill Level* | 2 | Spezialisten im Bereich Netzwerkkommunikation, mögliche Belohnung durch Diskreditierung des Dienstes, Angriff eines weiteren Systems außerhalb von unserem, wenige Konkurrenten, wenige Ziele | |
| *Motive* | 4 | | |
| *Opportunity* | 3 | | |
| *Size* | 2 | | |
| *Vulnerability* | *Value* | *Comment* | |
| *Ease of discovery* | 5 | mäßig schwierig zu entdecken, schwer ausnutzbar, versteckte Schwachstelle, keine Erkennung | |
| *Ease of exploit* | 3 | | |
| *Awareness* | 4 | | |
| *Intrusion* | 9 | | |
| *Likelihood* | 4.00 | MEDIUM | |
| *Technical Impact* | *Value* | *Comment* | |
| *Loss of Confidentiality* | 1 | keine sensiblen Daten enthüllen, viele Daten leicht beschädigt, wenige sekundäre Dienste unterbrochen, möglicherweise nachvollziehbar | |
| *Loss of Integrity* | 5 | | |
| *Loss of Availability* | 1 | | |
| *Loss of Accountability* | 7 | | |
| *Technical Impact* | 3.50 | MEDIUM | |
| *Business Impact* | *Value* | *Comment* | |
| *Financial damage* | 7 | signifikante Auswirkungen auf jährlichen Gewinn, Beschädigung des Markennamens, keine Verletzung, keine sensiblen Daten | |
| *Reputation damage* | 9 | | |
| *Non-compliance* | 1 | | |
| *Privacy Violation* | 1 | | |
| *Business Impact* | 4.50 | MEDIUM | |
| *Gesamtrisiko* | *technisch* | Medium | |
| | *geschäftlich* | Medium | |

**Tabelle B.44:** Risk Rating für Bedrohung Nr. 44

| | | |
|---|---|---|
| *Nummer* | 45 | |
| *Beschreibung* | Manipulierte Aggregation der Daten (ad-hoc) | |
| *Bedrohung* | Wertverlust der aggregierten Daten durch mangelnde Integrität | |
| *Kategorie* | Spoofing Attacks | |
| *Angreifer* | Konkurrent des Dienstes | |
| *Schnittstelle* | Dienste | |
| *Entwickler, Methode* | CP, Brainstorming | |
| *Threat Agent* | *Value* | *Comment* |
| *Skill Level* | 3 | Softwareentwickler, eigene Software oder Manipulation vorhandener nötig, muss vor Ort sein, wenige, außer Schaden am Dienst, Beeinträchtigung anderer Teilnehmer, keinen Nutzen, wenige |
| *Motive* | 4 | |
| *Opportunity* | 5 | |
| *Size* | 2 | |
| *Vulnerability* | *Value* | *Comment* |
| *Ease of discovery* | 5 | mäßig schwierig zu entdecken, schwer ausnutzbar, versteckte Schwachstelle, dokumentiert aber ohne Kontrolle |
| *Ease of exploit* | 3 | |
| *Awareness* | 4 | |
| *Intrusion* | 8 | |
| *Likelihood* | 4.25 | MEDIUM |
| *Technical Impact* | *Value* | *Comment* |
| *Loss of Confidentiality* | 1 | keine sensiblen Daten enthüllen, viele Daten leicht beschädigt, wenige sekundäre Dienste unterbrochen, möglicherweise nachvollziehbar |
| *Loss of Integrity* | 5 | |
| *Loss of Availability* | 1 | |
| *Loss of Accountability* | 7 | |
| *Technical Impact* | 3.50 | MEDIUM |
| *Business Impact* | *Value* | *Comment* |
| *Financial damage* | 7 | signifikante Auswirkungen auf jährlichen Gewinn, Beschädigung des Markennamens, keine Verletzung, keine sensiblen Daten |
| *Reputation damage* | 9 | |
| *Non-compliance* | 1 | |
| *Privacy Violation* | 1 | |
| *Business Impact* | 4.50 | MEDIUM |
| *Gesamtrisiko* | *technisch*<br>*geschäftlich* | Medium<br>Medium |

**Tabelle B.45:** Risk Rating für Bedrohung Nr. 45

| | | | |
|---|---|---|---|
| *Nummer* | 46 | | |
| *Beschreibung* | Gezielt Dienste mit hohen Datenaufkommen/Ressourcenverbrauch anbieten, um dadurch dem Netzwerk und anderen Diensten zu schaden | | |
| *Bedrohung* | Dienstleistungsausfall durch Verbrauchen von Ressourcen | | |
| *Kategorie* | Denial of Service (DoS) | | |
| *Angreifer* | Konkurrent des Dienstes/Konkurrenten im Bereich Fahrzeugkommunikation | | |
| *Schnittstelle* | Dienste | | |
| *Entwickler, Methode* | CP, Brainstorming | | |
| *Threat Agent* | *Value* | *Comment* | |
| *Skill Level* | 3 | Softwareentwickler im Bereich Mobileanwendungsentwicklung und Webservices, Anwendung und Dienst müssen verfügbar sein auf den genutzten Plattformen, geringe Anzahl von Konkurrenten | |
| *Motive* | 4 | | |
| *Opportunity* | 4 | | |
| *Size* | 2 | | |
| *Vulnerability* | *Value* | *Comment* | |
| *Ease of discovery* | 5 | mäßig schwierig zu entdecken, schwer ausnutzbar, offensichtliche Schwachstelle, keine Erkennung | |
| *Ease of exploit* | 3 | | |
| *Awareness* | 6 | | |
| *Intrusion* | 9 | | |
| *Likelihood* | 4.50 | MEDIUM | |
| *Technical Impact* | *Value* | *Comment* | |
| *Loss of Confidentiality* | 1 | keine sensiblen Daten enthüllen, wenige Daten leicht beschädigt, umfassend primäre Dienste unterbrochen, vollständig nachvollziehbar | |
| *Loss of Integrity* | 1 | | |
| *Loss of Availability* | 7 | | |
| *Loss of Accountability* | 1 | | |
| *Technical Impact* | 2.50 | LOW | |
| *Business Impact* | *Value* | *Comment* | |
| *Financial damage* | 7 | signifikante Auswirkungen auf jährlichen Gewinn, Unterstellung von böswilligem Handeln, keine Verletzung, keine sensiblen Daten | |
| *Reputation damage* | 5 | | |
| *Non-compliance* | 1 | | |
| *Privacy Violation* | 1 | | |
| *Business Impact* | 3.50 | MEDIUM | |
| *Gesamtrisiko* | *technisch* | | Low |
| | *geschäftlich* | | Medium |

**Tabelle B.46:** Risk Rating für Bedrohung Nr. 46

| | | | |
|---|---|---|---|
| *Nummer* | 47 | | |
| *Beschreibung* | Falsche Informationen aussenden, um Teilnehmer zu beeinflussen und möglicherweise schädliche Reaktionen der Nutzer oder von sicherheitskritischen Anwendungen zu provozieren | | |
| *Bedrohung* | Schaden am Leib und Leben der Nutzer | | |
| *Kategorie* | Spoofing Attacks | | |
| *Angreifer* | Softwareentwickler im Bereich Kommunikationstechnologien | | |
| *Schnittstelle* | ONP Message Filter | | |
| *Entwickler, Methode* | CP, Brainstorming | | |
| *Threat Agent* | *Value* | *Comment* | |
| *Skill Level* | 3 | Softwareentwickler, Unfälle provozieren bis hin zu Sabotage, vor Ort sein, eigene Software nötig, wenige | |
| *Motive* | 9 | | |
| *Opportunity* | 5 | | |
| *Size* | 2 | | |
| *Vulnerability* | *Value* | *Comment* | |
| *Ease of discovery* | 7 | einfach zu entdecken, schwer ausnutzbar, offensichtliche Schwachstelle, dokumentiert aber ohne Kontrolle | |
| *Ease of exploit* | 3 | | |
| *Awareness* | 6 | | |
| *Intrusion* | 8 | | |
| *Likelihood* | 5.38 | MEDIUM | |
| *Technical Impact* | *Value* | *Comment* | |
| *Loss of Confidentiality* | 1 | keine sensiblen Daten enthüllen, alle Daten vollständig beschädigt, wenige primäre Dienste unterbrochen, vollständig anonym | |
| *Loss of Integrity* | 9 | | |
| *Loss of Availability* | 3 | | |
| *Loss of Accountability* | 9 | | |
| *Technical Impact* | 5.50 | MEDIUM | |
| *Business Impact* | *Value* | *Comment* | |
| *Financial damage* | 9 | drohender Bankrott, Beschädigung des Markennamens, hochkarätige Verletzung, keine sensiblen Daten | |
| *Reputation damage* | 9 | | |
| *Non-compliance* | 7 | | |
| *Privacy Violation* | 1 | | |
| *Business Impact* | 6.50 | HIGH | |
| *Gesamtrisiko* | *technisch* | | Medium |
| | *geschäftlich* | | High |

**Tabelle B.47:** Risk Rating für Bedrohung Nr. 47

| | | | |
|---|---|---|---|
| *Nummer* | 48 | | |
| *Beschreibung* | Durch manipulierte Softwarekomponenten andere Softwarekomponenten (z.B. Anwendungen) beeinflussen | | |
| *Bedrohung* | Verlust der Integrität durch Verfälschung von Informationen, Verlust der Vertraulichkeit | | |
| *Kategorie* | Spoofing Attacks | | |
| *Angreifer* | Softwareentwickler im Bereich Mobile-Anwendungsentwicklung | | |
| *Schnittstelle* | Alle innerhalb des ONP fähigen Gerätes | | |
| *Entwickler, Methode* | CP, Brainstorming | | |
| *Threat Agent* | *Value* | *Comment* | |
| *Skill Level* | 2 | Entwickler für Mobileanwendungen, Auslösen von Fehlfunktionen durch manipulierte Daten, eigene Anwendung/Schwachstelle in Anwendung/Schwachstelle im Betriebssystem, wenige | |
| *Motive* | 4 | | |
| *Opportunity* | 6 | | |
| *Size* | 3 | | |
| *Vulnerability* | *Value* | *Comment* | |
| *Ease of discovery* | 5 | mäßig schwierig zu entdecken, mittelschwer ausnutzbar, offensichtliche Schwachstelle, dokumentiert aber ohne Kontrolle | |
| *Ease of exploit* | 4 | | |
| *Awareness* | 6 | | |
| *Intrusion* | 8 | | |
| *Likelihood* | 4.75 | MEDIUM | |
| *Technical Impact* | *Value* | *Comment* | |
| *Loss of Confidentiality* | 1 | keine sensiblen Daten enthüllen, alle Daten vollständig beschädigt, wenige primäre Dienste unterbrochen, vollständig anonym | |
| *Loss of Integrity* | 9 | | |
| *Loss of Availability* | 3 | | |
| *Loss of Accountability* | 9 | | |
| *Technical Impact* | 5.50 | MEDIUM | |
| *Business Impact* | *Value* | *Comment* | |
| *Financial damage* | 3 | geringe Auswirkungen auf jährlichen Gewinn, Beschädigung des Markennamens, hochkarätige Verletzung, keine sensiblen Daten | |
| *Reputation damage* | 9 | | |
| *Non-compliance* | 7 | | |
| *Privacy Violation* | 1 | | |
| *Business Impact* | 5.00 | MEDIUM | |
| *Gesamtrisiko* | *technisch* | | Medium |
| | *geschäftlich* | | Medium |

**Tabelle B.48:** Risk Rating für Bedrohung Nr. 48

| | | |
|---|---|---|
| *Nummer* | 49 | |
| *Beschreibung* | Anwendungen aus dem AppStore stehlen (z.B. Anwendung einmal bezahlen und dann von Gerät zu Gerät kopieren) | |
| *Bedrohung* | Kostenlose Nutzung von kostenpflichtigen Diensten | |
| *Kategorie* | Berechtigungsmissbrauch | |
| *Angreifer* | Böswilliger Nutzer mit geringen technischen Fähigkeiten | |
| *Schnittstelle* | AS-WS2OR | |
| *Entwickler, Methode* | CP, Attack Trees | |
| *Threat Agent* | *Value* | *Comment* |
| *Skill Level* | 7 | Geringe Fähigkeiten nötig, Anwendung einmal kaufen und auf allen Geräten nutzen, man muss registriert sein und einmal bezahlen, also ein Nutzer sein |
| *Motive* | 9 | |
| *Opportunity* | 7 | |
| *Size* | 6 | |
| *Vulnerability* | *Value* | *Comment* |
| *Ease of discovery* | 7 | einfach zu entdecken, schwer ausnutzbar, offensichtliche Schwachstelle, nur durch aktive Erkennung |
| *Ease of exploit* | 3 | |
| *Awareness* | 6 | |
| *Intrusion* | 1 | |
| *Likelihood* | 5.75 | MEDIUM |
| *Technical Impact* | *Value* | *Comment* |
| *Loss of Confidentiality* | 1 | keine sensiblen Daten enthüllen, wenige Daten leicht beschädigt, wenige sekundäre Dienste unterbrochen, möglicherweise nachvollziehbar |
| *Loss of Integrity* | 1 | |
| *Loss of Availability* | 1 | |
| *Loss of Accountability* | 7 | |
| *Technical Impact* | 2.50 | LOW |
| *Business Impact* | *Value* | *Comment* |
| *Financial damage* | 3 | geringe Auswirkungen auf jährlichen Gewinn, geringer Schaden, keine Verletzung, keine sensiblen Daten |
| *Reputation damage* | 1 | |
| *Non-compliance* | 1 | |
| *Privacy Violation* | 1 | |
| *Business Impact* | 1.50 | LOW |
| *Gesamtrisiko* | *technisch* | Low |
| | *geschäftlich* | Low |

**Tabelle B.49:** Risk Rating für Bedrohung Nr. 49

# C. Quellcode der Simulation

```
#!/usr/bin/python3
# -*- coding: utf-8 -*-

"""
simulation-repsys.py - Reputation System Simulation for CPS (M/VANETs)

Scenario:
  This simulation shows how to interconnect simulation model elements
  together using :class:`~simpy.resources.store.Store` for one-to-one,
  and many-to-one asynchronous processes. For one-to-many a simple
  BroadCastPipe class is constructed from Store.

Simulation By:
  Christoph Ponikwar

"""
import os
import random
import simpy
import csv
from datetime import datetime
import numpy as np

# RANDOM_SEED = 42
# RANDOM_SEED = 1337
RANDOM_SEED = np.pi
# SIM_TIME = 100000
# SIM_TIME = 1000
RANDOM_SEED = np.pi
# SIM_TIME = 1000000
# SIM_TIME = 100000
# SIM_TIME = 1000
SIM_TIME = 100
# SIM_TIME = 1
RESULTS = []
RESULTS_GOOD_GEN = []
RESULTS_BAD_GEN = []
RESULTS_RND_GEN = []
RESULTS_RND_GEN_FIFTY = []


```

```
class Message(object):
    """ This class defines :class:`Message`, a representation of a PDU (packet
    data unit), with different properties, i.e. meta data and payload
    information. Properties defined as class variables, which should be used to
    define the properties of an instance of this class, when instantiated.

    """
    # meta Layer 1
    meta_l1_cellular = "cellular"
    meta_l1_ad_hoc = "ad-hoc"
    # List for Rnd
    meta_l1_list = [meta_l1_cellular, meta_l1_ad_hoc]
    # meta Layer 1 signal strength
    meta_l1_ad_hoc_sig_str_na = 0
    meta_l1_ad_hoc_sig_str_same = 1
    meta_l1_ad_hoc_sig_str_half = 5
    meta_l1_ad_hoc_sig_str_min = 10
    # List for Rnd
    meta_l1_ad_hoc_list = [meta_l1_ad_hoc_sig_str_na,
                           meta_l1_ad_hoc_sig_str_same,
                           meta_l1_ad_hoc_sig_str_half,
                           meta_l1_ad_hoc_sig_str_min]

    # meta Layer 2 Type
    meta_l2_cellular_type_na = "central"
    meta_l2_ad_hoc_type_service = "service"
    meta_l2_ad_hoc_type_safety = "safety"
    # List for Rmd
    meta_l2_type_list = [meta_l2_cellular_type_na,
                         meta_l2_ad_hoc_type_service,
                         meta_l2_ad_hoc_type_safety]

    # meta Layer 2 ID
    meta_l2_cellular_mac_na = "MAC NA"
    meta_l2_ad_hoc_mac_valid = "valid MAC"
    # 00:00:00:00:00:00 / ff:ff:ff:ff:ff:ff
    meta_l2_ad_hoc_mac_invalid = "invalid MAC"
    # List for Rmd
    meta_l2_mac_list = [meta_l2_cellular_mac_na,
                        meta_l2_ad_hoc_mac_valid,
                        meta_l2_ad_hoc_mac_invalid]

    # meta Layer 3
    meta_l3_cellular_addr_valid = "valid IPv6/v4"
    meta_l3_cellular_addr_invalid = "invalid IPv6/v4"
    meta_l3_ad_hoc_addr_valid = "valid addr"
    meta_l3_ad_hoc_addr_invalid = "invalid addr"
    # List for Rmd
    meta_l3_addr_list = [meta_l3_cellular_addr_valid,
                         meta_l3_cellular_addr_invalid,
                         meta_l3_ad_hoc_addr_valid,
                         meta_l3_ad_hoc_addr_invalid]

    # info role
```

```
    info_role_central_service = "service role"
    info_role_rsu = "RSU"
    info_role_vehicle = "vehicle"
    info_role_mobile = "mobile"
    info_role_emergency_vehicle = "emergency vehicle"
    # List for Rmd
    info_role_list = [info_role_central_service,
                      info_role_rsu,
                      info_role_vehicle,
                      info_role_mobile,
                      info_role_emergency_vehicle]

    # info identity
    info_id_weak_valid = "valid weak id"
    info_id_weak_invalid = "invalid weak id"
    # decentral (rsu) signed pki (id-based crypto)
    info_id_mid_valid = "valid mid id"
    info_id_mid_invalid = "invalid mid id"
    # central signed pki ( special one or TLS Certs)
    info_id_strong_valid = "valid strong id"
    info_id_strong_invalid = "invalid strong id"
    # List for Rmd
    info_id_list = [info_id_weak_valid,
                    info_id_weak_invalid,
                    info_id_mid_valid,
                    info_id_mid_invalid,
                    info_id_strong_valid,
                    info_id_strong_invalid]

    # info position
    info_position_ad_hoc_range_na = meta_l1_ad_hoc_sig_str_na
    info_position_ad_hoc_range_same = meta_l1_ad_hoc_sig_str_same
    info_position_ad_hoc_range_half = meta_l1_ad_hoc_sig_str_half
    info_position_ad_hoc_range_max = meta_l1_ad_hoc_sig_str_min
    # List for Rmd
    info_position_list = [info_position_ad_hoc_range_na,
                          info_position_ad_hoc_range_same,
                          info_position_ad_hoc_range_half,
                          info_position_ad_hoc_range_max]

    # three basic use cases: ($node = rsu, vehicle, mobile)
    # - $node sends service infos
    # - $node sends safety infos
    # - infos from central service
    info_type_service = meta_l2_ad_hoc_type_service
    info_type_safety = meta_l2_ad_hoc_type_safety
    info_type_central = meta_l2_cellular_type_na
    # List for Rmd
    info_type_list = [info_type_service,
                      info_type_safety,
                      info_type_central]

    # State of information, valid via verfication or invalid or not verifiable
    info_state_valid = "info valid"
```

```
    info_state_invalid = "info invalid"
    info_state_not_verifiable = "info not verifiable"
    # List for Rmd
    info_stat_list = [info_state_valid,
                      info_state_invalid,
                      info_state_not_verifiable]

    msg_trust_good = "trustworthy"
    msg_trust_bad = "malicious"

    def __init__(self, time, gen_id, meta_l1, meta_l1_sig_str,
                 meta_l2_type, meta_l2_id, meta_l3, info_role,
                 info_id, info_position, info_type, info_state,
                 msg_str):
        self.time = time
        self.gen_id = gen_id
        self.meta_l1 = meta_l1
        self.meta_l1_sig_str = meta_l1_sig_str
        self.meta_l2_type = meta_l2_type
        self.meta_l2_id = meta_l2_id
        self.meta_l3 = meta_l3
        self.info_role = info_role
        self.info_id = info_id
        self.info_position = info_position
        self.info_type = info_type
        self.info_state = info_state
        self.msg_str = msg_str

        if (self.meta_l2_id == Message.meta_l2_ad_hoc_mac_invalid or
           self.meta_l3 == Message.meta_l3_ad_hoc_addr_invalid or
           self.meta_l3 == Message.meta_l3_cellular_addr_invalid or
           self.info_id == Message.info_id_weak_invalid or
           self.info_id == Message.info_id_mid_invalid or
           self.info_id == Message.info_id_strong_invalid or
           self.info_state == Message.info_state_invalid or
           self.meta_l1_sig_str != self.info_position or
           self.meta_l2_type != self.info_type):
            self.msg_trust = Message.msg_trust_bad
        else:
            self.msg_trust = Message.msg_trust_good


class BroadcastPipe(object):
    """ This class defines :class:`BroadcastPipe`, a broadcast pipe that allows
    one process (producer) to send messages to many processes (consumers).

    This simulates the broadcast nature of communication in M/VANETs.

    This wraps some instances of :class:`~simpy.resources.store.Store`, which
    are created each time the :meth:`get_msg_out_conn()` is invoked.
    Each :class:`~simpy.resources.store.Store` instances acts as a
    buffer where messages are queued up by the producers for the consumers to
    pick them up. This also allows asynchronous behaviour for consumers and
    producers. The :class:`~simpy.resources.store.Store` class can take
```

```
    arbitrary values to store. When using the :meth:`put_msg(msg)` every
    instance of :class:`~simpy.resources.store.Store` gets a new message added
    to it.


    The parameters are used to create a new
    :class:`~simpy.resources.store.Store` instance each time
    :meth:`get_msg_out_conn()` is called.

    """
    def __init__(self, env, capacity=simpy.core.Infinity):
        self.env = env
        self.capacity = capacity
        self.pipes = []

    # `put` is used as method name, because the default Store has this method,
    # too. Which allows the BroadcastPipe to be used interchangeable with a
    # simple Store.
    def put(self, msg):
        """Broadcast a *msg* to all receivers."""
        if not self.pipes:
            raise RuntimeError('There are no output pipes,' +
                               'no consumers.')
        events = [store.put(msg) for store in self.pipes]
        return self.env.all_of(events)  # Condition event for all "events"

    def get_msg_out_conn(self):
        """Get a new output connection for this broadcast pipe.

        The return value is a :class:`~simpy.resources.store.Store`.

        """
        pipe = simpy.Store(self.env, capacity=self.capacity)
        self.pipes.append(pipe)
        return pipe


# Generators start here!
def get_random_message_helper(name, env):
    """Helper method for generating a random message."""
    msg = Message(env.now, '%s' % (name),
                  random.choice(Message.meta_l1_list),
                  random.choice(Message.meta_l1_ad_hoc_list),
                  random.choice(Message.meta_l2_type_list),
                  random.choice(Message.meta_l2_mac_list),
                  random.choice(Message.meta_l3_addr_list),
                  random.choice(Message.info_role_list),
                  random.choice(Message.info_id_list),
                  random.choice(Message.info_position_list),
                  random.choice(Message.info_type_list),
                  random.choice(Message.info_stat_list),
                  ' says hello at %d' % (env.now))
    return msg
```

```

def msg_gen_random(name, env, out_pipe):
    """A process which randomly generates messages.

    The parameters are used to set the name of the generator, environment and
    output pipeline that acts as a :class:`~simpy.resources.store.Store`
    instance when used :meth:`~simpy.resources.store.Store.put()` is called.

    """
    while True:
        msg = get_random_message_helper(name, env)
        # put is a predefined method of stores so it works with BroadcastPipe
        # or with a default store
        out_pipe.put(msg)

        # wait for next transmission, i.e. every step in the simulation
        yield env.timeout(1)


def msg_gen_random_good(name, env, out_pipe):
    """A process which randomly generates honest messages.

    The parameters are used to set the name of the generator, environment and
    output pipeline that acts as a :class:`~simpy.resources.store.Store`
    instance when used :meth:`~simpy.resources.store.Store.put()` is called.

    """
    while True:
        msg = get_random_message_helper(name, env)
        if msg.msg_trust == Message.msg_trust_good:
            # put is a predefined method of stores so it works with
            # BroadcastPipe or with a default store
            out_pipe.put(msg)

            # wait for next transmission, i.e. every step in the simulation
            yield env.timeout(1)


def msg_gen_random_bad(name, env, out_pipe):
    """A process which randomly generates faulty messages.

    The parameters are used to set the name of the generator, environment and
    output pipeline that acts as a :class:`~simpy.resources.store.Store`
    instance when used :meth:`~simpy.resources.store.Store.put()` is called.

    """
    while True:
        msg = get_random_message_helper(name, env)
        if msg.msg_trust == Message.msg_trust_bad:
            # put is a predefined method of stores so it works with
            # BroadcastPipe or with a default store
            out_pipe.put(msg)

            # wait for next transmission, i.e. every step in the simulation
```

```
                yield env.timeout(1)


def msg_gen_random_fifty_fifty(name, env, out_pipe):
    """A process which randomly generates messages, with a 50/50 chance of good
    and bad Messages.

    The parameters are used to set the name of the generator, environment and
    output pipeline that acts as a :class:`~simpy.resources.store.Store`
    instance when used :meth:`~simpy.resources.store.Store.put()` is called.

    """
    while True:
        msg = get_random_message_helper(name, env)
        if random.choice([True, False]):
            while msg.msg_trust != Message.msg_trust_good:
                msg = get_random_message_helper(name, env)
            # put is a predefined method of stores so it works with
            # BroadcastPipe or with a default store
            out_pipe.put(msg)

            # wait for next transmission, i.e. every step in the simulation
            yield env.timeout(1)

        else:
            while msg.msg_trust != Message.msg_trust_bad:
                msg = get_random_message_helper(name, env)
            # put is a predefined method of stores so it works with
            # BroadcastPipe or with a default store
            out_pipe.put(msg)

            # wait for next transmission, i.e. every step in the simulation
            yield env.timeout(1)


def msg_gen_all_msg(name, env, out_pipe):
    """A process which generates all possible messages.

    The parameters are used to set the name of the generator, environment and
    output pipeline that acts as a :class:`~simpy.resources.store.Store`
    instance when used :meth:`~simpy.resources.store.Store.put()` is called.

    """
    for a in Message.meta_l1_list:
        for b in Message.meta_l1_ad_hoc_list:
            for c in Message.meta_l2_type_list:
                for d in Message.meta_l2_mac_list:
                    for e in Message.meta_l3_addr_list:
                        for f in Message.info_role_list:
                            for g in Message.info_id_list:
                                for h in Message.info_position_list:
                                    for i in Message.info_type_list:
                                        for j in Message.info_stat_list:
                                            msg = Message(env.now,
```

```
                                                                  '%s' % (name),
                                                                  a, b, c, d, e,
                                                                  f, g, h, i, j,
                                                                  ' says hi')
                                                    out_pipe.put(msg)
    # when generated
    yield env.timeout(1)


# Generators end here!


# Rputation System starts here!
def calulate_trust_score(msg):
    """
    The main reputation system function that calculates the overall trust score
    depending on the different steps in the trust assessment pipeline:
    # Pipeline:
    # \_/=> Context / Kontext
    #   \_/=> Role / Rollen
    #     \_/=> Identity / Identität
    #       \_/=> Experience / Erfahrung
    #         \_/=> Information
    # ======/_/=> Trust Score

    """
    trust_score = 0
    # three basic use cases for data dissemination:
    # ($node = rsu, vehicle, mobile)
    # - $node sends service infos
    # - $node sends safety infos
    # - infos from central service

    # Context
    context_dict = dict()
    # meta Layer 1
    context_dict[Message.meta_l1_cellular] = 0.75
    context_dict[Message.meta_l1_ad_hoc] = 0.5
    # meta Layer 2 Type
    context_dict[Message.meta_l2_cellular_type_na] = 0.5
    context_dict[Message.meta_l2_ad_hoc_type_service] = 0.5
    context_dict[Message.meta_l2_ad_hoc_type_safety] = 0.75

    trust_score_context = (context_dict[msg.meta_l1] +
                           context_dict[msg.meta_l2_type]) / 2

    # Role
    trust_score_role = 0
    role_dict = dict()

    role_central_dict = dict()
    role_rsu_dict = dict()
    role_emergency_dict = dict()
    role_vehicle_dict = dict()
```

```
    role_mobile_dict = dict()

    role_dict[Message.info_role_central_service] = role_central_dict
    role_dict[Message.info_role_rsu] = role_rsu_dict
    role_dict[Message.info_role_emergency_vehicle] = role_emergency_dict
    role_dict[Message.info_role_vehicle] = role_vehicle_dict
    role_dict[Message.info_role_mobile] = role_mobile_dict

    # central service role
    role_central_dict[Message.info_id_strong_valid] = 1
    role_central_dict[Message.info_id_mid_valid] = 0.5
    role_central_dict[Message.info_id_weak_valid] = 0.25
    role_central_dict[Message.info_id_strong_invalid] = 0
    role_central_dict[Message.info_id_mid_invalid] = 0
    role_central_dict[Message.info_id_weak_invalid] = 0

    # RSU role
    role_rsu_dict[Message.info_id_strong_valid] = 1
    role_rsu_dict[Message.info_id_mid_valid] = 0.75
    role_rsu_dict[Message.info_id_weak_valid] = 0
    role_rsu_dict[Message.info_id_strong_invalid] = 0
    role_rsu_dict[Message.info_id_mid_invalid] = 0
    role_rsu_dict[Message.info_id_weak_invalid] = 0

    # Emergency Vehicle role
    role_emergency_dict[Message.info_id_strong_valid] = 1
    role_emergency_dict[Message.info_id_mid_valid] = 0.75
    role_emergency_dict[Message.info_id_weak_valid] = 0
    role_emergency_dict[Message.info_id_strong_invalid] = 0
    role_emergency_dict[Message.info_id_mid_invalid] = 0
    role_emergency_dict[Message.info_id_weak_invalid] = 0

    # vehicle role
    role_vehicle_dict[Message.info_id_strong_valid] = 1
    role_vehicle_dict[Message.info_id_mid_valid] = 0.75
    role_vehicle_dict[Message.info_id_weak_valid] = 0.25
    role_vehicle_dict[Message.info_id_strong_invalid] = 0
    role_vehicle_dict[Message.info_id_mid_invalid] = 0
    role_vehicle_dict[Message.info_id_weak_invalid] = 0

    # mobile role
    role_mobile_dict[Message.info_id_strong_valid] = 1
    role_mobile_dict[Message.info_id_mid_valid] = 0.5
    role_mobile_dict[Message.info_id_weak_valid] = 0.25
    role_mobile_dict[Message.info_id_strong_invalid] = 0
    role_mobile_dict[Message.info_id_mid_invalid] = 0
    role_mobile_dict[Message.info_id_weak_invalid] = 0

    trust_score_role = role_dict[msg.info_role][msg.info_id]

    # Identity
    identiy_dict = dict()
    # meta Layer 2 ID
    identiy_dict[Message.meta_l2_cellular_mac_na] = 0.5
```

```
    identiy_dict[Message.meta_l2_ad_hoc_mac_valid] = 1
    # 00:00:00:00:00:00 / ff:ff:ff:ff:ff:ff
    identiy_dict[Message.meta_l2_ad_hoc_mac_invalid] = 0
    # meta Layer 3
    identiy_dict[Message.meta_l3_cellular_addr_valid] = 1
    identiy_dict[Message.meta_l3_cellular_addr_invalid] = 0
    identiy_dict[Message.meta_l3_ad_hoc_addr_valid] = 0.75
    identiy_dict[Message.meta_l3_ad_hoc_addr_invalid] = 0
    # info identity
    identiy_dict[Message.info_id_weak_valid] = 0.5
    identiy_dict[Message.info_id_weak_invalid] = 0
    # decentral (rsu) signed pki (id-based crypto)
    identiy_dict[Message.info_id_mid_valid] = 0.75
    identiy_dict[Message.info_id_mid_invalid] = 0
    # central signed pki ( special one or TLS Certs)
    identiy_dict[Message.info_id_strong_valid] = 1
    identiy_dict[Message.info_id_strong_invalid] = 0

    # get partial trust values
    part_ts_meta_l2_id = identiy_dict[msg.meta_l2_id]
    part_ts_meta_l3 = identiy_dict[msg.meta_l3]
    part_ts_info_id = identiy_dict[msg.info_id]

    identity_weight = 0
    if part_ts_meta_l2_id == 0:
        identity_weight += 1
    if part_ts_meta_l3 == 0:
        identity_weight += 1
    if part_ts_info_id == 0:
        identity_weight += 1

    id_weight = (3 + identity_weight)
    trust_score_identity = (identiy_dict[msg.meta_l2_id] +
                            identiy_dict[msg.meta_l3] +
                            identiy_dict[msg.info_id]) / id_weight

    # Experience
    # not implemented, better suited for a simulation that supports a more
    # detailed model for the communication specialities in M/VANETs.
    trust_score_experience = 'NA'

    # Information
    info_dict = dict()
    # info position
    # check if signal strength is in line with the measured
    positional_trust = 0
    if msg.info_position == msg.meta_l1_sig_str:
        positional_trust = 1

    # check if message type of layer 2 is consistent with information type
    type_consistency_trust = 0
    if(msg.meta_l2_type == msg.info_type):
        type_consistency_trust = 1

```

```
    # trust scores for the state of an information
    info_dict[Message.info_state_valid] = 1
    info_dict[Message.info_state_not_verifiable] = 0.5
    info_dict[Message.info_state_invalid] = 0

    # get partial trust values
    part_ts_info_state = info_dict[msg.info_state]

    info_weight = 0
    if part_ts_info_state == 0:
        info_weight += 1
    if type_consistency_trust == 0:
        info_weight += 1
    if positional_trust == 0:
        info_weight += 1

    trust_score_info = (part_ts_info_state + type_consistency_trust +
                        positional_trust) / (3 + info_weight)

    # Combination
    # final balancing and calculation of trust score
    # split up calualtion just for code beauty
    # Sum it all up!
    trust_score = (trust_score_context + trust_score_role +
                   trust_score_identity + trust_score_info)
    # Divide it and incorporate the weight
    trust_score = trust_score / (4 + info_weight + identity_weight)

    # result list containing the overall score and each individual score
    result = [trust_score, trust_score_context, trust_score_role,
              trust_score_identity, trust_score_experience,
              trust_score_info]
    return result

# Reputation System ens here!


# Consumer start here!
def msg_consume_until_empty(name, env, in_pipe):
    """A process which consumes messages until no message are available."""
    while True:
        # Get event for message pipe
        msg = yield in_pipe.get()

        # consume it
        print('at time %d: %s received message: %s.' %
              (env.now, name, msg.gen_id + msg.msg_str))

        result = [name, msg.time, msg.gen_id, msg.meta_l1,
                  msg.meta_l1_sig_str, msg.meta_l2_type, msg.meta_l2_id,
                  msg.meta_l3, msg.info_role, msg.info_id,
                  msg.info_position, msg.info_type, msg.info_state,
                  msg.msg_str, msg.msg_trust]
        trust_list = calulate_trust_score(msg)
```

```
        result.extend(trust_list)
        result.extend([RANDOM_SEED])
        RESULTS.append(result)
        if msg.gen_id == 'Generator Random':
            RESULTS_RND_GEN.append(result)
        if msg.gen_id == 'Generator Random Good':
            RESULTS_GOOD_GEN.append(result)
        if msg.gen_id == 'Generator Random Bad':
            RESULTS_BAD_GEN.append(result)
        if msg.gen_id == 'Generator Random Fifty':
            RESULTS_RND_GEN_FIFTY.append(result)
# Consumer ends here!


# Helper Function to record simulation results
def write_results(path, dt, rt, results):
    """
    Function for writing results no matter what result type it is.
    """
    with open(path + '/result-' + dt + '-' + rt + '.csv', 'w',
              newline='') as csvfile:
        writer = csv.writer(csvfile, csv.unix_dialect)
        csv_header = ['Consumer', 'Time', 'Generator', 'L1',
                      'L1 Signal Strength', 'L2 Type', 'L2 ID', 'L3',
                      'Role', 'ID', 'Position', 'Type', 'State',
                      'Message', 'Message Trust', 'Trust Score',
                      'TS Context', 'TS Role', 'TS ID', 'TS EXP',
                      'TS Info', 'Random Seed']
        writer.writerow(csv_header)
        writer.writerows(results)

# Setup and start the simulation
print('MONITOR Simulation Setup. Start...')
random.seed(RANDOM_SEED)
env = simpy.Environment()
bc_pipe = BroadcastPipe(env)

env.process(msg_gen_random('Generator Random', env, bc_pipe))
env.process(msg_gen_random_good('Generator Random Good', env, bc_pipe))
env.process(msg_gen_random_bad('Generator Random Bad', env, bc_pipe))
env.process(msg_gen_random_fifty_fifty('Generator Random Fifty',
                                       env, bc_pipe))
# env.process(msg_gen_all_msg('Generator All Msg', env, bc_pipe))
env.process(msg_consume_until_empty('Consumer RepSys',
                                    env, bc_pipe.get_msg_out_conn()))

print('MONITOR Simulation Setup. Done.')
print('Run Simulation...')
env.run(until=SIM_TIME)
print('Done.')
# Reporting aka. Logging aka. Dumping Simulation Data
print('Record Simultion Reults...')
dt = datetime.now()
dt = dt.strftime("%Y%m%d%H%M")
```

```
# create result directory
results_dir = './results-' + dt
os.mkdir(results_dir)

write_results(results_dir, dt, 'all', RESULTS)
write_results(results_dir, dt, 'rnd', RESULTS_RND_GEN)
write_results(results_dir, dt, 'good', RESULTS_GOOD_GEN)
write_results(results_dir, dt, 'bad', RESULTS_BAD_GEN)
write_results(results_dir, dt, 'rnd-fifty', RESULTS_RND_GEN_FIFTY)
print('Done. Thank You! Bye.')
```

# Abbildungsverzeichnis

# Tabellenverzeichnis

# Quellenverzeichnis